\documentclass[11pt,letterpaper]{article}
\usepackage[letterpaper,margin=1in]{geometry}
\usepackage[T1]{fontenc}
\usepackage{lmodern}
\usepackage{amsmath,amssymb,amsthm}
\usepackage{booktabs,array}
\usepackage{tikz}
\usetikzlibrary{decorations.pathreplacing,calc,shapes.misc}
\usepackage{array}
\usepackage[hidelinks]{hyperref}
\usepackage[capitalise,noabbrev]{cleveref}
\usepackage{thmtools}

\newif\ifanon
\ifdefined\anonbuild\anontrue\fi
\newcommand{\AN}[2]{\ifanon#1\else#2\fi}
\newcommand{\citeA}{\ifanon\cite{anon2026bpe}\else\cite{baliga2026bpe}\fi}
\newcommand{\citeAat}[1]{\ifanon\cite[#1]{anon2026bpe}\else\cite[#1]{baliga2026bpe}\fi}
\newcommand{\prevRC}{the earlier short-token preprint}
\newcommand{\prevOV}{the earlier overlap-count preprint}

\newcommand{\CkCross}{(2.292256,0.442317)}
\newcommand{\CkCurve}{(0.301030,0.500000) (0.477121,0.481481) (0.602060,0.468750) (0.698970,0.461120) (0.778151,0.456104) (0.845098,0.452568) (0.903090,0.449944) (0.954243,0.447920) (1.000000,0.446313) (1.041393,0.445006) (1.079181,0.443922) (1.113943,0.443008) (1.146128,0.442228) (1.176091,0.441554) (1.230449,0.440449) (1.255273,0.439990) (1.301030,0.439212) (1.322219,0.438879) (1.361728,0.438302) (1.397940,0.437818) (1.431364,0.437407) (1.462398,0.437053) (1.491362,0.436745) (1.531479,0.436352) (1.568202,0.436023) (1.602060,0.435744) (1.633468,0.435504) (1.672098,0.435232) (1.698970,0.435057) (1.732394,0.434854) (1.770852,0.434639) (1.806180,0.434458) (1.838849,0.434303) (1.875061,0.434145) (1.908485,0.434010) (1.939519,0.433894) (1.973128,0.433777) (2.008600,0.433664) (2.045323,0.433555) (2.079181,0.433463) (2.110590,0.433384) (2.146128,0.433301) (2.178977,0.433231) (2.214844,0.433159) (2.247973,0.433098) (2.283301,0.433038) (2.315970,0.432987) (2.350248,0.432937) (2.385606,0.432890) (2.419956,0.432848) (2.453318,0.432809) (2.487138,0.432774) (2.521138,0.432740) (2.556303,0.432709) (2.589950,0.432681) (2.624282,0.432654) (2.658011,0.432630) (2.692847,0.432607) (2.726727,0.432586) (2.761176,0.432567) (2.795185,0.432549) (2.829304,0.432533) (2.863323,0.432518) (2.897627,0.432504) (2.931458,0.432491) (2.965672,0.432479) (3.000000,0.432468)}
\newcommand{\CkDots}{0.301030/0.500000,0.477121/0.481481,0.602060/0.468750,0.778151/0.456104,1.000000/0.446313,1.146128/0.442228,1.477121/0.436894,2.000000/0.433690,2.292256/0.433024}
\newcommand{\CkFourteen}{0.442228}
\newcommand{\CkThirteen}{0.443008}
\newcommand{\CollMu}{1}
\newcommand{\CollOPT}{3}
\newcommand{\CollOPTfresh}{2}
\newcommand{\CtrlPartialHits}{35}
\newcommand{\CtrlPartialN}{40{,}000}
\newcommand{\CtrlTightHits}{45}
\newcommand{\CtrlTightN}{30{,}000}
\newcommand{\DecMu}{1}
\newcommand{\DecOPT}{4}
\newcommand{\DecOPTfresh}{3}
\newcommand{\MOneF}{1{,}000{,}000{,}000}
\newcommand{\MOneG}{44{,}210{,}475{,}533}
\newcommand{\MOneGuards}{56{,}955{,}365{,}441}
\newcommand{\MOneLength}{258{,}121{,}206{,}416}
\newcommand{\MOneRatio}{44210475533/100000000000}
\newcommand{\MOneRatioDec}{0.44210475533}
\newcommand{\MOneTies}{19}
\newcommand{\MonoAfter}{2}
\newcommand{\MonoBefore}{1}
\newcommand{\NonUDInputs}{7{,}135}
\newcommand{\NonUDMinRatio}{2/3}
\newcommand{\NonUDRatioChecks}{15{,}086}
\newcommand{\NonUDShadowChecks}{90{,}516}
\newcommand{\NonUDShadowViol}{1{,}392}

\newcommand{\OvAudBreaks}{1{,}775}

\newcommand{\OvFourC}{0.404069}
\newcommand{\OvFourCapprox}{0.40407}
\newcommand{\OvFourClo}{0.404068}
\newcommand{\OvFullCells}{217}
\newcommand{\OvFullMinima}{100, 100, 50, 69, 45046, 50}
\newcommand{\OvFullRows}{42}
\newcommand{\OvFullSlopes}{110{,}026, 181{,}019, 305{,}228, 110{,}457, 55{,}013, 500{,}000}

\newcommand{\OvHandCells}{32}
\newcommand{\OvHandCoeffs}{768}
\newcommand{\OvHandMinMargin}{2}
\newcommand{\OvHandOneSlacks}{13{,}155, 13{,}155, 7{,}555, 26{,}082{,}155, 57{,}293{,}796, 7{,}555}
\newcommand{\OvHandRows}{33}
\newcommand{\OvHandSlopes}{108{,}407, 188{,}716, 305{,}523, 111{,}046, 54{,}203, 500{,}000}
\newcommand{\OvInvSixThousand}{1.667\cdot10^{-4}}

\newcommand{\OvLBfull}{19477143/50000000}
\newcommand{\OvLBfullDec}{0.38954286}
\newcommand{\OvLBfullShort}{0.38954}
\newcommand{\OvLBhand}{194477/500000}
\newcommand{\OvLBhandDec}{0.388954}

\newcommand{\OvPsiFiveHalves}{1.590\cdot10^{-4}}
\newcommand{\OvPsiThirteenFifths}{-3.59\cdot10^{-5}}
\newcommand{\OvPsiTwelveFifths}{-4.65\cdot10^{-5}}

\newcommand{\OvRelaxGain}{0.398707}
\newcommand{\OvRelaxMass}{0.3806}
\newcommand{\OvRelaxRows}{75}

\newcommand{\OvTwoRowDec}{0.3786}
\newcommand{\OvTwoRowMinima}{$0$, $1/1750$, $1/700$, $23/625$, $9/560$, $9/560$}

\newcommand{\RateCharges}{158{,}770}
\newcommand{\RateFlat}{4{,}511}
\newcommand{\RateInst}{36{,}522}
\newcommand{\RateIntervals}{181{,}797}
\newcommand{\RateMutants}{5}
\newcommand{\RatioCols}{5}
\newcommand{\RatioKhead}{$k=2$ & $k=3$ & $k=4$ & $k=5$ & $k=6$}

\newcommand{\RatioMinTwo}{3/4}

\newcommand{\SidInputs}{1{,}453}
\newcommand{\SidRatio}{114{,}465}
\newcommand{\SidShadow}{460{,}874}

\newcommand{\TabOvFull}{\texttt{S} & $1.15$ & $1.19$ & $1$ & $1$ & 200{,}291 & \texttt{C} & $1.53$ & $1.63$ & $1$ & $1$ & 886{,}978 \\
\texttt{C} & $1.21$ & $1.23$ & $1$ & $1$ & 808{,}856 & \texttt{J} & $1.53$ & $1.63$ & $1$ & $1.125$ & 992{,}247 \\
\texttt{H} & $1.21$ & $1.23$ & $1$ & $1.025$ & 1{,}130{,}659 & \texttt{C} & $1.63$ & $1.76$ & $1$ & $1$ & 1{,}020{,}595 \\
\texttt{J} & $1.21$ & $1.23$ & $1$ & $1.025$ & 845{,}605 & \texttt{J} & $1.63$ & $1.76$ & $1$ & $1.125$ & 1{,}147{,}035 \\
\texttt{J} & $1.21$ & $1.23$ & $1$ & $1.05$ & 1{,}829{,}099 & \texttt{C} & $1.76$ & $1.9$ & $1$ & $1$ & 949{,}739 \\
\texttt{H} & $1.23$ & $1.25$ & $1$ & $1.025$ & 6{,}842{,}627 & \texttt{J} & $1.76$ & $1.9$ & $1$ & $1.125$ & 1{,}046{,}752 \\
\texttt{H} & $1.23$ & $1.25$ & $1$ & $1.05$ & 1{,}121{,}167 & \texttt{C} & $1.9$ & $2.05$ & $1$ & $1$ & 910{,}761 \\
\texttt{C} & $1.25$ & $1.275$ & $1$ & $1$ & 390{,}300 & \texttt{J} & $1.9$ & $2.05$ & $1$ & $1.125$ & 1{,}089{,}658 \\
\texttt{H} & $1.25$ & $1.275$ & $1$ & $1.075$ & 1{,}258{,}185 & \texttt{C} & $2.05$ & $2.2$ & $1$ & $1$ & 828{,}625 \\
\texttt{C} & $1.275$ & $1.3$ & $1$ & $1$ & 256{,}134 & \texttt{J} & $2.05$ & $2.2$ & $1$ & $1.125$ & 1{,}080{,}169 \\
\texttt{H} & $1.275$ & $1.3$ & $1$ & $1.075$ & 27{,}476 & \texttt{C} & $2.2$ & $2.4$ & $1$ & $1$ & 949{,}476 \\
\texttt{C} & $1.3$ & $1.32$ & $1$ & $1$ & 192{,}341 & \texttt{J} & $2.2$ & $2.4$ & $1$ & $1.15$ & 476{,}712 \\
\texttt{C} & $1.32$ & $1.35$ & $1$ & $1$ & 556{,}949 & \texttt{J} & $2.2$ & $2.4$ & $1.075$ & $1.1$ & 104{,}173 \\
\texttt{H} & $1.32$ & $1.35$ & $1$ & $1.075$ & 2{,}206{,}808 & \texttt{J} & $2.2$ & $2.4$ & $1.15$ & $1.025$ & 729{,}830 \\
\texttt{C} & $1.35$ & $1.39$ & $1$ & $1$ & 709{,}300 & \texttt{C} & $2.4$ & $2.58$ & $1$ & $1$ & 731{,}347 \\
\texttt{J} & $1.35$ & $1.39$ & $1$ & $1.1$ & 2{,}447{,}383 & \texttt{J} & $2.4$ & $2.58$ & $1.075$ & $1.1$ & 767{,}832 \\
\texttt{C} & $1.39$ & $1.45$ & $1$ & $1$ & 621{,}339 & \texttt{J} & $2.4$ & $2.58$ & $1.25$ & $1.025$ & 240{,}246 \\
\texttt{J} & $1.39$ & $1.45$ & $1$ & $1.1$ & 347{,}881 & \texttt{C} & $2.58$ & $2.72$ & $1$ & $1$ & 498{,}111 \\
\texttt{J} & $1.39$ & $1.45$ & $1$ & $1.125$ & 271{,}840 & \texttt{J} & $2.58$ & $2.72$ & $1.25$ & $1.025$ & 725{,}105 \\
\texttt{C} & $1.45$ & $1.53$ & $1$ & $1$ & 777{,}940 & \texttt{J} & $2.58$ & $2.72$ & $1.4$ & $1.025$ & 4{,}321 \\
\texttt{J} & $1.45$ & $1.53$ & $1$ & $1.125$ & 836{,}188 & \texttt{V} & $2.72$ & $2.85$ & $1$ & $1$ & 96{,}206 \\}
\newcommand{\TabOvHand}{$1$ & $1.08$ & 9{,}258 &  &  \\
$1.16$ & $1.2$ & 3{,}226 &  &  \\
$1.2$ & $1.22$ & 6{,}999 &  &  \\
$1.22$ & $1.24$ & 4{,}964 &  & \texttt{H}, $1.025$, 3{,}236 \\
$1.24$ & $1.26$ &  & 4{,}266 & \texttt{J}, $1.025$, 25{,}559 \\
$1.26$ & $1.29$ &  & 7{,}141 & \texttt{H}, $1.025$, 44{,}843 \\
$1.29$ & $1.32$ &  & 6{,}894 & \texttt{H}, $1.05$, 32{,}087 \\
$1.32$ & $1.35$ &  & 6{,}489 & \texttt{H}, $1.075$, 28{,}223 \\
$1.35$ & $1.4$ &  & 7{,}441 & \texttt{H}, $1.1$, 19{,}819 \\
$1.4$ & $1.49$ &  & 9{,}120 & \texttt{J}, $1.1$, 8{,}887 \\
$1.49$ & $1.6$ &  & 10{,}123 & \texttt{J}, $1.1$, 10{,}976 \\
$1.6$ & $1.74$ &  & 11{,}493 & \texttt{J}, $1.1$, 13{,}460 \\
$1.74$ & $1.9$ &  & 11{,}415 & \texttt{J}, $1.125$, 13{,}786 \\
$1.9$ & $2.08$ &  & 11{,}157 & \texttt{J}, $1.125$, 14{,}103 \\
$2.08$ & $2.28$ &  & 10{,}569 & \texttt{J}, $1.15$, 13{,}565 \\
$2.28$ & $2.49$ &  & 9{,}528 & \texttt{J}, $1.15$, 12{,}603 \\
$2.49$ & $2.68$ & 836 & 6{,}265 & \texttt{J}, $1.15$, 8{,}839 \\
$2.68$ & $2.83$ &  & 1{,}784 &  \\}
\newcommand{\TabOvHandSign}{right end & $1.025$ & $1.05$ & $1.075$ & $1.08$ & $1.1$ & $1.125$ & $1.15$ & $1.16$ & $1.2$ & $1.22$ & $1.24$ & $1.26$ & $1.29$ & $1.32$ & $1.35$ & $1.4$ \\
floor & 37 & 9 & 2 & 5 & 5 & 64 & 63 & 123 & 7 & 9 & 9 & 20 & 12 & 14 & 15 & 19 \\
\midrule
right end & $1.49$ & $1.6$ & $1.74$ & $1.9$ & $2.05$ & $2.08$ & $2.1$ & $2.15$ & $2.2$ & $2.25$ & $2.28$ & $2.3$ & $2.49$ & $2.68$ & $2.83$ & $3$ \\
floor & 19 & 24 & 32 & 43 & 57 & 75 & 75 & 265 & 582 & 400 & 97 & 97 & 123 & 123 & 151 & 12{,}849 \\}
\newcommand{\TabOvRelax}{$\ty{lu}$ & $2.6$ & 405 \\
$\ty{lu}$ & $2$ & 411 \\
$\ty{lu}$ & $1.4$ & 325 \\
$\ty{sn}$ & $1$ & 992 \\
$\ty{sn}$ & $1.23$ & 1{,}268 \\
$\ty{sh}$ & $2$ & 310 \\
$\ty{sh}$ & $1.5$ & 95 \\}
\newcommand{\TabOvShas}{\texttt{certificate.py} & \texttt{d759eb90365772334dc81f9e425a71f6991212dfcbb4b92941a06065802c4d5b} \\
\texttt{hand.json} & \texttt{ca5a8972c5487bd49eadf9f4b25be8385c42d544b16d5793040278081fd366ac} \\
\texttt{weights.json} & \texttt{35d5ac5261703c0fe41815536dc475316084faef09eee9768e6173a20d8b6a69} \\
\texttt{certificate.json} & \texttt{3c6c71db289e8924e4b557ade3e38ca221c5042ab65aefc2027819be3171da9e} \\
\texttt{relaxation.py} & \texttt{df3e08ab3d0752f14bd9b02fe576d27dc785202eb2d4d58b411e41fb2bbc0d46} \\
\texttt{relaxation.json} & \texttt{9c51a446a943c72cbf3a6454a2cb058e6eb0b7b8a9a8fc1d9cd50fe002cf2bd2} \\}

\newcommand{\TabPointwise}{Exhaustive: two letters, length $\le 6$, $\le 4$ rules in $\sigma$, $\le 4$ prefix rules & 7{,}439{,}505 & 0 & 133{,}318 \\
Exhaustive: two letters, length $\le 7$, $\le 3$ rules in $\sigma$, $\le 3$ prefix rules & 2{,}904{,}972 & 0 & 102{,}621 \\
Exhaustive: two letters, length $\le 8$, $\le 3$ rules in $\sigma$, $\le 3$ prefix rules & 12{,}720{,}185 & 0 & 330{,}645 \\
Exhaustive: three letters, length $\le 6$, $\le 3$ rules in $\sigma$, $\le 3$ prefix rules & 9{,}973{,}377 & 0 & 1{,}715{,}270 \\
Planted repeated words with random trees (4 seeds) & 1{,}200{,}000 & 0 & 118{,}533 \\
Random strings and random partial sequences (4 seeds) & 400{,}000 & 0 & 48{,}250 \\
Hand-built hard instances (4 strings) & 340{,}000 & 0 & min.\ slack 0 \\}
\newcommand{\TabRatio}{two & $\le 10$ & 1{,}022 & $2$--$6$ & 0 & 0 & $3/4$ & $2/3$ & $5/7$ & $7/9$ & $7/8$ \\
three & $\le 9$ & 4{,}924 & $2$--$6$ & 0 & 0 & $3/4$ & $2/3$ & $5/7$ & $3/4$ & $7/8$ \\
three & $\le 10$ & 14{,}766 & $2$--$6$ & 0 & 0 & $3/4$ & $2/3$ & $5/7$ & $3/4$ & $7/9$ \\
four & $\le 8$ & 3{,}770 & $2$--$5$ & 0 & 0 & $3/4$ & $2/3$ & $5/7$ & $6/7$ & -- \\}
\newcommand{\TabUpper}{2 & 3 & 4 & 55 & 11 & 4 & 12 & 12 & $1$ \\
3 & 5 & 9 & 170 & 23 & 216 & 36 & 45 & $4/5$ \\
4 & 10 & 16 & 535 & 31 & 331{,}776 & 104 & 160 & $13/20$ \\
5 & 13 & 25 & 1{,}104 & 37 & 24{,}883{,}200{,}000 & 210 & 325 & $42/65$ \\
14 & 50{,}000 & 196 & 25{,}071{,}451 & 143 & all & 4{,}334{,}708 & 9{,}800{,}000 & $154811/350000$ \\}
\newcommand{\UpAlphabet}{143}
\newcommand{\UpBlocks}{3571,\allowbreak 3061,\allowbreak 2624,\allowbreak 2249,\allowbreak 1927,\allowbreak 1652,\allowbreak 1416,\allowbreak 1214,\allowbreak 1040,\allowbreak 892,\allowbreak 764,\allowbreak 655,\allowbreak 562,\allowbreak 481}
\newcommand{\UpCap}{6{,}746}
\newcommand{\UpCutoff}{49{,}805}
\newcommand{\UpD}{14}
\newcommand{\UpDD}{390{,}488}
\newcommand{\UpF}{50{,}000}
\newcommand{\UpG}{4{,}334{,}708}
\newcommand{\UpGains}{50{,}001,\allowbreak 42{,}859,\allowbreak 36{,}737,\allowbreak 31{,}489,\allowbreak 26{,}991,\allowbreak 23{,}137,\allowbreak 19{,}833,\allowbreak 17{,}001,\allowbreak 14{,}573,\allowbreak 12{,}493,\allowbreak 10{,}709,\allowbreak 9{,}181,\allowbreak 7{,}871,\allowbreak 6{,}747}
\newcommand{\UpIncMid}{99{,}988}
\newcommand{\UpIncSide}{7{,}142}
\newcommand{\UpJ}{115}
\newcommand{\UpK}{196}
\newcommand{\UpLength}{25{,}071{,}451}
\newcommand{\UpOPT}{9{,}800{,}000}
\newcommand{\UpRatio}{154811/350000}
\newcommand{\UpRatioDec}{0.4423}
\newcommand{\UpResidual}{5{,}784}
\newcommand{\UpShaShort}{e9f1a00bac1edcb2}
\newcommand{\UpTrigram}{3{,}571}
\newcommand{\WtEncN}{1{,}744}
\newcommand{\WtGridBN}{3{,}915}
\newcommand{\WtGridN}{5{,}470}
\newcommand{\WtLetters}{28}
\newcommand{\WtMinD}{$1$, $3/4$, $4/5$}
\newcommand{\WtMinR}{$1/2$, $3/5$, $5/7$}

\newcommand{\WtWordTypes}{3{,}136}

\declaretheorem[name=Theorem]{theorem}
\declaretheorem[name=Lemma,sibling=theorem]{lemma}
\declaretheorem[name=Corollary,sibling=theorem]{corollary}
\declaretheorem[name=Proposition,sibling=theorem]{proposition}
\declaretheorem[name=Definition,sibling=theorem,style=definition]{definition}
\declaretheorem[name=Remark,sibling=theorem,style=remark]{remark}

\newcommand{\OPT}{\mathrm{OPT}}
\newcommand{\BPE}{\mathrm{BPE}}
\newcommand{\cnt}{\mathrm{cnt}}
\newcommand{\raw}{\mathrm{raw}}
\newcommand{\jn}{\mathrm{jn}}
\newcommand{\str}[1]{\texttt{#1}}
\newcommand{\tok}[1]{[\texttt{#1}]}
\newcommand{\cc}{c_\infty}
\newcommand{\Fo}{\mathcal F}
\newcommand{\Uside}{U_{\mathrm{side}}}
\newcommand{\Nmix}{N^{\mathrm{mix}}}
\newcommand{\Nlong}{N^{\mathrm{long}}}
\newcommand{\Lx}{\mathcal L}
\newcommand{\Mx}{\mathcal M}
\newcommand{\gb}{\bar\gamma}
\newcommand{\ty}[1]{\mathsf{#1}}
\newcommand{\kd}[1]{\texttt{#1}}

\hypersetup{pdftitle={Greedy BPE with Replacement Counts Saves at Least (1-(1-2/k)\textasciicircum k)/2 of the Optimum, Which Is Tight as k Grows}}

\title{Greedy BPE with Replacement Counts Saves at Least\\ $\frac12\bigl(1-(1-\frac2k)^k\bigr)$ of the Optimum, Which Is Tight
as $k\to\infty$}
\ifanon
\author{Anonymous submission}
\else
\author{Kundan Baliga\\{\small University of Illinois Urbana-Champaign}\\{\small\texttt{kbaliga2@illinois.edu}}
\and Aiden Xie\\{\small Georgia Institute of Technology}\\{\small\texttt{axie82@gatech.edu}}
\and Aditya Byju\\{\small Georgia Institute of Technology}\\{\small\texttt{abyju6@gatech.edu}}}
\hypersetup{pdfauthor={Kundan Baliga, Aiden Xie, Aditya Byju}}
\fi
\date{}

\begin{document}
\maketitle

\begin{abstract}
Byte-Pair Encoding (BPE) builds the tokenisers of language models by repeatedly merging a most frequent adjacent
pair of symbols. Kozma and Voderholzer showed that $k$ greedy merges save, in the worst case, between $1/3$ and
$\frac12(1-e^{-2})\approx0.4323$ of what the best $k$ merges save, and asked for the exact constant. We answer this for
the rule they analyse, which ranks a pair by the replacements its merge makes: on every string and word table, at every
budget $k$ and for every tie-breaking rule, greedy saves at least $\frac12\bigl(1-(1-\frac2k)^k\bigr)$ of the optimum,
so $\frac12(1-e^{-2})$ is the exact worst case, approached as $k\to\infty$ on growing alphabets but never attained. The
proof rests on a shadow inequality: after $i$ steps the optimum is at most $2G_i+k\,g_{i+1}$, where $G_i$ is greedy's
saving so far and $g_{i+1}$ its next gain. A second theorem shows that on strings greedy saves at least
$\frac12(1-e^{-2})+10^{-9}k^{-1/2}$ of the optimum; with our construction, the worst case at budget $k$ exceeds $\frac12(1-e^{-2})$ by
$\Theta(k^{-1/2})$. For the overlapping counts of deployed trainers, every run saves more than $\OvLBfullShort$ of the
optimum (by an exact computer check); the exact constant for that rule is open.
\end{abstract}

\section{Introduction}\label{sec:intro}

Every large language model reads text through a tokeniser, and the vocabularies of many current models are built by
Byte-Pair Encoding (BPE) \cite{gage1994,sennrich2016bpe}. The trainer starts from the text as a sequence of characters
or bytes. It finds a most frequent adjacent pair of symbols, replaces its occurrences by one new symbol, and repeats
until it has made $k$ merges; the new symbols join the vocabulary. Each replacement shortens the training text by one
symbol. The total shortening, the \emph{saving} or \emph{utility}, measures how well the vocabulary compresses the text it
was learned from, and it is the quantity that BPE's greedy choices aim to maximise.

Greedy merges can interfere: a merge made early can destroy occurrences of pairs that a better vocabulary would have
used. How much can this cost? A constant-factor guarantee answers the question for every corpus at once. If greedy always
saves at least a fraction $\alpha$ of what the best $k$ merges save, then no input, however adversarial, makes the
trainer lose more than a fraction $1-\alpha$ of the achievable compression; and an input on which greedy saves only
$\alpha$ shows that no better guarantee exists.

\paragraph{The question.} Zouhar et al.~\cite{zouhar2023formal} began the formal study of BPE as an optimisation problem,
and Kozma and Voderholzer \cite{kozma2026esa,kozma2026arxiv} made the question precise. They compare greedy with the
best sequence of $k$ \emph{partial} merges, each of which may replace any set of non-overlapping occurrences of one pair.
Write $\BPE(s,k)$ for the utility of greedy on a string $s$ with budget $k$, and $\OPT(s,k)$ for this optimum. Computing
$\OPT$ is APX-complete \cite[Theorem~1]{kozma2026esa}, \cite{kozma2026arxiv}, so approximation is the natural lens, and the question is the value
of
\[
\rho\;=\;\inf\frac{\BPE(s,k)}{\OPT(s,k)},
\]
the infimum taken over strings $s$, budgets $k$ with $\OPT(s,k)>0$, and the tie-breaking rule of greedy. We write $\rho_k$
for the same infimum at a fixed budget $k$, and $\cc=\frac12(1-e^{-2})\approx0.43233$. Kozma and Voderholzer proved
$\rho\ge1/3$. The extended version of their paper (arXiv:2411.08671v2 \cite{kozma2026arxiv}) constructs strings, with
budgets that grow without bound, on which greedy gets at most $\cc+\varepsilon$ of the optimum; ``we need no assumption
on tie-breaking'' \cite[Appendix~B]{kozma2026arxiv}. Their Theorem~2 reads $0.3333\le\alpha\le0.4324$ ($0.4324$ is $\cc$
rounded up), and they add: ``Closing this gap is an intriguing open question.'' \cite{kozma2026arxiv}. Independently,
Yang \cite{yang2026} gave another family with the same limit, and proved the matching lower bound against optima whose
tokens have at most three letters \cite[Theorem~1 and Corollary~1]{yang2026}. Kozma and Voderholzer also give a different
algorithm, EvenOdd, with ratio $\frac12$ for the partial-merge problem \cite[Theorem~4]{kozma2026arxiv}; it builds only
two-letter tokens, and the question here is about greedy itself (\cref{sec:related}).

\paragraph{Main result.} We close the gap for the rule that Kozma and Voderholzer analyse: greedy that ranks a pair by the
number of replacements its merge makes. Write
\[
c_k=\tfrac12\bigl(1-(1-\tfrac2k)^k\bigr),\qquad\text{so } c_1=1,\ c_2=\tfrac12,\ c_3=\tfrac{13}{27},\ \text{and } c_k
\text{ decreases to }\cc .
\]

\medskip\noindent\begin{center}
\fbox{\begin{minipage}{0.95\textwidth}
\textbf{Main theorem} (\cref{thm:main,thm:table,thm:exact}). For every string and every word table with positive real
weights, every budget $k$ and every greedy run with any tie-breaking,
\[
\BPE\;\ge\;c_k\cdot\OPT .
\]
Since $c_k>\cc$ for $k\ge2$ and matching inputs exist, the exact worst case is $\rho=\cc$, approached only as
$k\to\infty$ and never attained.

\smallskip
\textbf{Shadow inequality} (\cref{cor:rc}). For every greedy run and every $i\ge0$,
\[
\OPT\;\le\;2G_i+k\,g_{i+1},
\]
where $G_i$ is the saving of greedy's first $i$ merges and $g_{i+1}$ the saving of the next one.

\smallskip
\textbf{The finite-budget gap} (\cref{thm:rate,cor:rate}). For every string, every budget $k\ge2$ and every greedy run,
$\BPE\ge(\cc+10^{-9}k^{-1/2})\OPT$. With the construction of \cref{thm:upper}, the worst case at budget $k$ is
$\rho_k=\cc+\Theta(k^{-1/2})$, while $c_k-\cc$ is only of order $1/k$. The content is the order: for $k\le10^{16}$ the
bound $c_k$ of the main theorem is the larger one.
\end{minipage}}
\end{center}
\medskip

The shadow inequality is the heart of the paper, and the main theorem follows from it in three lines. No assumption is
made on the alphabet, on the lengths of the optimal tokens or on tie-breaking.

The second main result concerns finite budgets. The guarantee $c_k$ exceeds $\cc$ by about $e^{-2}/k$, but the
near-worst inputs of \cref{thm:upper} at budget $k=d^2$ have ratio about $c_d$, which exceeds $\cc$ by about $e^{-2}/\sqrt k$. \Cref{thm:rate}
shows that the construction has the true order: on strings, every greedy run saves at least $(\cc+10^{-9}k^{-1/2})\OPT$.
So $\rho_k-\cc=\Theta(k^{-1/2})$. This bound is asymptotic: for $k\le10^{16}$ the guarantee $c_k$ is larger. For
$k\ge4\cdot10^{10}$ the proof gives the stronger bound $(c_k+1.5\cdot10^{-9}k^{-1/2})\OPT$, so $c_k$ is not tight at any
such budget (\cref{cor:rate}). The constant $10^{-9}$ is far from the upper constant, which is about $e^{-2}$ on square budgets; we have not tried
to optimise it. The proof follows the windows of the shadow argument over short intervals of steps: when greedy's gains fall, the
windows that shared pairs must lose them, and few greedy steps can destroy many windows at once.

\Cref{tab:results} summarises all results. The secondary ones are the following. The constant is also exact on word tables, the weighted lists of distinct words that
real trainers process (\cref{thm:table}). It survives when merged symbols are identified by their spelling, as in
subword-nmt and Hugging Face, provided the input is uniquely decodable (\cref{prop:identity}). The matching inputs need
only $\Theta(\sqrt k)$ letters, and at $k=\UpK$ an exact certificate on a byte alphabet gives ratio $\UpRatio$ (\cref{sec:upper}).
Finally, the trainers rank pairs by overlapping counts, not by replacements; for that rule we prove a bracket below
$\cc$ but leave the exact constant open (\cref{sec:overlap}).

\paragraph{What this means for deployed trainers.} subword-nmt and Hugging Face \cite{sennrich2016bpe,subwordnmt,hftokenizers}
rank a pair by its occurrences counted with overlaps. If no tokenisation along a run has three equal adjacent tokens,
the two rules score every pair alike, the run is a replacement-count run, and the main theorem applies to it as it is
(\cref{sec:limits}). Runs of three equal symbols, such as three dots, spaces or zeros, are common in real text, so this
condition can fail; then, for the deployed rule, only the bracket of \cref{sec:overlap} is proved. All guarantees
assume that the trainer takes all $k$ merges; a trainer that stops at a minimum pair frequency, as subword-nmt does by
default, keeps only an additive guarantee (\cref{sec:limits}). For symbols identified by spelling, as in these trainers, every guarantee here assumes that the corpus is
uniquely decodable (\cref{prop:identity}), which fails, for example, if the text contains the trainer's end-of-word
marker. The comparison is with the best $k$ partial merges, and every lower bound holds equally
against the best $k$ full merges, a vocabulary that a trainer could actually deploy. What is measured is the compression
of the training corpus itself, not the encoding of new text or the quality of a model. Finally, our near-worst inputs use
alphabets that grow with $k$. As word tables they fit a byte alphabet only for budgets below $129^2=16{,}641$, and even
there the proved bound on their ratio, $\cc+O(1/\sqrt K)$ (\cref{thm:table}(b)), approaches $\cc$ slowly; at byte-level
budgets of the size used for large models it is open how close to $\cc$ greedy can be pushed. The guarantee $c_k$ itself holds at every budget and on every alphabet.

\begin{table}[t]
\centering\small
\setlength{\tabcolsep}{4pt}
\renewcommand{\arraystretch}{1.15}
\begin{tabular}{@{}>{\raggedright}p{2.75cm}>{\raggedright}p{2.95cm}>{\raggedright}p{3.2cm}>{\raggedright}p{1.6cm}>{\raggedright}p{2.25cm}>{\raggedright\arraybackslash}p{2.25cm}@{}}
\toprule
Rule, input & Lower bound & Upper bound & Status & Where & How\\
\midrule
Replacement counts, strings & $c_k\cdot\OPT$ at every $k$, so ratio $>\cc$ & ratio $\le\cc+\varepsilon$ for large $k$
\cite{kozma2026arxiv}, also with $\Theta(\sqrt k)$ letters; $\UpRatio\approx\UpRatioDec$ at $k=\UpK$ & exact: $\cc$, not
attained & \cref{thm:main,thm:upper,thm:exact} & by hand; $k=\UpK$ numbers by exact arithmetic\\
Replacement counts, strings, budget $k$ & $(\cc+10^{-9}k^{-1/2})\OPT$ & ratio $\le\cc+8k^{-1/2}$ at every $k\ge2$ &
$\rho_k-\cc=\Theta(k^{-1/2})$; constant open & \cref{thm:rate,cor:rate} & by hand\\
Replacement counts, word tables & $c_k\cdot\OPT$ at every $k$ & $\cc+O(1/\sqrt K)$ at every budget $K\ge64$, over
$2\lfloor\sqrt K\rfloor$ letters (within bytes only for $K<16{,}641$) & exact: $\cc$, not attained & \cref{thm:table} & by hand\\
Replacement counts, symbols identified by spelling & as for strings and tables (\cref{thm:rate}: strings only), on uniquely decodable inputs & as for
strings and tables & exact: $\cc$, not attained & \cref{prop:identity} & by hand\\
\addlinespace
Overlap counts (deployed trainers) & $\lambda_*\cdot\OPT$ with $\lambda_*>\OvLBfullShort$, at every $k$; $\frac25\OPT$ if
optimal tokens have $\le3$ letters & $\limsup$ of the ratio $\le\Lambda_4\approx\OvFourCapprox<\cc$, with $\Theta(k)$ letters & open &
\cref{prop:overlap} & $\lambda_*$ by an exact computer check; $\frac7{18}$ by a finite calculation, checkable by hand in principle\\
\bottomrule
\end{tabular}
\caption{Results. ``Ratio'' is the worst case of $\BPE/\OPT$ over inputs and tie-breaking. All rows cover every
tie-breaking rule; the overlap-count row holds for strings, word tables and, on uniquely decodable inputs,
string-identified symbols. ``By hand'' means that the proof can be checked on paper. Two entries use exact
arithmetic by computer: the numbers of the certificate at $k=\UpK$, and the finite sign check behind $\lambda_*$, whose
$\OvFullRows$ rows are printed in \cref{app:parta}. No proof uses a computer search (\cref{app:evidence}).}
\label{tab:results}
\end{table}

\paragraph{Relation to earlier preprints.} \AN{Four unrefereed 2026 preprints treat the same question.
\Cref{tab:companions} lists the results of three of them, \citeA{}, a preprint on optima with short tokens and a preprint on overlap counts (the latter two are superseded by this paper and no longer distributed), and
their status here. The fourth, \cite{prior2026fixed}, concerns GreedTok and fixed alphabets; it
also states the isolation property of \cref{lem:iso}.}{This paper supersedes three unrefereed 2026 preprints by the same group: \citeA{}, a preprint on optima with short tokens and a preprint on overlap counts (the latter two are no longer distributed). \Cref{tab:companions} lists their results as proved, implied, improved or reproved here, and this paper is
intended as their refereed venue of record; the results of \citeA{} marked there as cited remain available in that preprint, and one overlap-count result, the last row of the table, is not included here. A fourth preprint by the
present authors, \cite{prior2026fixed}, on GreedTok and fixed alphabets, remains a separate companion; it also states the isolation property of
\cref{lem:iso}.} New in this paper are the
general proof of the shadow inequality (\cref{sec:proof}), by a method different from the partial proofs of
\prevRC{} and without a computer; the finite-budget lower bound of \cref{thm:rate}; the construction with $\Theta(\sqrt k)$ letters and the certificate at
$k=\UpK$; the word-table results; \cref{prop:identity}; and all three parts of \cref{prop:overlap}. Against optima whose
tokens have at most three letters, the bound $c_k\OPT$ is due to Yang \cite[Theorem~1]{yang2026} (stated there for one tie rule; the proof covers all); the new content of
\cref{thm:main} is that no restriction on the optimum is needed. No theorem proved here depends on these preprints.
We cite without reproof only $\rho_2=3/4$ and $\rho_3\le7/12$ from \citeA{} and the fixed-alphabet theorem of
\cite{prior2026fixed}.

\begin{table}[t]
\centering\small
\renewcommand{\arraystretch}{1.12}
\begin{tabular}{@{}>{\raggedright}p{7.4cm}>{\raggedright}p{2.6cm}>{\raggedright\arraybackslash}p{5.4cm}@{}}
\toprule
Result & First appeared & Status here\\
\midrule
The shadow inequality, as a conjecture, with the recurrence and the constant $c_k$ & short-token preprint &
proved in general (\cref{cor:rc,thm:main})\\
The shadow inequality when some optimal sequence has tokens of at most seven letters, and for $i\le3$, partly by
computer; $\rho_3\ge13/27$, $\rho_4\ge15/32$ & short-token preprint & implied by \cref{cor:rc,thm:main}, by hand\\
$\rho\ge0.391$, and the one-third barrier for pair-packing analyses & \citeA{} & bound improved to $\cc$
(\cref{thm:exact}); barrier not included\\
$\rho_2=3/4$ and $\rho_3\le7/12$ & \citeA{} & cited (\cref{sec:k2,sec:limits}), not reproved\\
Overlap counts: every run saves more than $0.3341104$ of the optimum & overlap-count preprint & improved to
$\lambda_*>\OvLBfullShort$ (\cref{prop:overlap}(a))\\
Overlap counts: the odd-run argument (the tail pair of an odd run survives) & overlap-count preprint &
reproved and used for the marks of \cref{lem:suffix}; the segment bounds of \cref{prop:overlap}(a) use its notions\\
Overlap counts: the exact guarantee of the first two steps & overlap-count preprint & not included\\
\bottomrule
\end{tabular}
\caption{Results of three earlier preprints on the same question and their status in this paper.}
\label{tab:companions}
\end{table}

\paragraph{Roadmap.} \Cref{sec:overview} explains the four main ideas with figures. \Cref{sec:prelim} sets up the model.
\Cref{sec:warmup} works the argument by hand on a two-merge example and shows why simpler arguments fail.
\Cref{sec:proof} proves the shadow inequality and the main theorem, \cref{sec:upper} gives the matching construction,
and \cref{sec:rate} determines the order of the gap at finite budgets.
\Cref{sec:table,sec:overlap} treat word tables, string-identified symbols and overlapping counts, \cref{sec:related}
places the result among analyses of greedy algorithms, and \cref{sec:limits} collects the limitations. The appendices
hold the constructions, the remaining proofs, worked examples, and a description of the released checking scripts.

\section{Technical overview}\label{sec:overview}

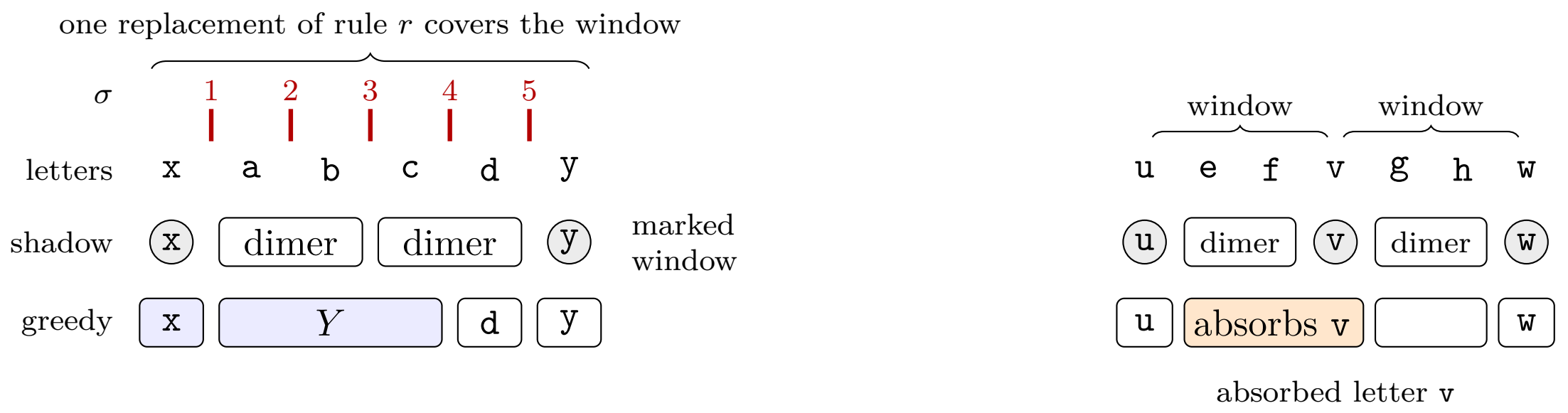
\begin{figure}[t]
\centering
\begin{tikzpicture}[x=0.74cm,y=0.75cm,every node/.style={font=\small}]
  \foreach \i/\l in {1/x,2/a,3/b,4/c,5/d,6/y} {\node at (\i,0) {\str{\l}};}
  \node[anchor=east,font=\scriptsize] at (0.45,0) {letters};
  \node[anchor=east,font=\scriptsize] at (0.45,-0.9) {shadow};
  \draw[fill=gray!15] (1,-0.9) circle (0.27); \node at (1,-0.9) {\str x};
  \draw[rounded corners=2pt] (1.6,-1.2) rectangle (3.4,-0.6); \node at (2.5,-0.9) {dimer};
  \draw[rounded corners=2pt] (3.6,-1.2) rectangle (5.4,-0.6); \node at (4.5,-0.9) {dimer};
  \draw[fill=gray!15] (6,-0.9) circle (0.27); \node at (6,-0.9) {\str y};
  \node[anchor=east,font=\scriptsize] at (0.45,-1.9) {greedy};
  \draw[rounded corners=2pt,fill=blue!8] (0.6,-2.2) rectangle (1.4,-1.6); \node at (1,-1.9) {\str x};
  \draw[rounded corners=2pt,fill=blue!8] (1.6,-2.2) rectangle (4.4,-1.6); \node at (3,-1.9) {$Y$};
  \draw[rounded corners=2pt] (4.6,-2.2) rectangle (5.4,-1.6); \node at (5,-1.9) {\str d};
  \draw[rounded corners=2pt] (5.6,-2.2) rectangle (6.4,-1.6); \node at (6,-1.9) {\str y};
  \node[anchor=east,font=\scriptsize] at (0.45,0.9) {$\sigma$};
  \foreach \c/\n in {1.5/1,2.5/2,3.5/3,4.5/4,5.5/5} {\draw[very thick,red!70!black] (\c,0.35) -- (\c,0.75);
     \node[font=\scriptsize,red!70!black] at (\c,0.98) {\n};}
  \draw[decorate,decoration={brace,amplitude=4pt}] (0.75,1.25) -- (6.25,1.25)
     node[midway,above=4pt,font=\scriptsize,align=center] {one replacement of rule $r$ covers the window};
  \node[font=\scriptsize,align=left,anchor=west] at (6.6,-0.9) {marked\\window};
  \begin{scope}[xshift=9.2cm]
  \foreach \i/\l in {1/u,2/e,3/f,4/v,5/g,6/h,7/w} {\node at (\i*0.8,0) {\str{\l}};}
  \draw[fill=gray!15] (0.8,-0.9) circle (0.27); \node at (0.8,-0.9) {\str u};
  \draw[rounded corners=2pt] (1.3,-1.2) rectangle (2.7,-0.6); \node[font=\scriptsize] at (2.0,-0.9) {dimer};
  \draw[fill=gray!15] (3.2,-0.9) circle (0.27); \node at (3.2,-0.9) {\str v};
  \draw[rounded corners=2pt] (3.7,-1.2) rectangle (5.1,-0.6); \node[font=\scriptsize] at (4.4,-0.9) {dimer};
  \draw[fill=gray!15] (5.6,-0.9) circle (0.27); \node at (5.6,-0.9) {\str w};
  \draw[rounded corners=2pt] (0.45,-2.2) rectangle (1.15,-1.6); \node at (0.8,-1.9) {\str u};
  \draw[rounded corners=2pt,fill=orange!20] (1.3,-2.2) rectangle (3.55,-1.6); \node at (2.4,-1.9) {absorbs \str v};
  \draw[rounded corners=2pt] (3.7,-2.2) rectangle (5.1,-1.6);
  \draw[rounded corners=2pt] (5.25,-2.2) rectangle (5.95,-1.6); \node at (5.6,-1.9) {\str w};
  \draw[decorate,decoration={brace,amplitude=3pt}] (0.9,0.4) -- (3.1,0.4) node[midway,above=2pt,font=\scriptsize] {window};
  \draw[decorate,decoration={brace,amplitude=3pt}] (3.3,0.4) -- (5.5,0.4) node[midway,above=2pt,font=\scriptsize] {window};
  \node[font=\scriptsize,align=center] at (3.2,-2.75) {absorbed letter \str v};
  \end{scope}
\end{tikzpicture}
\caption{The accounting behind $U-\sum_rm_r\le2G_i$. Left: windows run between consecutive single letters of the shadow
(circles). This window has $\nu=2$ dimers and $2\nu+1=5$ gaps, all of them junctions of $\sigma$ (numbered ticks), so it
is marked; its first two greedy tokens $(\str x,Y)$ form a pair shared by all marked windows of the same rule, which pays
for one junction, and the other $2\nu=4$ junctions, two per dimer, are left unpaid. Right: the letter \str v is single in
the shadow but lies inside a longer greedy token; this absorbed letter spoils the two windows it ends and is paid for by
one saved symbol of greedy (\cref{lem:lost}).}
\label{fig:accounting}
\end{figure}

\paragraph{The shadow inequality, and why the factor is $2$.} Fix greedy's tokenisation $s_i$ after $i$ steps, with
saving $G_i$, and an optimal sequence $\sigma$ of $k$ partial merges. Each replacement that $\sigma$ makes joins two
symbols across one boundary, its \emph{junction}; there are $U=\OPT$ junctions. All replacements of one rule $r$ of
$\sigma$ spell the same word $w_r$. Suppose that $m_r$ of them contain, wholly inside, the same adjacent pair $(X,Y)$ of
greedy's current tokens. These $m_r$ occurrences of $XY$ are disjoint, so the next greedy merge makes at least $m_r$
replacements: $m_r\le g_{i+1}$. Hence the shadow inequality follows from
\begin{equation}\label{eq:pointwise}
U-\sum_r m_r\;\le\;2G_i ,
\end{equation}
which says that greedy's saving pays, at two junctions per saved symbol, for every junction that no common pair
witnesses. To find common pairs we compare greedy with its \emph{shadow}: the run that replays, from the start, only
those of greedy's merges that join two letters. The shadow consists of single letters and two-letter \emph{dimers}, and
it saves $M\le G_i$ symbols. A \emph{window} runs from one single letter of the shadow to the next
(\cref{fig:accounting}); consecutive windows share their end letter. A window with $\nu$ dimers has $2\nu+1$ gaps. If all of them are junctions, the window lies
inside one replacement of $\sigma$; we call it \emph{marked} and let it pay for one junction. Either way at most $2\nu$
of its junctions are left unpaid, two per dimer, so the unpaid junctions number at most $2M$. This is where the factor
$2$ comes from, and it cannot be lowered: on $s=\str{aabaab}$ with $k=2$, after the one full merge of $(\str b,\str a)$
(a prefix of full merges, though not a greedy one) we have $G=1$ and $\mu=1$, while merging $\str{ab}$ and then $\str a\tok{ab}$ saves $\OPT=4=2G+k\mu$.

Two things remain. First, a marked window must produce a pair shared by all marked windows of its rule. The shadow is
\emph{rigid}: inside any occurrence of a window's word it has its dimers at the same offsets (\cref{lem:rigid}). So all
marked windows of rule $r$ sit at the same offsets in $w_r$ and spell the same word. By \emph{isolation}
(\cref{lem:iso}), greedy tokenises a stretch between two of its surviving token boundaries exactly as it would
tokenise that stretch alone. So all these windows carry the same greedy tokens, and their first two tokens are a common
pair. Second, this needs the window's end letters to be single letters of greedy too. Greedy's other merges can swallow
a single letter of the shadow; each such \emph{absorbed} letter spoils at most the two windows it ends, and greedy
pays for these letters: their number $L$ is at most $G_i-M$ (\cref{lem:lost}). Altogether
$U-\sum_rm_r\le2M+2(G_i-M)=2G_i$. The argument compares two tokenisations at a fixed prefix and never follows the run step
by step; \cref{sec:warmup} shows why the natural per-replacement charging fails.

\paragraph{From the inequality to $c_k$.} Write $D_i=\OPT-2G_i$ for greedy's deficit after $i$ steps. The shadow
inequality says $g_{i+1}\ge D_i/k$, and $D_{i+1}=D_i-2g_{i+1}$, so
\[
D_{i+1}\;\le\;\Bigl(1-\frac2k\Bigr)D_i,\qquad D_k\;\le\;\Bigl(1-\frac2k\Bigr)^k\OPT,\qquad
G_k\;\ge\;\tfrac12\Bigl(1-\bigl(1-\tfrac2k\bigr)^k\Bigr)\OPT=c_k\OPT .
\]
At $k=2$, for example, $g_1\ge\OPT/2$ already, so $G_2\ge G_1\ge\frac12\OPT$. This recurrence is the greedy template for
maximising a set function with curvature $2$ \cite{conforti1984} (\cref{sec:related}). The merges of BPE form a sequence,
not a set, so the template does not apply directly; all of the difficulty lies in the inequality.
\Cref{fig:overview}(a) plots $c_k$.

\paragraph{Near-worst inputs with $\Theta(\sqrt k)$ letters.} Greedy loses when one of its merges destroys two merges
of the optimum. In a block $X_i\,Y_j\,X_{j+t}\,Y_i$ of four distinct letters the optimum merges the two \emph{side} pairs
$X_iY_j$ and $X_{j+t}Y_i$ and saves two symbols, while greedy merges the \emph{middle} pair $Y_jX_{j+t}$, saves one,
and leaves no side pair in the block (\cref{fig:overview}(b)). The construction of \cref{sec:upper} uses this mechanism
of Kozma and Voderholzer and of Yang. The optimum merges the $k=d^2$ side pairs, each with $F$ occurrences. Greedy
spends $d$ phases on the $d$ middle pairs of each level $t$. Each middle replacement removes two side occurrences, so the
counts available to later phases shrink by the factor $1-2/d$ per phase, and greedy saves about
$d\sum_{t<d}F(1-\frac2d)^t=c_d\,kF$. Guard letters between the blocks keep greedy inside them. They come from a reusable
palette of $O(\sqrt k)$ letters, chosen so that no pair with a guard is frequent, so the whole string has
$\Theta(\sqrt k)$ letters. Some growth is unavoidable: on a fixed alphabet the ratio tends to $1$ as $k$ grows
\cite{prior2026fixed}.

\paragraph{What changes for overlapping counts.} The trainers rank a pair by its occurrences counted with overlaps; on
\str{xxx} the pair $(\str x,\str x)$ scores $2$ but its merge saves $1$ (\cref{fig:overview}(c)), so the shadow
inequality, which still holds, bounds greedy's next gain only up to a factor $2$. Refunds for letters consumed between
non-junctions, detection of self steps by earlier shadows and an alternating-path argument give linear bounds at every
score level; averaged with explicit weights, they show that every run saves at least $\lambda_*>\OvLBfullShort$ of
the optimum by an exact computer check, and more than $\frac7{18}$ by a finite calculation that is long but checkable by
hand in principle. A family whose blocks \str{xxx} score $2$ but save $1$, over $\Theta(k)$ letters, shows that for
large budgets the worst case is at most $\Lambda_4\approx\OvFourCapprox<\cc$, so on such alphabets the deployed rule is
strictly worse than replacement counts at every large budget (\cref{sec:overlap}).

\begin{figure}[t]
\centering
\begin{minipage}[b]{0.39\textwidth}\centering
\begin{tikzpicture}[x=1.15cm,y=11cm,every node/.style={font=\scriptsize}]
  \draw[->] (0,0.32) -- (3.15,0.32) node[right=1pt] {$k$};
  \draw[->] (0,0.32) -- (0,0.525);
  \foreach \k/\lab in {1/1,10/10,100/100,1000/1000} {\draw ({log10(\k)},0.32) -- ({log10(\k)},0.315) node[below] {\lab};}
  \foreach \y in {0.40,0.45,0.50} {\draw (0,\y) -- (-0.04,\y) node[left] {\y};}
  \draw (0,0.33333) -- (-0.04,0.33333) node[left] {$\frac13$};
  \fill[gray!15] (0,0.33333) rectangle (3.1,0.43233);
  \node[gray!80!black,align=center,font=\tiny] at (1.75,0.375) {gap left open by\\Kozma--Voderholzer};
  \draw[densely dashed] (0,0.43233) -- (3.1,0.43233) node[pos=0.97,above] {$\cc$};
  \draw[thick,blue!70!black] plot coordinates {\CkCurve};
  \foreach \x/\y in \CkDots {\fill[blue!70!black] (\x,\y) circle (0.9pt);}
  \node[blue!70!black,anchor=west] at ({log10(2)+0.05},0.5) {$c_2=\frac12$};
  \node[blue!70!black] at (1.25,0.465) {$c_k$};
  \draw \CkCross node[cross out,draw,inner sep=1.4pt] {} node[above=1pt] {$k=\UpK$};
\end{tikzpicture}
\par\smallskip (a) the guarantee $c_k$
\end{minipage}\hfill
\begin{minipage}[b]{0.33\textwidth}\centering
\begin{tikzpicture}[x=0.95cm,y=0.75cm,every node/.style={font=\small}]
  \foreach \i/\l in {1/{X_i},2/{Y_j},3/{X_{j+t}},4/{Y_i}} {\node at (\i,0) {$\l$};}
  \draw[thick,blue!70!black] (0.75,0.35) -- (0.75,0.6) -- (2.25,0.6) -- (2.25,0.35);
  \draw[thick,blue!70!black] (2.75,0.35) -- (2.75,0.6) -- (4.25,0.6) -- (4.25,0.35);
  \node[blue!70!black,font=\scriptsize] at (2.5,1.0) {optimum: two side merges, saves $2$};
  \draw[thick,orange!80!black] (1.75,-0.35) -- (1.75,-0.6) -- (3.25,-0.6) -- (3.25,-0.35);
  \node[orange!80!black,font=\scriptsize,align=center] at (2.5,-1.2) {greedy: middle merge, saves $1$,\\no side pair left};
\end{tikzpicture}
\par\smallskip (b) one block of the construction
\end{minipage}\hfill
\begin{minipage}[b]{0.24\textwidth}\centering
\begin{tikzpicture}[x=0.75cm,y=0.75cm,every node/.style={font=\small}]
  \foreach \i in {1,2,3} {\node at (\i,0) {\str x};}
  \draw[thick] (1,0.3) to[bend left=50] node[above,font=\scriptsize] {1} (2,0.3);
  \draw[thick] (2,0.3) to[bend left=50] node[above,font=\scriptsize] {2} (3,0.3);
  \node[font=\scriptsize] at (2,1.25) {score $2$};
  \draw[->] (2,-0.35) -- (2,-0.85);
  \node at (1.5,-1.2) {$\tok{xx}$}; \node at (3,-1.2) {\str x};
  \node[font=\scriptsize] at (2,-1.75) {saves $1$};
\end{tikzpicture}
\par\smallskip (c) a self-pair run
\end{minipage}
\caption{(a) The guarantee $c_k$ of the main theorem against the budget (logarithmic scale), with its limit $\cc$; the
shaded band is the gap $[\frac13,\cc]$ between the bounds of Kozma and Voderholzer. The cross is the exact ratio
$\UpRatio$ of the certificate at $k=\UpK$. (b) Greedy's middle merge in a block of the matching construction blocks both
side merges of the optimum. (c) Overlapping counts score the run \str{xxx} twice, but its merge saves one symbol.}
\label{fig:overview}
\end{figure}
\section{Model and notation}\label{sec:prelim}

\paragraph{Strings, cuts and tokens.} A string $s=s[1]\cdots s[n]$ is a sequence of \emph{letters}. A \emph{cut} is an
index $c\in\{0,1,\dots,n\}$, the boundary between positions $c$ and $c+1$; cuts $0$ and $n$ are the \emph{end cuts}. For
cuts $a<b$ we write $(a,b]$ for the positions $a+1,\dots,b$ and $s(a,b]$ for the word they spell. A \emph{tokenisation}
of $s$ partitions the positions into intervals, the \emph{tokens}, each carrying a symbol; a token of one position is a
\emph{single letter}. A cut \emph{survives} if it is a boundary between tokens (end cuts always survive), and is
\emph{erased} otherwise. The \emph{utility} of a tokenisation is its number of erased cuts, $n$ minus the number of tokens.

\paragraph{Merges.} A merge of an ordered pair $(X,Y)$ of symbols replaces occurrences of $X$ immediately followed by $Y$
by a symbol $Z$ that spells the concatenation of the words of $X$ and $Y$. In the proofs $Z$ is a \emph{fresh} symbol,
new at every merge; the string-identified model of the trainers is treated in \cref{sec:identity} and gives the
same results on uniquely decodable inputs. A \emph{full} merge scans the tokens
from left to right and replaces each occurrence whose first token was not used by the previous replacement (so $\str{aaa}$
becomes $Z\str a$). A \emph{partial} merge replaces an arbitrary, possibly empty, set of pairwise non-overlapping
occurrences. Each replacement erases exactly one cut. For a tokenisation $t$, $\cnt_t(X,Y)$ is the number of replacements
made by the full merge of $(X,Y)$, and $\mu(t)=\max_{(X,Y)}\cnt_t(X,Y)$, with $\mu(t)=0$ if $t$ has one token. The
\emph{raw} count $\raw_t(X,Y)$ is the number of adjacent token pairs of $t$ with symbols $X,Y$, overlaps allowed (so
$\raw_{\str{aaa}}(\str a,\str a)=2$); it is used only for overlapping counts.

\begin{lemma}[counts]\label{lem:cnt}
$\cnt_t(X,Y)$ is the largest number of pairwise non-overlapping occurrences of $XY$ in $t$.
\end{lemma}

\begin{proof}
For $X\ne Y$ two occurrences cannot overlap, since a shared token would carry both symbols, and the scan replaces all of
them. For $X=Y$ the occurrences lie in maximal runs of $X$; a run of $\ell$ tokens holds at most $\lfloor\ell/2\rfloor$
non-overlapping occurrences, and the scan replaces exactly that many.
\end{proof}

\begin{lemma}[monotonicity]\label{lem:mono}
With fresh symbols, under a partial or full merge, the maximum over all pairs of the number of non-overlapping
occurrences of the pair does not increase. In particular $\mu$ does not increase along any sequence of full merges.
\end{lemma}

\begin{proof}
Two adjacent old tokens after the merge were adjacent before it, so a pair of old symbols keeps a subset of its
occurrences. Non-overlapping occurrences of a pair with the fresh symbol use distinct fresh tokens (a pair of two fresh
tokens uses two of them), and there are as many of these as the merge made replacements, at most the old maximum
(\cref{lem:cnt}).
\end{proof}

The proof uses freshness, and the partial-merge statement fails for string-identified symbols (\cref{app:identity}).
Wherever a bound on the replacements of an arbitrary partial merge is needed, we use instead the following cap, which
holds for fresh symbols and, on uniquely decodable inputs, for string-identified ones.

\begin{lemma}[initial cap]\label{lem:cap}
Every replacement of a merge of $(X,Y)$ erases an original cut whose two letters are the last letter of $X$ and the first
letter of $Y$, and different replacements of one partial merge erase cuts in pairwise disjoint pairs of positions. So a
partial merge, made after any history of merges, makes at most $\mu(s)$ replacements, where $s$ is the original string.
\end{lemma}

\begin{proof}
A replacement joins two adjacent tokens; the cut between them is an original cut, and the letters on its two sides are
the last letter of the left token and the first letter of the right one. The replaced occurrences are non-overlapping,
so the two positions around the erased cuts are disjoint. These cuts are thus non-overlapping occurrences in $s$ of one
pair of letters, at most $\mu(s)$ of them by \cref{lem:cnt}.
\end{proof}

\paragraph{Greedy.} Greedy BPE starts from $s_0$, the string as a tokenisation into single letters, and at step $i$
performs the full merge of a pair with $\cnt_{s_{i-1}}=\mu(s_{i-1})$, ties broken arbitrarily, producing $s_i$. Its gains are
$g_i=\mu(s_{i-1})$, its partial sums $G_i=g_1+\dots+g_i$ ($G_0=0$), and $\BPE(s,k)=G_k$. A run that reaches a single token
continues with empty steps and zero gains. Every statement about greedy holds for every tie-breaking rule. This is the
rule of Kozma and Voderholzer, whose BPE performs full merges, ``always choosing a pair whose replacement maximizes
compression utility'', with replacements from left to right ($\str{aaa}\to\str{Xa}$) \cite[Section~1]{kozma2026arxiv};
the utility of a merge is our $\cnt$.

\paragraph{The optimum and its forest.} A \emph{partial $k$-sequence} $\sigma=(Q_1,\dots,Q_k)$ applies $k$ partial merges
in turn; $Q_r$ merges the pair $(X_r,Y_r)$ into the fresh symbol $Z_r$, and may replace no occurrence (for instance when
the pair does not occur). Its utility $\mathrm{util}(\sigma)$ is the utility of its final tokenisation, and
$\OPT(s,k)$ is the largest utility of a partial $k$-sequence \cite{kozma2026esa}; empty merges make it nondecreasing in
$k$. The optimum $\OPT^m(s,k)$ over sequences of full merges satisfies $\BPE\le\OPT^m\le\OPT$, so every lower bound below
also holds against $\OPT^m$. Every replacement made by $Q_r$ is a \emph{node} of \emph{label} $r$; its \emph{span} is the
interval of positions it covers, its two \emph{children} are the tokens it joins, and its \emph{junction} $\jn(v)$ is the
cut between them. The nodes form a forest $\Fo$ over the positions of $s$. We use three facts.
\begin{itemize}
\item[(F1)] Distinct nodes have distinct junctions, and $\mathrm{util}(\sigma)$ is the number of nodes. If $v$ is the root of
a tree with span $(a,b]$, the junctions of the nodes of this tree are exactly the cuts $a+1,\dots,b-1$.
\item[(F2)] \emph{Rule determinism.} All nodes of label $r$ spell the same word $w_r$, and their junctions lie at the same
offset in it.
\item[(F3)] Spans are laminar, and nodes of one label have pairwise disjoint spans.
\end{itemize}
(F1) holds because a replacement erases its junction and erased cuts stay erased; the cuts inside a root's span are erased,
and the node that erased one lies below the root. (F2) holds because each fresh symbol spells a fixed word. (F3) holds because
a node's span is a token when the node is created and stays inside a token afterwards, and the nodes of one label are
created by one partial merge from non-overlapping occurrences.

\paragraph{Witnesses.} Let $t$ be a tokenisation of $s$ (later, greedy's current one). For a label $r$ and a pair of
symbols $(X,Y)$, let $N_r(X,Y;t)$ be the number of label-$r$ nodes whose span contains, wholly inside the span, two
adjacent tokens of $t$ with symbols $X$ and $Y$. Then $m_r(t)=\max_{(X,Y)}N_r(X,Y;t)$, and $m_r=0$ for labels without
nodes.

\begin{lemma}\label{lem:witness}
$m_r(t)\le\mu(t)$ for every label $r$.
\end{lemma}

\begin{proof}
The witnessing occurrences of $XY$ lie in pairwise disjoint spans (F3), so they do not overlap; apply \cref{lem:cnt}.
\end{proof}

Word tables, the input of real trainers, are defined in \cref{sec:table}. \Cref{tab:notation} collects the notation.

\begin{table}[!htb]
\centering\small
\begin{tabular}{@{}>{\raggedright}p{3.9cm}p{11.6cm}@{}}
\toprule
Symbol & Meaning\\
\midrule
$s$, $n$, $s(a,b]$ & the input string, its length, the word on positions $a+1,\dots,b$\\
cut $c$; survives, erased & boundary after position $c$; is, is not a boundary between tokens\\
$\cnt_t(X,Y)$, $\raw_t(X,Y)$, $\mu(t)$ & replacements of the full merge of $(X,Y)$ on $t$; adjacencies of $X,Y$ with overlaps;
the largest $\cnt_t$\\
$s_i$, $g_i$, $G_i$; $\BPE$ & greedy's tokenisation after $i$ steps, its $i$-th gain, $g_1+\dots+g_i$; $G_k$\\
$\OPT$, $\OPT^m$ & best utility of $k$ partial merges; of $k$ full merges\\
$\sigma$, $\Fo$, $U$; label $r$, $w_r$ & a partial merge sequence (usually optimal), its forest of nodes, its utility; a rule
of $\sigma$ and the word its nodes spell\\
junction $\jn(v)$; $m_r(t)$ & cut erased by node $v$; largest number of label-$r$ nodes containing one common pair of
tokens of $t$\\
shadow, dimer; $G$, $M$, $L$ & replay of a prefix's letter-pair merges, a two-letter shadow token; utility of the prefix, of the
shadow; number of absorbed letters\\
window, gap; $\nu$, $P$ & stretch between consecutive shadow single letters, a cut inside it; number of dimers in a window;
number of marked windows\\
$c_k$, $\cc$; $\rho$, $\rho_k$ & $\frac12\bigl(1-(1-\frac2k)^k\bigr)$, $\frac12(1-e^{-2})$; worst-case ratio over all budgets,
at budget $k$\\
$\mathcal T$, $(\omega_j,\phi_j)$, $D_t$, $R_t$ & word table, its rows (word, weight), weighted counts $\cnt$ and $\raw$\\
$o_i$, $O_i$; $\lambda_*$, $\lambda_{\rm h}$, $\Lambda_4$ & gain and saving of overlap greedy; its lower bounds (exact
certificate, hand calculation) and upper bound (\cref{prop:overlap})\\
$d$, $F$; $h_t$, $w_t$, $\delta_t$ & parameters of the constructions; levels and multiplicities of \cref{app:upper}\\
\bottomrule
\end{tabular}
\caption{Notation used in the body. Symbols local to one appendix proof are defined there.}
\label{tab:notation}
\end{table}
\section{Warm-up: two merges by hand, and why simpler arguments fail}\label{sec:warmup}

Everything in this section can be checked with pencil and paper; scripts re-check every number (\cref{app:evidence}).

\subsection{Two merges, worked by hand}\label{sec:k2}

Take $s=\str{aaabaab}$ and $k=2$. The pairs $\str{aa}$ (positions $1$--$2$ and $5$--$6$) and $\str{ab}$ (positions $3$--$4$
and $6$--$7$) both have count $2$, and $\str{ba}$ has count $1$. If greedy merges $\str{aa}$ first, it obtains
$s_1=\tok{aa}\,\str a\,\str b\,\tok{aa}\,\str b$, in which every adjacent pair occurs once; so $g_1=2$, $g_2=1$ and
$\BPE=3$. (Merging $\str{ab}$ first leads to a saving of $4$.) The partial $2$-sequence that merges $\str{ab}$ (both
copies) and then $(\str a,\tok{ab})$ (both copies) has utility $4$. No $2$-sequence does better: all nodes of one label
have the same letter pair at their junction (F2), in disjoint spans (F3), so each label has at most $\mu(s)=2$ nodes.
Hence $\OPT(s,2)=4$, and this run has ratio $3/4$, the exact worst case at budget two \citeAat{Theorem~22}; the bound of
\cref{thm:main} at $k=2$ is $c_2=1/2$.

\emph{The recurrence.} At $i=0$ the shadow inequality reads $\OPT\le2\cdot0+2\,\mu(s)=4$, with equality. So
$g_1\ge\OPT/2$, and every greedy run saves at least $2=c_2\OPT$, as the recurrence of \cref{sec:overview} predicts.

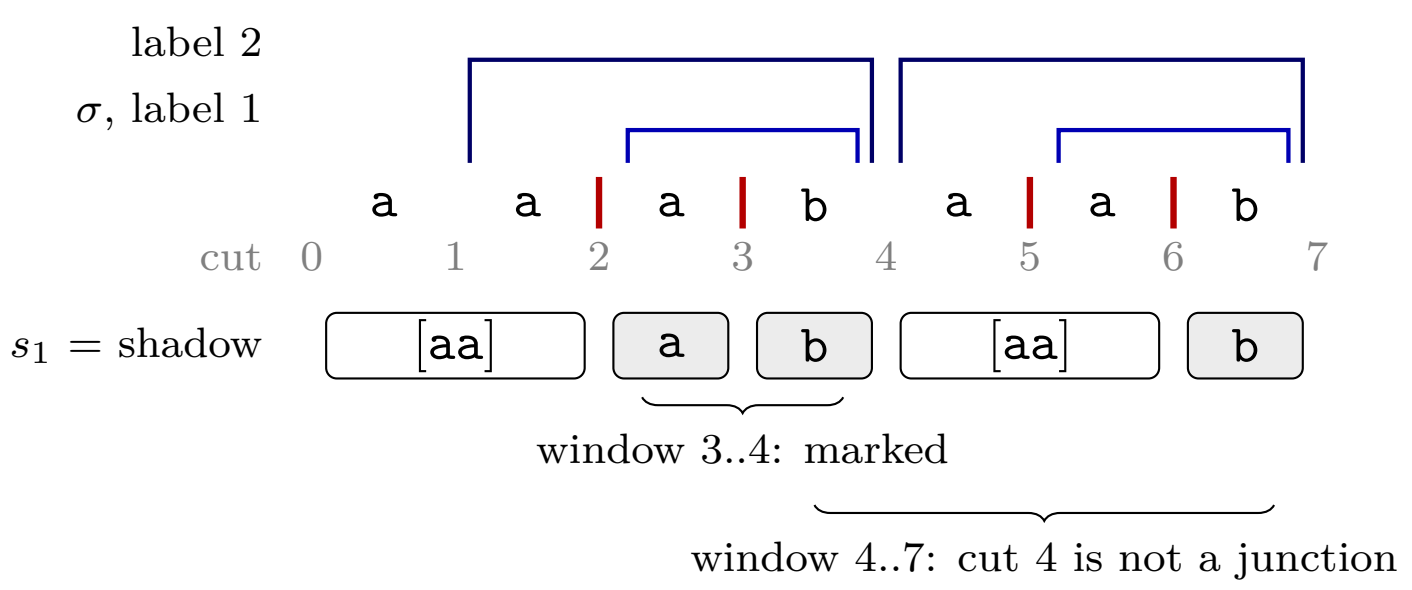
\begin{figure}[!htb]
\centering
\begin{tikzpicture}[x=0.95cm,y=0.62cm,every node/.style={font=\small}]
  \foreach \i/\l in {1/a,2/a,3/a,4/b,5/a,6/a,7/b} {\node at (\i,0) {\str{\l}};}
  \foreach \i in {0,...,7} {\node[font=\scriptsize,gray] at (\i+0.5,-0.55) {\i};}
  \node[anchor=east,font=\scriptsize,gray] at (0.3,-0.55) {cut};
  \node[anchor=east,font=\scriptsize] at (0.3,1.0) {$\sigma$, label 1};
  \draw[thick,blue!70!black] (2.7,0.45) -- (2.7,0.8) -- (4.3,0.8) -- (4.3,0.45);
  \draw[thick,blue!70!black] (5.7,0.45) -- (5.7,0.8) -- (7.3,0.8) -- (7.3,0.45);
  \node[anchor=east,font=\scriptsize] at (0.3,1.75) {label 2};
  \draw[thick,blue!40!black] (1.6,0.45) -- (1.6,1.55) -- (4.4,1.55) -- (4.4,0.45);
  \draw[thick,blue!40!black] (4.6,0.45) -- (4.6,1.55) -- (7.4,1.55) -- (7.4,0.45);
  \foreach \c in {2,3,5,6} {\draw[very thick,red!70!black] (\c+0.5,-0.25) -- (\c+0.5,0.3);}
  \node[anchor=east,font=\scriptsize] at (0.3,-1.5) {$s_1$ = shadow};
  \draw[rounded corners=2pt] (0.6,-1.85) rectangle (2.4,-1.15); \node at (1.5,-1.5) {$\tok{aa}$};
  \draw[rounded corners=2pt,fill=gray!15] (2.6,-1.85) rectangle (3.4,-1.15); \node at (3,-1.5) {\str a};
  \draw[rounded corners=2pt,fill=gray!15] (3.6,-1.85) rectangle (4.4,-1.15); \node at (4,-1.5) {\str b};
  \draw[rounded corners=2pt] (4.6,-1.85) rectangle (6.4,-1.15); \node at (5.5,-1.5) {$\tok{aa}$};
  \draw[rounded corners=2pt,fill=gray!15] (6.6,-1.85) rectangle (7.4,-1.15); \node at (7,-1.5) {\str b};
  \draw[decorate,decoration={brace,amplitude=3pt,mirror}] (2.8,-2.05) -- (4.2,-2.05)
     node[midway,below=3pt,font=\scriptsize,align=center] {window $3..4$: marked};
  \draw[decorate,decoration={brace,amplitude=3pt,mirror}] (4.0,-3.2) -- (7.2,-3.2)
     node[midway,below=3pt,font=\scriptsize,align=center] {window $4..7$: cut $4$ is not a junction};
\end{tikzpicture}
\caption{The two-merge example $s=\str{aaabaab}$ after greedy's first merge $\str{aa}$. Above: the optimal sequence
$\sigma$ (label $1$ merges $\str{ab}$, label $2$ merges $\str a\tok{ab}$) and its four junctions (thick ticks). Below:
greedy's tokens, which here equal the shadow; shaded tokens are the shadow's single letters. The marked window carries
the pair $(\str a,\str b)$, the witness counted in $m_1$.}
\label{fig:k2}
\end{figure}

\emph{The shadow argument after the first step} (\cref{fig:k2}). Label $1$ ($\str{ab}$) has nodes on positions $3$--$4$
and $6$--$7$; label $2$ ($\str a\tok{ab}$) on $2$--$4$ and $5$--$7$. In $s_1$ the span $3$--$4$ contains the tokens
$\str a,\str b$, and the span $6$--$7$ contains no two whole tokens, so $m_1=1$. The span $2$--$4$ contains the pair
$(\str a,\str b)$ and the span $5$--$7$ the pair $(\tok{aa},\str b)$, so $m_2=1$. Thus $U-\sum_rm_r=2\le2G_1=4$, and the
shadow inequality holds: $4\le2\cdot2+2\cdot1$. The only merge of the prefix joins two letters, so the shadow equals the
actual run: $M=2$ dimers, no absorbed letters ($L=0$), single letters at positions $3$, $4$ and $7$. The window from $3$
to $4$ has one gap, cut $3$, which is the junction of the label-$1$ node on $3$--$4$; it is marked, and the mark goes to
that node. The window from $4$ to $7$ is not marked, because cut $4$ is not a junction. The junctions are cuts
$2,3,5,6$, so $U-P=3\le2M=4$ as \cref{lem:windows} promises, and the kept mark is the witness counted in $m_1$.

\subsection{Why simpler arguments fail}

\paragraph{Full merges are essential.}\label{sec:labz}
Take $s=\str{labz|labz}$ and the sequence that merges $\str{ab}$, then $\str l\tok{ab}$, then $\tok{lab}\str z$, in
both copies; $U=6$, which is optimal for $k=3$ since no pair of $s$ occurs more than twice. Now tokenise $s$ by merging
$\str{ab}$ \emph{only in the first copy}: the tokens are
$\str l\,\tok{ab}\,\str z\,\str|\,\str l\,\str a\,\str b\,\str z$, with utility $1$. Label $\str{ab}$ has a witness only in
the second copy; label $\str l\tok{ab}$ sees $(\str l,\tok{ab})$ in the first copy and $(\str l,\str a),(\str a,\str b)$ in
the second; the root sees $(\str l,\tok{ab}),(\tok{ab},\str z)$ and $(\str l,\str a),(\str a,\str b),(\str b,\str z)$. No
pair is shared, so $\sum_rm_r=3$ and $U-\sum_rm_r=3>2\cdot1$: inequality \eqref{eq:pointwise} fails. With the full merge
of $\str{ab}$ both copies read $\str l\,\tok{ab}\,\str z$, the utility is $2$, $\sum_rm_r=0+2+2=4$, and $6-4\le4$.
Each copy alone is the output of full merges (merge $\str{ab}$; merge nothing). What fails is the combination, which
no single sequence of full merges produces. So an argument that certifies one copy of a word at a time, without using
that all copies are tokenised by the same full merges, cannot prove \eqref{eq:pointwise}; our proof uses the full-merge
semantics through isolation (\cref{lem:iso}).

\paragraph{The natural per-replacement charging fails.}
Following the run and charging each greedy replacement two units for the witnesses it removes does not work: on
$\str{abaaa}$ with $k=5$, one replacement of a legal greedy run removes three witnesses while $2G$ grows by two
(\cref{app:threewit}).
Inequality \eqref{eq:pointwise} survives there only on slack accumulated at earlier steps, which is why our proof
compares two tokenisations at a fixed prefix instead.
\section{The shadow argument}\label{sec:proof}

Throughout this section $p_1,\dots,p_i$ are full merges on $s$ with arbitrary pairs (not necessarily greedy's). The
\emph{actual} tokenisation is their output, and $G$ is its utility. A merge is a \emph{letter-pair merge} if both of its
operands are letters of $s$.

\subsection{Isolation}

A full scan of $(X,Y)$ has a state before each token $\tau$ it examines: $\tau$ is either \emph{free}, or was
\emph{consumed} as the second token of the replacement just made. At a free token the scan replaces $\tau$ and its
successor if they spell $XY$, and the successor is then consumed; at a consumed token it moves on.

\begin{lemma}[isolation]\label{lem:iso}
If cuts $a<b$ both survive after $p_1,\dots,p_i$, then the tokens inside $(a,b]$ are exactly, symbol by symbol, the tokens
produced by $p_1,\dots,p_i$ on the string $s(a,b]$ alone.
\end{lemma}

\begin{proof}
Erased cuts stay erased, so $a$ and $b$ survive after every step $j\le i$. Induct on $j$: assume that after step $j-1$
the tokens inside $(a,b]$ are those of the isolated run, and consider the full merge $p_j=(X,Y)$. Let $\tau_1$ be the
first token after cut $a$ and $\tau_\ell$ the last token before cut $b$.
\begin{itemize}
\item \emph{Entry.} The whole scan reaches $\tau_1$ in the free state: $\tau_1$ could be consumed only by a replacement of
the last token before $a$ together with $\tau_1$, and that replacement would erase cut $a$. The isolated scan starts at
$\tau_1$, also free.
\item \emph{Inside.} From $\tau_1$ on, both scans examine the same tokens (by the induction hypothesis) in the same state,
so they make the same decision at every token whose successor also lies in $(a,b]$.
\item \emph{Exit.} At $\tau_\ell$ the isolated scan stops. The whole scan does not replace $\tau_\ell$ together with the
next token, since that would erase cut $b$.
\end{itemize}
So the two scans make the same replacements inside $(a,b]$, and the claim holds after step $j$.
\end{proof}

The entry step is what makes the lemma hold for pairs $(X,X)$: merging $(\str a,\str a)$ on $\str{baaa}$ gives
$\str b\,\tok{aa}\,\str a$, and the tokens on $(1,4]$ are $\tok{aa}\,\str a$, as on $\str{aaa}$ alone. The hypothesis
that $a$ survives is needed: on $\str{aaa}$ the merge erases cut $1$, and the tokens meeting $(1,3]$ are not the single
token $\tok{aa}$ of the run on $\str{aa}$.

\subsection{The shadow}

\begin{definition}[shadow]\label{def:shadow}
The \emph{shadow} of $p_1,\dots,p_i$ is the output of the subsequence of its letter-pair merges, in their order and with
their output symbols, run on $s$ from the start. Its tokens are single letters and two-letter tokens, called
\emph{dimers}. Let $M$ be the number of dimers (the utility of the shadow), and let $L$ be the number of positions that
are single letters in the shadow but lie in a token of at least two letters in the actual tokenisation (\emph{absorbed}
letters).
\end{definition}

Every merge of the shadow has two letters as operands and acts on single letters only, so the shadow never creates a
token longer than two, and a dimer is never merged again.

\begin{figure}[t]
\centering
\begin{tikzpicture}[x=0.95cm,y=0.8cm,every node/.style={font=\small}]
  \foreach \i/\l in {1/x,2/a,3/b,4/c,5/d,6/y,7/c,8/d} {\node at (\i,0) {\str{\l}};}
  \foreach \i in {1,...,8} {\node[font=\scriptsize,gray] at (\i,-0.45) {\i};}
  \node[anchor=east,font=\scriptsize] at (0.3,1.0) {actual};
  \draw[rounded corners=2pt] (0.6,0.65) rectangle (1.4,1.35); \node at (1,1) {\str x};
  \draw[rounded corners=2pt] (1.6,0.65) rectangle (4.4,1.35); \node at (3,1) {$Y=\tok{abc}$};
  \draw[rounded corners=2pt] (4.6,0.65) rectangle (5.4,1.35); \node at (5,1) {\str d};
  \draw[rounded corners=2pt] (5.6,0.65) rectangle (6.4,1.35); \node at (6,1) {\str y};
  \draw[rounded corners=2pt] (6.6,0.65) rectangle (8.4,1.35); \node at (7.5,1) {$Z=\tok{cd}$};
  \node[anchor=east,font=\scriptsize] at (0.3,-1.2) {shadow};
  \draw[rounded corners=2pt,fill=gray!12] (0.6,-1.55) rectangle (1.4,-0.85); \node at (1,-1.2) {\str x};
  \draw[rounded corners=2pt] (1.6,-1.55) rectangle (3.4,-0.85); \node at (2.5,-1.2) {$X=\tok{ab}$};
  \draw[rounded corners=2pt] (3.6,-1.55) rectangle (5.4,-0.85); \node at (4.5,-1.2) {$Z=\tok{cd}$};
  \draw[rounded corners=2pt,fill=gray!12] (5.6,-1.55) rectangle (6.4,-0.85); \node at (6,-1.2) {\str y};
  \draw[rounded corners=2pt] (6.6,-1.55) rectangle (8.4,-0.85); \node at (7.5,-1.2) {$Z=\tok{cd}$};
  \foreach \c in {0.5,1.5,5.5,6.5,8.5} {\draw[densely dashed,gray] (\c,-1.8) -- (\c,1.6);}
  \draw[decorate,decoration={brace,amplitude=4pt,mirror}] (0.7,-1.95) -- (6.3,-1.95)
     node[midway,below=4pt,font=\scriptsize] {window from \str x to \str y: two shadow dimers, five gaps};
\end{tikzpicture}
\caption{The shadow on $s=\str{xabcdycd}$ for the full merges $\str{ab}\to X$, $X\str c\to Y$, $\str{cd}\to Z$. The
shadow replays only the letter-pair merges $\str{ab}\to X$ and $\str{cd}\to Z$. Dashed lines are the cuts that survive
in both runs; the block between cuts $1$ and $5$ has two tokens in each run, and the actual run's inner boundary
(cut $4$) lies inside the shadow dimer $Z$ (\cref{lem:lost}). The shaded letters are the shadow's single letters, and
they are single letters of the actual run too. So every copy of the six letters \str{xabcdy} whose end letters are
single in the actual run is tokenised as $\str x,Y,\str d,\str y$ (\cref{lem:iso}), and all such copies share the pair
$(\str x,Y)$, although the shadow pair there is $(\str x,X)$.}
\label{fig:shadow}
\end{figure}
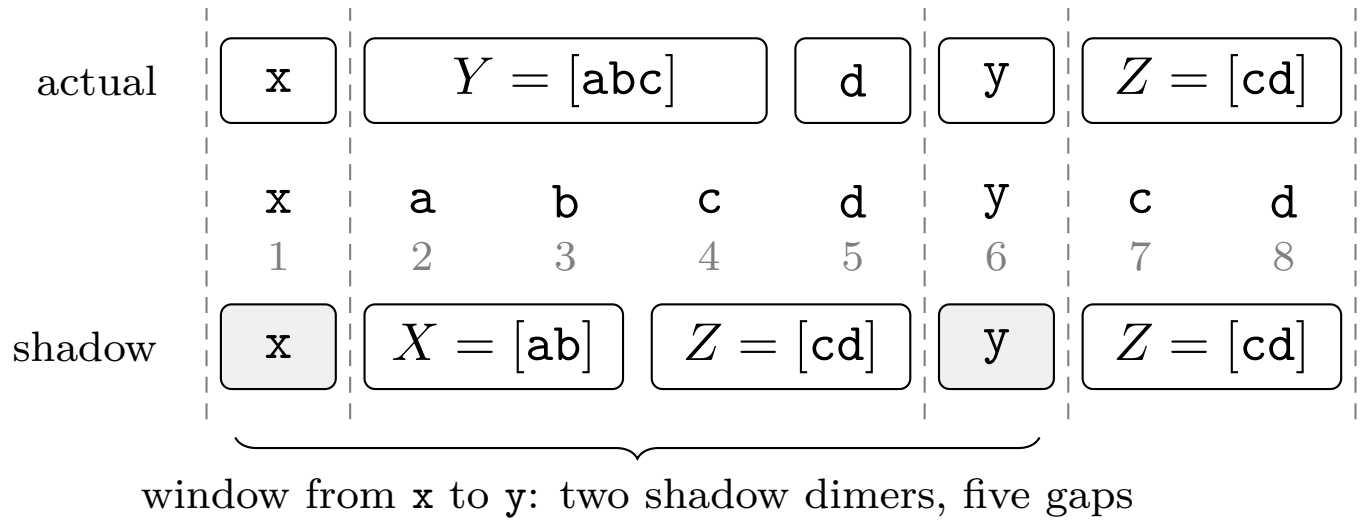

\begin{lemma}[absorbed letters are paid for]\label{lem:lost}
$L\le G-M$.
\end{lemma}

\begin{proof}
Cut $s$ at every cut that survives in both the actual tokenisation and the shadow, including the end cuts, into
\emph{blocks}. Every token of either tokenisation lies inside one block, so $G$, $M$ and $L$ are sums over the blocks.
The shadow is itself the output of full merges, so \cref{lem:iso} applies to both runs: on each block, the actual run and
the shadow agree with $p_1,\dots,p_i$, respectively with its letter-pair subsequence, run on the block alone. Fix a block,
and run the shadow in step with the prefix, doing nothing at the steps that are not letter-pair merges.

\emph{Case 1: no replacement of the actual run inside the block has an operand of two or more letters.} A step that is
not a letter-pair merge has an operand of at least two letters, so it makes no replacement in the block. The letter-pair
steps then act on identical tokenisations of the block in both runs, so by induction over the steps the two runs
coincide on the block. Every boundary between their tokens inside the block would then be a common surviving cut; since
blocks are cut at every common surviving cut, the block is a single token, the same in both runs (a single letter or a
dimer). It contributes $0$ to $L$ and $0$ to $G-M$.

\emph{Case 2: otherwise.} Let $p_j$ be the first step that makes, inside the block, a replacement with an operand $E$ of
at least two letters. The steps before $j$ satisfy the hypothesis of case~1 on the block, so the argument of case~1,
applied to them, shows that the two runs coincide on the block before step $j$. Hence $E$, a token of the actual run
before step $j$, is a token of the shadow at that time; it has at least two letters, so it is a dimer, and since dimers
are never merged again in the shadow, $E$ is a dimer of the final shadow. The actual run merges $E$ further at step $j$,
so the interior cut of $E$ is erased in the actual run. Let the block contain $u$ shadow single letters, $\nu$ dimers and
$a$ actual tokens. The $a-1$ boundaries between actual tokens inside the block are not common cuts, so each lies strictly
inside a dimer of the shadow, and distinct boundaries lie in distinct dimers (a dimer has one interior cut). The interior
cut of $E$ is erased in the actual run, so $E$ is none of these dimers. The block thus has at least $(a-1)+1$ dimers,
the $a-1$ that contain a boundary and $E$, so $\nu\ge a$. The block has $u+2\nu$ positions, so
its contribution to $G-M$ is $(u+2\nu-a)-\nu=u+\nu-a\ge u$, while at most $u$ of its letters are absorbed. Summing over the
blocks proves the lemma.
\end{proof}

\subsection{Windows between shadow single letters}

Let $x_1<\dots<x_\xi$ be the positions of the shadow's single letters. The \emph{window} $W_j$ consists of the positions
$x_j,\dots,x_{j+1}$: a single letter, $\nu\ge0$ dimers, and a single letter. Its \emph{gaps} are the $2\nu+1$ cuts
$x_j,\dots,x_{j+1}-1$ between its two end letters. Fix a partial merge sequence $\sigma$ with forest $\Fo$ and utility
$U$. The window is \emph{marked} if each of its gaps is the junction of a node of $\Fo$, and the mark is attached to the
lowest node of $\Fo$ whose span contains both end letters. Let $P$ be the number of marked windows.

To compare the shadow at two occurrences of one word, we need to read off its dimers inside a window from the letters
of the window alone. The following matching process gives this local description. Call a cut $c$ \emph{eligible} if $(s[c],s[c+1])$ is the pair of a
letter-pair merge of the prefix, and let $\tau(c)$ be the first such merge. Process the eligible cuts in increasing order
of $(\tau(c),c)$, and \emph{accept} a cut if neither of its two positions belongs to an accepted cut so far. A position
$c$ belongs only to the cuts $c-1$ and $c$, so whether cut $c$ is accepted depends only on whether cut $c-1$ or cut $c+1$
was accepted before it.

\begin{lemma}[the shadow as a matching]\label{lem:matching}
The accepted cuts are exactly the interior cuts of the shadow's dimers.
\end{lemma}

\begin{proof}
Induct over the letter-pair merges of the prefix. Consider a merge of the letter pair $(x,y)$, possibly with $x=y$.
\begin{itemize}
\item \emph{Its first merge in the subsequence.} By induction, before this merge the accepted cuts are the interior cuts
of the current dimers; these are the cuts with an earlier $\tau$, which come first in the order. So the occurrences of
$(x,y)$ in the current shadow, pairs of adjacent single letters $x,y$, are the cuts with letters $(x,y)$ whose two
positions belong to no accepted cut. The scan examines these cuts from left to right and replaces one exactly when its
first position was not consumed by the replacement just before it, that is, when cut $c-1$ was not just accepted; the
second position is free because cut $c+1$ has not been examined yet. Processing the same cuts from left to right and
accepting a cut when both of its positions are free makes the same choices. For $x=y$ this is the pairing
$\str{aaa}\to Z\str a$: in each maximal run of single $x$'s the scan pairs the first two letters, then the next two, and
so on.
\item \emph{A later merge of the same pair.} It finds no occurrence. After the first merge no two single letters $x,y$
are adjacent: an occurrence left unreplaced was blocked by a replacement that used its first letter. For $x=y$ each run of
single $x$'s keeps at most its last letter, which is not adjacent to another single $x$. Single letters only disappear
in the shadow, so no occurrence can reappear. Correspondingly, no cut has $\tau(c)$ at a later merge of the same pair.
\qedhere
\end{itemize}
\end{proof}

\begin{lemma}[rigidity]\label{lem:rigid}
Let $W$ be a window on positions $\alpha,\dots,\beta$, and let $\alpha',\dots,\beta'$ be any positions of $s$ spelling the
same word, with $\beta'-\alpha'=\beta-\alpha$; the two sets of positions may overlap. Translate cuts by $\alpha'-\alpha$.
Then a cut with both of its positions in $W$ is accepted exactly when its translate is. In particular every dimer of $W$
is a dimer at the translated positions.
\end{lemma}

\begin{proof}
Call a cut \emph{internal} to $W$ if both its positions lie in $W$, that is, if it is a gap. A cut and its translate
have the same letters, so they are both eligible or both not, with the same $\tau$; and the order $(\tau(c),c)$ restricted
to the internal cuts is the same as the order of their translates. We prove the claim for the internal cuts in this
order. Recall that only the cuts $c\pm1$ can block a cut $c$.
\begin{itemize}
\item If $c$ is accepted for $W$, its positions lie in a dimer of $W$, so neither is $\alpha$ or $\beta$, which stay single
letters. Hence $\alpha<c$ and $c+1<\beta$, and both cuts $c-1$ and $c+1$ are internal. Each of them that comes before $c$
in the order was rejected for $W$, since otherwise it would block $c$; by induction its translate was rejected too. So the
translate of $c$ is not blocked and is accepted.
\item If an eligible $c$ is rejected for $W$, one of its positions belongs to a cut accepted before it. That position is
not $\alpha$ or $\beta$, so it lies strictly inside $W$, and the blocking cut has both positions in $W$: it is internal.
By induction its translate was accepted, and it blocks the translate of $c$.
\end{itemize}
The cuts outside $W$, such as $\alpha-1$ and $\beta$ and their translates, never decide anything here: an internal cut that
is accepted has only internal neighbours, and an internal cut that is rejected is rejected because of an internal
neighbour. The cut $\alpha'-1$ could block the translate of $\alpha$, but the cut $\alpha$ is never accepted for $W$, so
the claim needs nothing from it. In particular every dimer of $W$ is a dimer at the translated positions.
\end{proof}

Rigidity concerns only the translated positions $\alpha',\dots,\beta'$ and says nothing about the shadow outside them;
in \cref{lem:windows}(b) it is applied only to positions inside the span of one node.

\begin{lemma}[windows]\label{lem:windows}
\begin{enumerate}
\item[(a)] The marks are attached to distinct nodes, and $U-P\le2M$.
\item[(b)] Let $W$ and $W'$ be marked windows attached to nodes $v$ and $v'$ of the same label $r$. Then $W$ and $W'$
occupy the same positions within the spans of $v$ and $v'$, counted from the left end of the span.
\end{enumerate}
\end{lemma}

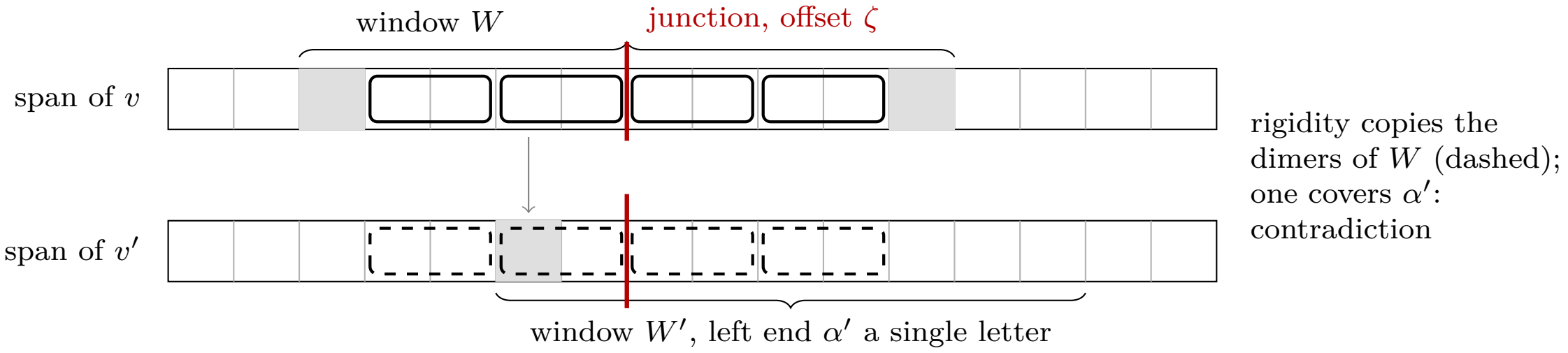
\begin{figure}[t]
\centering
\begin{tikzpicture}[x=0.62cm,y=0.72cm,every node/.style={font=\small}]
  \node[anchor=east,font=\scriptsize] at (-0.2,2) {span of $v$};
  \draw (0,1.6) rectangle (16,2.4);
  \foreach \i in {1,...,15} {\draw[gray!60] (\i,1.6) -- (\i,2.4);}
  \fill[gray!25] (2,1.6) rectangle (3,2.4); \fill[gray!25] (11,1.6) rectangle (12,2.4);
  \foreach \a in {3,5,7,9} {\draw[rounded corners=2pt,thick] (\a+0.08,1.7) rectangle (\a+1.92,2.3);}
  \draw[decorate,decoration={brace,amplitude=4pt}] (2,2.55) -- (12,2.55) node[pos=0.2,above=3pt,font=\scriptsize] {window $W$};
  \draw[very thick,red!70!black] (7,1.45) -- (7,2.75);
  \node[anchor=east,font=\scriptsize] at (-0.2,0) {span of $v'$};
  \draw (0,-0.4) rectangle (16,0.4);
  \foreach \i in {1,...,15} {\draw[gray!60] (\i,-0.4) -- (\i,0.4);}
  \fill[gray!25] (5,-0.4) rectangle (6,0.4);
  \draw[decorate,decoration={brace,amplitude=4pt,mirror}] (5,-0.55) -- (14,-0.55) node[midway,below=3pt,font=\scriptsize] {window $W'$, left end $\alpha'$ a single letter};
  \draw[very thick,red!70!black] (7,-0.75) -- (7,0.75);
  \foreach \a in {3,5,7,9} {\draw[rounded corners=2pt,thick,dashed] (\a+0.08,-0.3) rectangle (\a+1.92,0.3);}
  \draw[->,gray] (5.5,1.5) -- (5.5,0.5);
  \node[font=\scriptsize,align=left,anchor=west] at (16.3,1.0) {rigidity copies the\\ dimers of $W$ (dashed);\\ one covers $\alpha'$:\\ contradiction};
  \node[font=\scriptsize,red!70!black,anchor=west] at (7.1,3.05) {junction, offset $\zeta$};
\end{tikzpicture}
\caption{The offset argument of \cref{lem:windows}(b). Both spans spell $w_r$ and have their junction at the same offset
$\zeta$ (thick line), and both windows contain the two letters around it. If $W'$ starts later than $W$, its left end
$\alpha'$ (shaded) falls strictly inside the offsets of $W$. Rigidity transfers every dimer of $W$ to the same offsets in
the span of $v'$, so a dimer covers $\alpha'$, which is a single letter of the shadow.}
\label{fig:offsets}
\end{figure}

\begin{proof}
(a) If all gaps of a window are junctions, they are erased in the final tokenisation of $\sigma$, so the window lies in
one final token of $\sigma$, the span of a root of $\Fo$, and a lowest node $v$ containing both end letters exists.
Neither child of $v$ contains both end letters, so the junction of $v$ separates them and is one of the gaps of the
window. Different windows have disjoint sets of gaps, so their marks go to nodes with different junctions, hence to
different nodes. For the count, $U$ is the number of junctions (F1), and every junction is a cut in $\{1,\dots,n-1\}$.
Split these cuts into the gaps of the windows, the cuts before $x_1$, and the cuts from $x_\xi$ on; consecutive
windows share an end letter but no gap, so every cut in $\{1,\dots,n-1\}$ lies in exactly one of these parts. A window with $\nu$ dimers
has $2\nu+1$ gaps; it contributes at most $2\nu$ to $U-P$, since either it is marked or one of its gaps is not a junction.
Before $x_1$ there are only dimers, say $\nu$ of them, and exactly $2\nu$ cuts; likewise after $x_\xi$. If the shadow has no
single letter, $n=2M$ and there are $2M-1$ cuts. Every dimer is counted in exactly one of these parts, so $U-P\le2M$.

(b) Let $W$ and $W'$ be as in (b) (\cref{fig:offsets}). By (F2) the spans of $v$ and $v'$ spell the same word $w_r$ and their
junctions sit at the same offset $\zeta$ in it: the junction separates the letters at offsets $\zeta$ and $\zeta+1$. By (a) each
junction is a gap of its window, so in the coordinates of $w_r$ both windows contain the offsets $\zeta$ and $\zeta+1$. If both
ends of the two windows have the same offsets, the windows occupy the same positions, as claimed. Otherwise, swapping $W$
and $W'$ if needed, either the left end $\alpha'$ of $W'$ has an offset $l'$ strictly larger than the offset $l$ of the
left end of $W$, or the left ends agree and the right end of $W'$ has a strictly smaller offset than the right end of
$W$. We treat the first case; the second is the same argument applied to the right ends.
Since $l<l'\le\zeta$ and $\zeta+1$ lies in $W$, the offset $l'$ lies strictly inside the interval of offsets of $W$.
Inside the span of $v'$, the positions with the offsets of $W$ spell the word of $W$; by \cref{lem:rigid}, every position there
that has the offset of a position covered by a dimer of $W$ is covered by a dimer. Every position strictly inside a window
is covered by one of its dimers, so $\alpha'$ is covered by a dimer of the shadow. But $\alpha'$ is a single letter of the
shadow, the left end of $W'$. This contradiction proves (b).
\end{proof}

\subsection{The pointwise inequality}

\begin{theorem}[pointwise inequality]\label{thm:pointwise}
Let $p_1,\dots,p_i$ be full merges on $s$ with output $t$ and utility $G$, and let $\sigma$ be a partial merge sequence
with any number of rules and utility $U$. Then
\[
U-\sum_r m_r(t)\;\le\;2G .
\]
\end{theorem}

\begin{proof}
Call a marked window \emph{kept} if both of its end letters are single letters of $t$. An absorbed letter is an end of at
most two windows, so at least $P-2L$ marked windows are kept.

Fix a label $r$ and its kept windows. By \cref{lem:windows}(b) they occupy the same offsets in spans that spell $w_r$, so
they spell the same word. Let a kept window occupy positions $\alpha,\dots,\beta$. Its end letters are single letters of $t$,
so the cuts $\alpha-1$ and $\beta$ survive, and by \cref{lem:iso} the tokens of $t$ inside $(\alpha-1,\beta]$ are those
that $p_1,\dots,p_i$ produce on this word alone. So all kept windows of $r$ carry the same sequence of tokens of $t$. It has
at least two tokens, because $\alpha$ and $\beta$ are distinct single letters; its first two tokens form one pair $(X,Y)$,
wholly inside each window and hence inside the span of the node that carries the mark. The marks sit on distinct nodes
(\cref{lem:windows}(a)), so $m_r(t)\ge N_r(X,Y;t)$ is at least the number of kept windows of $r$. Summing over the labels
and using \cref{lem:windows}(a) and \cref{lem:lost},
\[
U-\sum_rm_r(t)\;\le\;U-(P-2L)\;\le\;2M+2L\;\le\;2M+2(G-M)\;=\;2G.
\]
\emph{The kept-window family.} \Cref{app:overlap} uses the following family of cuts that this proof produces. For each
kept window, on positions $\alpha,\dots,\beta$ and with its mark on a node of label $r$, take the cut $\alpha$ with label
$r$. Its left token is the single letter $X$ of $t$ at $\alpha$ and its right token is $Y$, the pair found above. These
cuts have four properties. (i) They are distinct cuts between consecutive tokens of $t$, since the cut $\alpha$ is a gap of
its window and distinct windows have disjoint sets of gaps. (ii) There are at least $P-2L\ge U-2G$ of them, by the
display. (iii) Their labels are labels of $\sigma$, and all cuts of one label $r$ are occurrences in $t$ of the one pair
$(X,Y)$ of $r$. (iv) The occurrences of one label are pairwise disjoint: each lies inside the span of the node that
carries its mark, these nodes are distinct, and nodes of one label have disjoint spans (F3).
\end{proof}

The witness pair $(X,Y)$ is a pair of actual tokens and can differ from the pair that the shadow shows in the same window
(\cref{fig:shadow}); tracking the shadow pair instead would not give a common witness. The factor $2$ cannot be lowered:
on $\str{aabaab}$, after the full merge of $(\str b,\str a)$, \cref{cor:rc} with $k=2$ holds with equality,
$\OPT=4=2\cdot1+2\cdot1$ (\cref{sec:overview}), and the checks of \cref{app:evidence} find many instances of equality in \cref{thm:pointwise}
with $G>0$.

\subsection{The shadow inequality and the main theorem}

\begin{corollary}[shadow inequality]\label{cor:rc}
Let $t$ be the output of any sequence of full merges on a string $s$, with utility $G$. Then $\OPT(s,k)\le2G+k\,\mu(t)$
for every budget $k$. In particular, for every greedy run and every $i\ge0$, $\OPT(s,k)\le2G_i+k\,g_{i+1}$.
\end{corollary}

\begin{proof}
Let $\sigma$ be an optimal partial $k$-sequence, so $U=\OPT(s,k)$, and apply \cref{thm:pointwise}. By
\cref{lem:witness}, $m_r(t)\le\mu(t)$ for each of the $k$ labels. For greedy take $t=s_i$; then $G=G_i$ and
$\mu(s_i)=g_{i+1}$.
\end{proof}

\begin{theorem}[main theorem]\label{thm:main}
For every string $s$, every budget $k\ge2$ and every greedy run with any tie-breaking rule,
\[
\BPE(s,k)\;\ge\;\frac{1-(1-2/k)^k}{2}\,\OPT(s,k)\;>\;\frac{1-e^{-2}}{2}\,\OPT(s,k)\quad\text{if }\OPT(s,k)>0 .
\]
For $k=1$, $\BPE(s,1)=\OPT(s,1)$, which is the case $c_1=1$.
\end{theorem}

\begin{proof}
Let $U=\OPT(s,k)$ and $\theta=1-2/k\ge0$. By \cref{cor:rc}, $g_{i+1}\ge(U-2G_i)/k$, so
$U-2G_{i+1}=(U-2G_i)-2g_{i+1}\le \theta\,(U-2G_i)$ for $0\le i<k$; this holds whatever the sign of $U-2G_i$. Since $\theta\ge0$ the
inequalities can be chained, and $U-2G_k\le \theta^k(U-2G_0)=\theta^kU$, that is, $G_k\ge\frac12(1-\theta^k)U$. Finally
$0\le \theta<e^{-2/k}$, so $\theta^k<e^{-2}$. For $k=1$, a single partial merge makes at most $\mu(s)$ replacements
(\cref{lem:cnt}), and greedy makes $\mu(s)$.
\end{proof}

At $k=3$ and $k=4$ the
theorem gives $13/27$ and $15/32$, the lower bounds on $\rho_3$ and $\rho_4$ of \prevRC. The
recurrence in the proof, which \prevRC{} derives from its conjecture, is the greedy template for submodular
maximisation with curvature $2$ (\cref{sec:related}); all of the difficulty lies in \cref{cor:rc}.

\section{The matching upper bound}\label{sec:upper}

The limit is due to Kozma and Voderholzer: their construction \cite[Appendix~B]{kozma2026arxiv} gives, for every
$\varepsilon>0$, a string and a budget with $\BPE/\OPT^m\le\cc+\varepsilon$ ``independently of tie-breaking'', for greedy
ranked by our $\cnt$; since $\OPT^m\le\OPT$, the bound also holds against the partial-merge optimum. Yang
\cite[Theorem~2]{yang2026} gives another family. Our self-contained construction (\cref{app:upper}) uses their
mechanism, a greedy merge whose two symbols each sit next to a merge of the optimum, with levels that shrink
geometrically and guard letters from a palette of size $O(\sqrt k)$ instead of fresh separators.

\begin{theorem}[matching construction]\label{thm:upper}
For every $\varepsilon>0$ there are a string $s$ and a budget $k$ such that every greedy run, with any tie-breaking, has
$\BPE(s,k)\le(\cc+\varepsilon)\OPT(s,k)$. One can take, for any $d\ge8$ with $c_d+d^{-2}+d^{-3}\le\cc+\varepsilon$, the
string of \cref{app:upper} with parameters $d$ and $F=d^3$, which has fewer than $18d=18\sqrt k$ letters, and $k=d^2$.
\end{theorem}

\begin{theorem}[the exact worst case]\label{thm:exact}
$\rho=\inf\BPE(s,k)/\OPT(s,k)=\cc$, the infimum over strings $s$, budgets $k\ge1$ with $\OPT(s,k)>0$, and
greedy runs. The infimum is not attained, and it remains the same if the tie-breaking rule is chosen in greedy's favour.
All of this holds equally against the full-merge optimum: $\inf\BPE(s,k)/\OPT^m(s,k)=\cc$, not attained.
\end{theorem}

\begin{proof}
Every ratio is at least $c_k>\cc$ for $k\ge2$ by \cref{thm:main}, and equals $1$ for $k=1$; so $\rho\ge\cc$ and no ratio
equals it. By \cref{thm:upper}, or by \cite[Appendix~B]{kozma2026arxiv}, every greedy run on suitable strings has ratio
below $\cc+\varepsilon$ for every $\varepsilon>0$. Against $\OPT^m$: since $\OPT^m\le\OPT$, every ratio
$\BPE/\OPT^m$ is at least $\BPE/\OPT\ge c_k$ (and $1$ at $k=1$); and the comparator of \cref{thm:upper} consists of full
merges, so its upper bound holds against $\OPT^m$ too (\cref{app:upper}).
\end{proof}

At $k=\UpK$ the construction gives a string of $\UpLength$ symbols over $\UpAlphabet$ letters, one byte per symbol, on
which every greedy run has ratio exactly $\UpRatio\approx\UpRatioDec$ against the exact partial-merge optimum
(\cref{app:upper}): below $\cc+0.01$, and above $\cc$, as every ratio must be. No budget $k\le13$ has a ratio below
$\cc+0.01$ on any string, since $c_k$ decreases in $k$ and $c_{13}\approx\CkThirteen>\cc+0.01$.

\paragraph{Alphabets.} Approaching $\cc$ needs alphabets that grow with $k$: on a fixed alphabet the ratio of every
greedy run, of either ranking rule, tends to $1$ as $k$ grows \cite[Theorem~5(b)]{prior2026fixed}, with a deficit of order
$1/\log k$ \cite{prior2026fixed}. Our near-worst inputs use $\Theta(\sqrt k)$ letters; the word tables of \cref{thm:table}(b) use
$2\lfloor\sqrt K\rfloor$, so they fit a byte alphabet only while $\lfloor\sqrt K\rfloor\le128$, that is, exactly when
$K<129^2=16{,}641$. For larger budgets over bytes, which include those of large language models, it is open how close
to $\cc$ the ratio can come; the lower bound $c_k$ holds there as everywhere. Neither construction shows that $c_k$ is the worst case at a fixed budget $k$; at $k=2$ it is not
(\cref{sec:k2}), and by \cref{cor:rate} it is not at any budget $k\ge4\cdot10^{10}$.

\section{The gap at finite budgets}\label{sec:rate}

At budget $k$ the main theorem leaves a gap between the guarantee $c_k$ and the inputs of \cref{thm:upper}. The guarantee
exceeds $\cc$ by about $e^{-2}/k$; the construction, at $k=d^2$, has ratio about $c_d$, which exceeds $\cc$ by about
$e^{-2}/\sqrt k$. The construction has the right order.

\begin{theorem}[the finite-budget gap]\label{thm:rate}
For every string $s$ over any finite alphabet, every budget $k\ge2$ and every greedy run with any tie-breaking rule
(fresh symbols),
\[
\BPE(s,k)\;\ge\;\Bigl(\cc+\frac{10^{-9}}{\sqrt k}\Bigr)\OPT(s,k).
\]
\end{theorem}

\begin{corollary}[the order of the gap]\label{cor:rate}
Consequently $10^{-9}k^{-1/2}\le\rho_k-\cc\le 8k^{-1/2}$ for every $k\ge2$, so $\rho_k-\cc=\Theta(k^{-1/2})$; the upper
bound follows from \cref{thm:upper} with $F=d^3$, $d=\lfloor\sqrt k\rfloor$, padded by at most $2d$ further merges, and
holds for every greedy run. Moreover $\rho_k\ge c_k+1.5\cdot10^{-9}k^{-1/2}>c_k$ for every $k\ge4\cdot10^{10}$.
\end{corollary}

The upper half of \cref{cor:rate} adds little to \cref{thm:upper,thm:table}: \cref{thm:upper} gives $\cc+O(k^{-1/2})$
at square budgets, and the same padding gives $\cc+O(K^{-1/2})$ at every budget $K\ge64$ in \cref{thm:table}(b); only the explicit constant $8$, for strings at every budget, is added here. The
new content is the lower half. The constant $10^{-9}$ is far from the constant of the construction, about $e^{-2}$ on
square budgets, and we have not tried to improve it; the leading constant of $\rho_k-\cc$ is open. Both proofs are in
\cref{app:rate}.

\paragraph{Why $c_k$ is not the truth.} Write $S_i=2G_i+k\,g_{i+1}-\OPT$, the slack of the shadow inequality at step $i$;
it is nonnegative by \cref{cor:rc}. Unrolling the recurrence of \cref{thm:main} exactly gives
\[
G_k\;=\;c_k\,\OPT+\frac1k\sum_{i=0}^{k-1}\Bigl(1-\frac2k\Bigr)^{k-1-i}S_i ,
\]
so the guarantee $c_k\OPT$ is approached only if the shadow inequality is nearly tight at almost every step. It is not.
At step $i$ there are at least $\OPT-2G_i$ kept windows (proof of \cref{thm:pointwise}), grouped by the labels of the
optimum; the windows of one label carry the same tokens, so they share a pair, and a label has at most $g_{i+1}$ of them.
Follow them over a short interval of $\ell$ steps during which greedy's gain falls from $u$ to $v$. At the end each
label keeps at most $v$ live windows, so at least $\OPT-2G_i-kv$ windows die, and this number is
$2\sum_jg_j-S_{i+\ell}$, the sum over the steps of the interval. A window dies only when a greedy merge erases one of
its two outer cuts or joins it into a single token. All live windows of a label end in the same token, so a merge of
$(X,Y)$ that erases many right outer cuts needs many windows ending in $X$. Over the whole run, and summed over all
symbols, the windows and tokens that can end in a symbol number at most $3\OPT$, so at most $k/100$ steps merge a pair
$(X,Y)$ such that more than $30\sqrt k\,\OPT/k$ windows or tokens end in $X$ and more than that many begin with $Y$, and
every other step kills little more than $g_j$ windows. Hence $S_{i+\ell}$ is at least about
$\sum_jg_j$, up to an error that is small when the gain falls slowly. Summed over the second quarter of the run this gives
$\sum_iS_i\ge\ell\,\OPT/100$, where the interval length $\ell$ is a fixed small fraction of $\sqrt k$, and the identity
turns this slack into the extra $10^{-9}k^{-1/2}\OPT$.

\begin{remark}[a run below $\cc+0.01$ at $k=100$]\label{rem:m1}
At $k=100$ there is a string over $20+\MOneGuards$ letters and a greedy run, with ties broken adversarially at
$\MOneTies$ of its steps, that saves $\MOneG$ cuts while $\OPT(s,100)=10^{11}$; its ratio $\MOneRatioDec$ is below
$\cc+0.01$. So the smallest budget at which some string and greedy run have ratio below $\cc+0.01$ lies between $14$
and $100$.
\end{remark}

The run of \cref{rem:m1} is one greedy history, verified by exact computation (\cref{app:m1}); it uses adversarial ties
and a distinct guard letter between consecutive blocks. The certificate at $k=\UpK$ is different in kind: it holds for
every greedy run and fits a byte alphabet.

\paragraph{Scope.} \Cref{thm:rate} is proved for strings with fresh symbols, the model of \cref{sec:proof}, against the
partial-merge optimum, so it also holds against $\OPT^m$. By \cref{prop:identity} it also holds for symbols identified
by their spelling on uniquely decodable strings (\cref{sec:table}). We state it for strings only, not for word tables.

\section{Word tables and string-identified symbols}\label{sec:table}

Real trainers do not see one string. subword-nmt and Hugging Face keep each distinct word of the corpus once with its
frequency and never merge across words \cite{sennrich2016bpe,hftokenizers}, and they identify a merged symbol with the string it
spells. Both features keep the exact constant, the second on uniquely decodable inputs (defined below).

\paragraph{Word tables.} A \emph{word table} $\mathcal T$ is a finite list of rows $(\omega_j,\phi_j)$, $j=1,\dots,m$:
a nonempty word $\omega_j$ over the alphabet and a weight $\phi_j>0$, the frequency of the word in the corpus (a positive
integer in practice; we allow positive reals). A state $t=(t_1,\dots,t_m)$ tokenises each row, starting with single
letters. A rule acts on one ordered pair of symbols in all rows at once and gives every replacement the same fresh
symbol; no replacement crosses a row boundary. A full rule scans each row from left to right as above; a partial rule
replaces any set of non-overlapping occurrences in each row, chosen row by row. The weighted counts are
\[
D_t(X,Y)=\sum_j\phi_j\,\cnt_{t_j}(X,Y),\qquad R_t(X,Y)=\sum_j\phi_j\,\raw_{t_j}(X,Y),\qquad \mu(t)=\max_{(X,Y)}D_t(X,Y),
\]
and the utility of a state is $\sum_j\phi_j(|\omega_j|-|t_j|)$, where $|t_j|$ is the number of tokens of row $j$.
Greedy performs at each step the full rule of a pair that maximises $D_t$; its gains, partial sums and
$\BPE(\mathcal T,k)$ are defined as for strings, and $\OPT(\mathcal T,k)$ is the largest utility of at most $k$ partial
rules. A string is a table with one row of weight $1$. Splitting a row into copies of its word, each choosing its
replacements for itself, does not change the optimum: for a fixed list of rules, the tokenisations a copy can reach
depend only on its word, so they are those its row can reach, and every copy can take the best of them. So a table with
integer weights has the same optimum as the corpus that lists each word as often as its weight.

Both halves of the exact worst case carry over to word tables, the input of real trainers.

\begin{theorem}[word tables]\label{thm:table}
\begin{enumerate}
\item[(a)] For every word table $\mathcal T$ with positive real weights, every budget $k\ge2$ and every greedy run with any
tie-breaking, $\BPE(\mathcal T,k)\ge c_k\,\OPT(\mathcal T,k)$, and $\OPT(\mathcal T,k)\le2G_i+k\,g_{i+1}$ for every
$i\ge0$. For $k=1$, $\BPE(\mathcal T,1)=\OPT(\mathcal T,1)$.
\item[(b)] The infimum of $\BPE(\mathcal T,k)/\OPT(\mathcal T,k)$ over tables, budgets $k\ge1$ with
$\OPT(\mathcal T,k)>0$ and greedy runs is $\cc$. It is not attained, it is unchanged if ties are broken in greedy's
favour, and it is approached along all budgets: for every $K\ge64$ there is a table of words of lengths $2$ and $4$ over
$2\lfloor\sqrt K\rfloor$ letters on which every greedy run of $K$ merges has ratio at most
$c_d+\frac{d+1}{d^3}+\frac{2(d^3+1)}{d^4}=\cc+O(1/\sqrt K)$, where $d=\lfloor\sqrt K\rfloor$.
\item[(c)] At $k=\UpK$ there is a table of $\WtWordTypes$ words of lengths $2$ and $4$ over $\WtLetters$ letters on
which every greedy run saves $\UpG$ and $\OPT(\mathcal T,\UpK)=\UpOPT$; the ratio is $\UpRatio\approx\UpRatioDec$.
\end{enumerate}
\end{theorem}

The proof is in \cref{app:table}. For (a), write the table out as one string, with $\phi_j$ copies of word $\omega_j$
separated by distinct letters, and apply \cref{thm:pointwise}; real weights follow by continuity. For (b) and (c), the
construction of \cref{app:upper} becomes a table once its guard letters are dropped: word boundaries do the guards'
job. By \cref{prop:identity}, all three parts hold for string-identified symbols on uniquely decodable tables.

\paragraph{String-identified symbols.}\label{sec:identity} subword-nmt and Hugging Face identify a symbol with the string
it spells, so two merges that split the same string differently produce the same symbol. Call an input \emph{uniquely
decodable} if no two distinct sequences of initial symbols that both occur consecutively in its rows spell the same
string; this holds when the initial symbols are single characters, also with word-final variants whose marker string
never occurs in the text. On such inputs identification by spelling changes nothing: the full-merge histories of the
two models correspond one to one with the same counts and gains, and the partial-merge optima coincide
(\cref{prop:identity} in \cref{app:identity}). So every bound of \cref{sec:proof,sec:upper,sec:rate} and of this section holds verbatim for string-identified symbols on
uniquely decodable inputs. The hypothesis cannot be dropped from the shadow inequality: with initial symbols $\str a$,
$\str b$, $\str{ab}$ and the rows $[\str a,\str{ab}]$ and $[\str b,\str a,\str a,\str b]$ of weight $1$, every pair occurs
once, so $\mu=\CollMu$, but two string-identified merges save $\CollOPT>2\cdot0+2\cdot\CollMu$. Whether \cref{thm:main}
itself survives without the hypothesis we do not know; a search of small non-decodable inputs found no violation of it
(\cref{app:evidence}).

\section{Overlapping counts, the rule trainers use}\label{sec:overlap}

subword-nmt \cite[Algorithm~1]{sennrich2016bpe} and Hugging Face \cite{hftokenizers} rank a pair by its occurrences
counted with overlaps and then replace from left to right (\cref{app:fidelity} gives the code). Call this rule \emph{overlap greedy}; it saves
$o_i$ at step $i$, and $O_i=o_1+\dots+o_i$. The two rules differ only on runs of three or more equal tokens, where a pair
$(X,X)$ scores up to twice what its merge saves. \Cref{cor:rc} holds for any prefix of full merges, so
$\OPT\le2O_i+k\,\mu(s_i)$, but now only $\mu(s_i)\le2o_{i+1}$; with the recurrence of \cref{thm:main} this gives only
$\frac12(1-(1-\frac1k)^k)$, which tends to $\frac12(1-e^{-1})<\frac13$ (for $k\le4$ this bound appears in
\prevRC). Let
$\bar\rho^{\mathrm{ov}}_k$ be the infimum over strings of the \emph{largest} ratio $O_k/\OPT$ of a legal overlap-greedy
run, so that ties are broken in greedy's favour.

\begin{proposition}[overlap greedy]\label{prop:overlap}
\begin{enumerate}
\item[(a)] Every overlap-greedy run, on every string or word table with positive real weights and at every budget $k$, has
\[
O_k\;\ge\;\lambda_*\OPT,\qquad \lambda_*=\OvLBfull=\OvLBfullDec,
\]
and a shorter calculation gives $O_k\ge\lambda_{\rm h}\OPT$ with $\lambda_{\rm h}=\OvLBhand=\OvLBhandDec>\frac7{18}$.
\item[(b)] $\limsup_{k\to\infty}\bar\rho^{\mathrm{ov}}_k\le \Lambda_4=\frac12-\frac14\sqrt{5/6}\,e^{-13/15}<\OvFourC$.
\item[(c)] Let $\sigma$ be any partial $k$-sequence on a string or word table, with utility $U$. Every overlap-greedy run of $k$ merges
has
\[
O_k\;\ge\;\frac25\Bigl(U-\sum_{T}\Bigl\lfloor\frac{|T|-2}2\Bigr\rfloor\Bigr),
\]
the sum over the tokens $T$ of the final tokenisation of $\sigma$ with $|T|\ge4$ letters (on a word table, each token
weighted by its row). In particular $O_k\ge\frac25U$
whenever the final tokens of $\sigma$ have at most three letters.
\end{enumerate}
\end{proposition}

The proofs are in \cref{app:overlap}, and \cref{sec:overview} sketches (a) and (b). Parts (b) and (c) are proved by hand. The
bound $\lambda_{\rm h}$ of part (a) rests on a finite exact calculation that can be checked by hand, though it is long:
one table of $\OvHandRows$ weights, and the signs of $\OvHandCoeffs$ Bernstein coefficients ($\OvHandCells$ cubic cells,
six step types, four coefficients each), each computed from four values of a sum over the rows; the smallest of them is
about $\OvHandMinMargin\cdot10^{-6}$. We also checked it by computer. The constant $\lambda_*$ is computer-certified:
it uses the $\OvFullRows$ weights printed in \cref{tab:ovfull} and a sign check of $\OvFullCells$ quadratic pieces in
exact rational arithmetic, done by a short released script. Neither proof uses a search. Part~(a) improves the bound $0.3341104$ of \prevOV. Part~(c) tracks a potential of the junctions of $\sigma$ whose two letters are
still single letters; it holds for every comparator, optimal or not, with no assumption on equal pairs, nested merges or
ties. Since $\Lambda_4<\cc$, part~(b) gives $\bar\rho^{\mathrm{ov}}_k<\cc$ for every sufficiently large budget $k$, while
$\rho_k\ge c_k>\cc$ for every $k\ge2$ by \cref{thm:main}. So at every sufficiently large budget, as worst-case ratios,
the deployed rule is strictly worse than replacement counts, even with its ties broken in its favour; the strings of
part~(b) use $\Theta(k)$ letters (\cref{app:fourphase}), and over a fixed alphabet such as bytes the comparison is open. The proved
bracket is $[\lambda_*,\Lambda_4]\approx[\OvLBfullShort,\OvFourCapprox]$, and $[\frac7{18},\Lambda_4]$ with the hand-checkable calculation alone.
By part~(c), a string on which some run saves less than $\frac25$ of the optimum has no optimal sequence whose final
tokens have at most three letters: the part of the bracket below $\frac25$ is due entirely to optimal tokens of four or
more letters. \textbf{The exact worst-case constant of overlap greedy is open.}

\paragraph{A certificate for a single run.} \Cref{cor:rc} needs only that the prefix consists of full merges, so it
applies to every prefix of an overlap-greedy run, or of any run of full merges:
\[
\OPT(s,k)\;\le\;\min_{0\le i\le k}\bigl(2O_i+k\,\mu(s_i)\bigr),
\]
where $\mu(s_i)$ is the largest replacement count of $s_i$. Here $O_i$ is the saving of the run's first $i$ steps. The right side is computed from the run alone, so a trainer
can certify after the fact that it saved at least $O_k/\min_i(2O_i+k\mu(s_i))$ of the optimum on its own corpus,
provided the corpus is uniquely decodable (\cref{prop:identity}); by \cref{lem:tablepw} the same holds on word tables,
where $O_i$ and $\mu$ are weighted.

\section{Related work}\label{sec:related}

\paragraph{Curvature and the work of Zouhar et al.} For a monotone submodular set function under a cardinality
constraint $k$, greedy achieves $1-1/e$ of the optimum \cite{nemhauser1978}, and $\frac1c(1-e^{-c})$ if the function has
total curvature $c$ \cite{conforti1984}; Zhang et al.~\cite{zhang2016string} extend these bounds to string-submodular
functions. \Cref{thm:main} is this template with coefficient $2$, $\OPT-2G_{i+1}\le(1-\frac2k)(\OPT-2G_i)$; the merges of
BPE form a sequence, not a set, and the content of the theorem is the inequality the template needs. Zouhar et al.\
prove a bound $\frac1{\sigma_{\mathrm Z}}(1-e^{-\sigma_{\mathrm Z}})$ for greedy BPE against the best full-merge sequence
of the same length, where $\sigma_{\mathrm Z}$ (their $\sigma$) is a total backward curvature adapted to merge sequences
\cite[Theorem~3.10 and Definition~3.9]{zouhar2023formal}, and write ``We do not have a formal bound for
$\sigma(\mu^\star)$'' \cite[Section~3.3]{zouhar2023formal}. Their proof uses $\sigma_{\mathrm Z}$ only through a step
inequality along greedy's own run \cite[Appendix~A]{zouhar2023formal}, which at $\sigma_{\mathrm Z}=2$ follows from
\cref{cor:rc}. So \cref{thm:main} is their curvature-$2$ recurrence proved unconditionally, against the partial-merge
optimum and for every tie-breaking. It does not bound $\sigma_{\mathrm Z}$ itself, which can exceed $2$
(\cref{app:threewit}).

\paragraph{A different algorithm.} Kozma and Voderholzer also give EvenOdd, ``a linear time $0.5$-\hspace{0pt}approximation for
OPE'' \cite[Theorem~4]{kozma2026arxiv}, the partial-merge problem. It takes the $k$ most frequent pairs of letters of
the input and merges a set of their occurrences that do not overlap, keeping at least half of them; so every token it
builds has two letters, and its merges are partial. Its guarantee exceeds $\cc$, but it does not compute the vocabulary
that BPE trainers deploy, and Kozma and Voderholzer note that BPE appears to do better on typical inputs and call
EvenOdd a proof of concept \cite[Section~1]{kozma2026arxiv}. Our question, like theirs about BPE, is how good the
greedy rule itself is.

\paragraph{Re-Pair and other objectives.} Re-Pair \cite{larsson2000repair} runs the same greedy process until no pair
repeats and is judged by the size of the resulting grammar \cite{charikar2005smallest}; its approximation ratio for
grammar size is $\Omega(\log n/\log\log n)$, already on binary strings \cite{hucke2017repair,bannai2021smallest}. For
compressed length greedy's ratio can grow linearly in $n$ \cite[Theorem~3]{kozma2026arxiv}. Tokenisers that select a
vocabulary and then segment the text, such as GreedTok \cite{lim2025partition} or the Unigram model
\cite{kudo2018subword}, optimise different objectives.

\section{Limitations and open problems}\label{sec:limits}

\paragraph{Replacement counts versus overlapping counts.} \Cref{thm:main,thm:exact,thm:table} concern greedy that ranks
pairs by replacements; for the overlap-count rule of the trainers, \cref{prop:overlap} gives only the bracket
$[\lambda_*,\Lambda_4]\approx[\OvLBfullShort,\OvFourCapprox]$, whose left end rests on an exact finite sign check by computer (with
the hand-checkable calculation alone, $[\frac7{18},\Lambda_4]$), and the bound $\frac25$ against optima whose final tokens have at most three letters. Whether the bound
$\frac25$ extends to all optima, which would place the constant in $[\frac25,\Lambda_4]$, is open. No reweighting
of the inequalities behind $\lambda_*$ gives it: a mixture of steps at seven (type, score) points meets all $\OvRelaxRows$ averaged
bounds of the two certificates while saving only $\OvRelaxGain\,\OPT$ (\cref{app:parta}). Other inequalities might exclude
such mixtures, and whether one is realised by a string is open. If at every
step the pair chosen by overlapping counts also makes the most replacements, as when no tokenisation along the run has
three equal adjacent tokens, the run is a replacement-count run with some tie-breaking, and \cref{thm:main,thm:table}
apply to it: the raw count and the replacement count of a pair differ only on runs of three or more equal tokens
(\cref{app:overlap}), as also observed in \citeAat{Section~8}. Three equal adjacent tokens, such as three dots, spaces
or zeros, break it.

\paragraph{What the model covers.} The word-table model with string-identified symbols matches the trainers in what it
merges: frequency-weighted pairs inside pre-tokenised words, full left-to-right scans, and symbols identified by their
spelling, on uniquely decodable inputs (which fail, for example, when the corpus contains a trainer's end-of-word marker
string). Whether \cref{thm:main} holds for string-identified symbols on inputs that are not uniquely decodable is open;
the shadow inequality itself can fail there (\cref{sec:identity}). \Cref{app:fidelity} checks the remaining details of subword-nmt and Hugging Face: a vocabulary-size budget
changes nothing on uniquely decodable inputs, while a few options lie outside the model. All results assume that the
trainer takes all $k$ merges. subword-nmt stops as soon as no pair has frequency at least \texttt{min\_frequency}, $2$ by
default, and Hugging Face's trainer has the same option, \texttt{min\_frequency} \cite{hftokenizers}; on a corpus in
which every pair occurs once, a cut-off of $2$ learns nothing while the optimum is positive, so no multiplicative
guarantee survives the cut-off. What survives is additive: a run stopped at a cut-off $f$ is a prefix of
a legal run whose remaining steps each save at most $f-1$ symbols, so it loses at most $f-1$ symbols per unused merge.

\paragraph{Alphabets, fixed budgets, and what is measured.} The worst case $\cc$ needs alphabets that grow with $k$
(\cref{sec:upper}), and the overlap family behind $\Lambda_4$ needs more: written as one string it puts a distinct
separator letter between any two consecutive blocks, and as a string or a word table it uses $\Theta(k)$ private letters
(\cref{app:fourphase}). At a fixed budget $k$ the worst case $\rho_k$ can lie above $c_k$, as it does at $k=2$ and, by
\cref{cor:rate}, at every $k\ge4\cdot10^{10}$. At $k=3$
the bounds are $13/27\le\rho_3\le7/12$ (the upper bound from \citeA{}), and to our knowledge the exact value of $\rho_k$ is open for every
$k\ge3$. Its order is known, $\rho_k-\cc=\Theta(k^{-1/2})$ (\cref{cor:rate}), but not its leading constant: $\sqrt k\,(\rho_k-\cc)$
lies between $10^{-9}$ and $8$ for every $k\ge2$, and on square budgets at most about $e^{-2}$. \Cref{thm:rate} is stated for
strings only (with fresh symbols, and with symbols identified by spelling on uniquely decodable strings); we do not claim it
for word tables. Over all strings, the smallest budget at which some string has ratio below $\cc+0.01$ for every
greedy run lies between $14$ (by \cref{thm:main}) and $\UpK$ (by the certificate of \cref{sec:upper}); if ties may be
broken against greedy, it lies between $14$ and $100$ (\cref{rem:m1}). All results are statements about worst-case ratios
over all inputs. They concern the tokens saved on the input on which the merges are learned, not the encoding of new
text.

\ifanon\else
\paragraph{Use of AI tools.} AI-assisted tools, including Claude (Anthropic) and GPT-6.2 Astra (OpenAI), were used
throughout the research process to support mathematical reasoning, computational experimentation, proof verification, and
manuscript preparation. The authors guided the research direction, directed and evaluated the central arguments, and
oversaw the computational and verification workflows. The authors take full responsibility for the results and
conclusions presented.
\fi

\bibliographystyle{plainurl}
\bibliography{refs}

\begin{thebibliography}{10}

\bibitem{prior2026fixed}
{\ifanon Anonymous\else Kundan Baliga, Aiden Xie, and Aditya Byju\fi}.
\newblock {GreedTok} is a 0.3287-approximation, and greedy tokenizers lose
  order $1/\log k$ on fixed alphabets.
\newblock Preprint, Zenodo, version 1.0, 2026.
\newblock \href {https://doi.org/10.5281/zenodo.24451478}
  {\path{doi:10.5281/zenodo.24451478}}.

\bibitem{baliga2026bpe}
Kundan Baliga, Aiden Xie, and Aditya Byju.
\newblock Greedy byte-pair encoding achieves 0.391 of the optimum, past the
  one-third barrier for packing-based analyses.
\newblock arXiv:2611.08671v2, 2026.
\newblock URL: \url{https://arxiv.org/abs/2611.08671v2}.

\bibitem{bannai2021smallest}
Hideo Bannai, Momoko Hirayama, Danny Hucke, Shunsuke Inenaga, Artur Je{\.z},
  Markus Lohrey, and Carl~Philipp Reh.
\newblock The smallest grammar problem revisited.
\newblock {\em IEEE Transactions on Information Theory}, 67(1):317--328, 2021.
\newblock arXiv:1908.06428.
\newblock \href {https://doi.org/10.1109/TIT.2020.3038147}
  {\path{doi:10.1109/TIT.2020.3038147}}.

\bibitem{charikar2005smallest}
Moses Charikar, Eric Lehman, Ding Liu, Rina Panigrahy, Manoj Prabhakaran, Amit
  Sahai, and Abhi Shelat.
\newblock The smallest grammar problem.
\newblock {\em IEEE Transactions on Information Theory}, 51(7):2554--2576,
  2005.
\newblock \href {https://doi.org/10.1109/TIT.2005.850116}
  {\path{doi:10.1109/TIT.2005.850116}}.

\bibitem{conforti1984}
Michele Conforti and G{\'e}rard Cornu{\'e}jols.
\newblock Submodular set functions, matroids and the greedy algorithm: Tight
  worst-case bounds and some generalizations of the {Rado--Edmonds} theorem.
\newblock {\em Discrete Applied Mathematics}, 7(3):251--274, 1984.
\newblock \href {https://doi.org/10.1016/0166-218X(84)90003-9}
  {\path{doi:10.1016/0166-218X(84)90003-9}}.

\bibitem{gage1994}
Philip Gage.
\newblock A new algorithm for data compression.
\newblock {\em The C Users Journal}, 12(2):23--38, 1994.

\bibitem{hucke2017repair}
Danny Hucke, Artur Je{\.z}, and Markus Lohrey.
\newblock Approximation ratio of {RePair}.
\newblock arXiv:1703.06061, 2017.
\newblock URL: \url{https://arxiv.org/abs/1703.06061}.

\bibitem{hftokenizers}
{Hugging Face}.
\newblock tokenizers.
\newblock Software repository, \url{https://github.com/huggingface/tokenizers}.
\newblock File \texttt{tokenizers/src/models/bpe/trainer.rs} at release
  v0.21.0; file \texttt{word.rs} in the same directory, retrieved 2026-11-22.

\bibitem{kozma2026esa}
L{\'a}szl{\'o} Kozma and Johannes Voderholzer.
\newblock Theoretical analysis of byte-pair encoding.
\newblock In {\em 34th Annual European Symposium on Algorithms (ESA 2026)},
  volume 388 of {\em Leibniz International Proceedings in Informatics
  (LIPIcs)}, pages 80:1--80:17. Schloss Dagstuhl -- Leibniz-Zentrum f{\"u}r
  Informatik, 2026.
\newblock \href {https://doi.org/10.4230/LIPIcs.ESA.2026.80}
  {\path{doi:10.4230/LIPIcs.ESA.2026.80}}.

\bibitem{kozma2026arxiv}
L{\'a}szl{\'o} Kozma and Johannes Voderholzer.
\newblock Theoretical analysis of byte-pair encoding.
\newblock arXiv:2411.08671v2, 2026.
\newblock Extended version of the ESA 2026 paper, version 2.
\newblock URL: \url{https://arxiv.org/abs/2411.08671v2}.

\bibitem{kudo2018subword}
Taku Kudo.
\newblock Subword regularization: Improving neural network translation models
  with multiple subword candidates.
\newblock In {\em Proceedings of the 56th Annual Meeting of the Association for
  Computational Linguistics (Volume 1: Long Papers)}, pages 66--75. Association
  for Computational Linguistics, 2018.
\newblock \href {https://doi.org/10.18653/v1/P18-1007}
  {\path{doi:10.18653/v1/P18-1007}}.

\bibitem{larsson2000repair}
N.~Jesper Larsson and Alistair Moffat.
\newblock Off-line dictionary-based compression.
\newblock {\em Proceedings of the IEEE}, 88(11):1722--1732, 2000.
\newblock \href {https://doi.org/10.1109/5.892708}
  {\path{doi:10.1109/5.892708}}.

\bibitem{lim2025partition}
Jia~Peng Lim, Shawn Tan, Davin Choo, and Hady~W. Lauw.
\newblock A partition cover approach to tokenization.
\newblock In {\em Advances in Neural Information Processing Systems (NeurIPS
  2025)}, 2025.
\newblock arXiv:2501.06246.

\bibitem{nemhauser1978}
George~L. Nemhauser, Laurence~A. Wolsey, and Marshall~L. Fisher.
\newblock An analysis of approximations for maximizing submodular set
  functions---{I}.
\newblock {\em Mathematical Programming}, 14(1):265--294, 1978.
\newblock \href {https://doi.org/10.1007/BF01588971}
  {\path{doi:10.1007/BF01588971}}.

\bibitem{subwordnmt}
Rico Sennrich.
\newblock subword-nmt.
\newblock Software repository, \url{https://github.com/rsennrich/subword-nmt}.
\newblock File \texttt{learn\_bpe.py}, retrieved 2026-11-22, SHA-256
  \texttt{5f897d9bbd6f34b6\ldots}.

\bibitem{sennrich2016bpe}
Rico Sennrich, Barry Haddow, and Alexandra Birch.
\newblock Neural machine translation of rare words with subword units.
\newblock In {\em Proceedings of the 54th Annual Meeting of the Association for
  Computational Linguistics (Volume 1: Long Papers)}, pages 1715--1725.
  Association for Computational Linguistics, 2016.
\newblock \href {https://doi.org/10.18653/v1/P16-1162}
  {\path{doi:10.18653/v1/P16-1162}}.

\bibitem{yang2026}
Hanyu Yang.
\newblock Byte-pair encoding can lose more than half of the optimal
  compression.
\newblock Preprint, Zenodo, version 1.1, 2026.
\newblock \href {https://doi.org/10.5281/zenodo.24301507}
  {\path{doi:10.5281/zenodo.24301507}}.

\bibitem{zhang2016string}
Zhenliang Zhang, Edwin K.~P. Chong, Ali Pezeshki, and William Moran.
\newblock String submodular functions with curvature constraints.
\newblock {\em IEEE Transactions on Automatic Control}, 61(3):601--616, 2016.
\newblock arXiv:1303.3018.
\newblock \href {https://doi.org/10.1109/TAC.2015.2440566}
  {\path{doi:10.1109/TAC.2015.2440566}}.

\bibitem{zouhar2023formal}
Vil{\'e}m Zouhar, Clara Meister, Juan Gastaldi, Li~Du, Tim Vieira, Mrinmaya
  Sachan, and Ryan Cotterell.
\newblock A formal perspective on byte-pair encoding.
\newblock In {\em Findings of the Association for Computational Linguistics:
  ACL 2023}, pages 598--614. Association for Computational Linguistics, 2023.
\newblock Theorem and definition numbers as in arXiv:2306.16837v3.
\newblock \href {https://doi.org/10.18653/v1/2023.findings-acl.38}
  {\path{doi:10.18653/v1/2023.findings-acl.38}}.

\end{thebibliography}

\appendix
\crefalias{section}{appendix}
\crefalias{subsection}{subappendix}

\section{Proof of the matching construction}\label{app:upper}

\paragraph{Parameters.} Take integers $d\ge2$ and $F\ge d$, and put $k=d^2$. Set $h_0=F$ and, for $0\le t<d$,
\[
w_t=\lfloor h_t/d\rfloor,\qquad h_{t+1}=h_t-2w_t,\qquad \delta_t=h_t+1-d\,w_t\quad(1\le \delta_t\le d).
\]
Assume (H): $w_t\ge1$ for $t\le d-2$, and $h_{d-1}\ge\max(1,w_0)$. All $h_t$ are nonnegative.

\paragraph{The string.} The letters are $X_i,Y_i$ for $i\in\mathbb Z_d$, and guard letters $H_0,H_1,\dots$ (below);
indices of letters are taken modulo $d$. The blocks are
\begin{center}\small
\begin{tabular}{@{}lll@{}}
\toprule
Block & Indices & Copies per index tuple\\
\midrule
$X_i\,Y_j\,X_{j+t}\,Y_i$ & $(t,j,i)\in\{0,\dots,d-1\}^3$ & $w_t$\\
$Y_j\,X_{j+t}$ & $(t,j)\in\{0,\dots,d-1\}^2$ & $\delta_t$\\
$X_i\,Y_j$ & $(i,j)\in\{0,\dots,d-1\}^2$ & $h_d$\\
\bottomrule
\end{tabular}
\end{center}
taken in this order, rows in the displayed order, index tuples in increasing lexicographic order, equal copies
consecutive, with one guard letter between consecutive blocks. The pairs $(X_i,Y_j)$ are the \emph{side} pairs and the
pairs $(Y_j,X_{j+t})$ the \emph{middle} pairs of level $t$.

\paragraph{Guards.} Let $\Delta=\max\{d(F-\sum_tw_t),\ \sum_t\delta_t\}$ and $J=2\lfloor(\Delta-1)/h_{d-1}\rfloor+1$. Each
letter starts at most $\Delta$ blocks and ends at most $\Delta$ blocks: $X_i$ starts $d(\sum_tw_t+h_d)=d(F-\sum_tw_t)$
blocks (using $h_d=F-2\sum_tw_t$) and $Y_j$ starts $\sum_t\delta_t$, and the end counts are the same with $X$ and $Y$
interchanged. The first $J$ boundaries (or all, if there are fewer) receive $H_0,\dots,H_{J-1}$ in order. Each later
boundary, with letter $a$ on its left and $b$ on its right, receives the guard $H_u$ of least index $u$ such that $aH_u$
and $H_ub$ have each occurred fewer than $h_{d-1}$ times so far. Before this boundary at most $\Delta-1$ boundaries have
$a$ on the left, so at most $\lfloor(\Delta-1)/h_{d-1}\rfloor$ guards are full for $a$, and likewise for $b$; as
$J>2\lfloor(\Delta-1)/h_{d-1}\rfloor$, a legal guard exists. So every ordered pair of a letter and a guard occurs at most
$h_{d-1}$ times. As long as merges stay inside blocks, a token next to a guard begins or ends with the original letter
next to that guard, so every pair of a token and a guard has count at most $h_{d-1}$ at every such step.

\paragraph{The comparator.} Merge the $d^2$ side pairs, for example in lexicographic order. For a side pair $(X_a,Y_b)$,
each level $t$ contributes $w_t$ occurrences as the left pair of the blocks $(t,b,a)$ and $w_t$ as the right pair of the
blocks $(t,a-t,b)$ (distinct positions, even when the two coincide as types), and the last row contributes $h_d$. These
occurrences are disjoint, and the side merges do not interfere, so every rule saves $2\sum_tw_t+h_d=F$ and
\begin{equation}\label{eq:upopt}
\OPT(s,k)\;\ge\;kF .
\end{equation}
The side merges are full merges, so also $\OPT^m(s,k)\ge kF$, and every upper bound on the ratio proved below holds
against the full-merge optimum as well.

\paragraph{Every greedy run.} We show that in phase $t=0,\dots,d-1$ greedy merges the $d$ middle pairs of level $t$ in
some order, each saving $h_t+1$. Suppose phases $0,\dots,t-1$ went so. In each earlier level $a$, each side pair has
lost its $w_a$ occurrences as a left pair and its $w_a$ occurrences as a right pair, so it has count
$F-2\sum_{a<t}w_a=h_t$. An unmerged middle pair of level
$a\ge t$ has count $dw_a+\delta_a=h_a+1$; different middle pairs occupy different blocks and do not affect each other's counts,
and $h_a+1<h_t+1$ for $a>t$ because $w_t\ge1$ for $t\le d-2$. The other adjacencies are of three kinds. A middle token
$Z=[Y_jX_v]$ of level $a$ has, inside a four-letter block, the neighbours $X_i$ and $Y_i$, and each pair $(X_i,Z)$ or
$(Z,Y_i)$ occurs $w_a\le w_0\le h_{d-1}$ times; in a two-letter block it has no neighbour inside the block. Two middle
tokens are never adjacent. Pairs with a guard occur at most $h_{d-1}$ times. Throughout phase $t$ the side counts only
fall, and $h_{d-1}\le h_t$. So the maximum count is $h_t+1$, attained exactly by the unmerged middle pairs of level
$t$, whose occurrences are disjoint; this holds after each merge of the phase, and after the last one every side pair has
lost its $2w_t$ occurrences of level $t$. The budget ends with the last phase. Every one of the $(d!)^d$ greedy histories
saves
\begin{equation}\label{eq:upG}
G_k=d\sum_{t=0}^{d-1}(h_t+1).
\end{equation}

\paragraph{Proof of \cref{thm:upper}.} Put $\theta=1-2/d$. Since $h_{t+1}=\theta h_t+2\{h_t/d\}$ with $0\le2\{h_t/d\}<2$, induction
gives $F\theta^t\le h_t\le F\theta^t+2\sum_{s<t}\theta^s<F\theta^t+d$, where the last step uses
$\sum_{s<t}\theta^s<1/(1-\theta)=d/2$. For $d\ge8$, $\theta^{d-1}\ge(3/4)^7>1/8$: for $d>2$,
$(d-1)\log \theta=-2\sum_{r\ge0}(d-1)^{-2r}/(2r+1)$ increases with $d$. If also $F\ge8d$, then $h_{d-1}>F/8\ge d$, so
$h_{d-1}\ge F/d\ge w_0$ and $h_t\ge h_{d-1}\ge d$ gives $w_t\ge1$ for all $t<d$; (H) holds. By \eqref{eq:upopt} and
\eqref{eq:upG},
\[
\frac{\BPE(s,k)}{\OPT(s,k)}\;\le\;\frac1{dF}\sum_{t<d}(h_t+1)\;<\;\frac1d\sum_{t<d}\theta^t+\frac{d+1}F\;=\;c_d+\frac{d+1}F ,
\]
since $\frac1d\sum_{t<d}\theta^t=\frac{1-\theta^d}{2}=c_d$. With $F=d^3$ the error is $d^{-2}+d^{-3}$, and $c_d\to\cc$. For the
alphabet, $\Delta\le dF$ (as $\sum_t\delta_t\le d^2\le dF$), so $(\Delta-1)/h_{d-1}<dF/(F/8)=8d$ and $J\le16d-1$; the string
uses at most $2d+J<18d$ letters. Conversely the $2d$ letters $X_i,Y_i$ are needed, so the alphabet has $\Theta(\sqrt k)$
letters. Every member of the family has ratio at least $c_d$: the largest pair count of $s$ is $F+1$, so by
\cref{lem:cap} every partial merge makes at most $F+1$ replacements, $\OPT\le k(F+1)$, and the ratio is at least
$(Fc_d+1)/(F+1)\ge c_d$ by $h_t\ge F\theta^t$. The sequence $c_d$ decreases, since $x\log(1-2/x)$ increases for $x>2$.

\paragraph{The certificate at $k=\UpK$.} Take $d=\UpD$, $F=\UpF$. The four-letter block multiplicities are
$(\UpBlocks)$, $h_{14}=\UpResidual$, $h_{13}=\UpCap$, $\Delta=\UpDD$ and $J=\UpJ$; the string has $\UpLength$ letters over
an alphabet of $\UpAlphabet$. Writing $X_i$, $Y_j$, $H_u$ as the bytes $i$, $14+j$, $28+u$, its SHA-256 begins
\texttt{\UpShaShort}. The phase gains $h_t+1$ are $(\UpGains)$, each taken $14$ times, so every greedy run saves
$\UpG$, and the side merges give $\Uside=196\cdot\UpF=\UpOPT$.

\emph{The partial optimum is exactly $\Uside$.} Every replacement of a partial merge turns an occurrence of the word spelled
by its output into one token, and the replacements of one rule use disjoint occurrences of that word. If the word
contains a guard, it contains a two-letter factor made of that guard and a neighbour, which occurs at most $h_{13}=\UpCap$
times. A guard-free word of three or more letters lies inside a four-letter block and contains a factor $X_iY_jX_v$ or
$Y_jX_vY_i$, which determines the block type and so occurs at most $w_0=\UpTrigram$ times. The remaining rules join two
letters and save at most $F+1=50{,}001$. To beat $\Uside$ with $196$ rules (pad with empty ones), each would have to save more
than $\Uside-195\cdot50{,}001=\UpCutoff$. Only the side pairs ($F$ occurrences) and the level-$0$ middle pairs ($F+1$) occur
that often, and two rules cannot use the same such pair, since together they would need more than its $F+1$
occurrences. Suppose $a$ of the rules are level-$0$ middle pairs and $196-a$ are side pairs, so $a$ side pairs are
omitted; the chosen pairs have $\Uside+a$ occurrences in all. Each level-$0$ middle pair meets two side occurrences in each of
its $14\cdot w_0$ four-letter blocks, $\UpIncMid$ meetings, and an omitted side pair takes part in only $2w_0=\UpIncSide$
meetings with level-$0$ middles. So at least $a\,(\UpIncMid-\UpIncSide)=2a\cdot13\cdot w_0$ meetings involve a chosen
middle occurrence and a chosen side occurrence, which share a letter, and in each at least one of the two is not
replaced. A missed occurrence accounts for at most two meetings (a middle occurrence has two sides, and a side
occurrence at most one middle), so the utility is at most $\Uside+a-13a\,w_0\le \Uside$, also when $a=0$. Hence
$\OPT(s,196)=\Uside$ and every greedy run has ratio exactly $\UpG/\UpOPT=\UpRatio$.

\emph{The first budget.} Since every member of the family has ratio at least $c_d$, and
$c_{13}\approx\CkThirteen>\cc+0.01>c_{14}\approx\CkFourteen$, the family has no ratio below $\cc+0.01$ for $d\le13$, and
$d=14$ has one. By \cref{thm:main} the same comparison excludes every budget $k\le13$ over all strings. The program
compares these rationals using Taylor bounds on $e^{-2}$.

\section{Proof of the finite-budget theorem}\label{app:rate}

Fix a string $s$, a budget $k\ge2$, a greedy run $s_0,\dots,s_k$ with gains $g_j$, and an optimal partial $k$-sequence
$\sigma$ with forest $\Fo$ and utility $U=\OPT(s,k)$; if $U=0$ there is nothing to prove. For each prefix $i\le k$ of the
run, the shadow, the windows, the marks and the kept windows are those of \cref{sec:proof} for greedy's first $i$ merges.
The gains do not increase (\cref{lem:mono}), and $G_k\le U$, since greedy's merges are themselves a partial
$k$-sequence. Put $f=U/k$, $\theta=1-2/k$, and
\[
S_i=2G_i+k\,g_{i+1}-U\qquad(0\le i\le k-1),
\]
which is nonnegative by \cref{cor:rc}. Since $g_{i+1}=(U-2G_i+S_i)/k$, the deficit $U-2G_i$ satisfies
$U-2G_{i+1}=\theta(U-2G_i)-2S_i/k$, and unrolling gives exactly
\begin{equation}\label{eq:slack}
G_k\;=\;c_k\,U+\frac1k\sum_{i=0}^{k-1}\theta^{k-1-i}S_i .
\end{equation}

\paragraph{Step 1: an inventory of window ends.} For a letter $A$, let $E_A$ and $B_A$ be the numbers of distinct windows,
identified by their positions, that are marked at some prefix $i\le k$ and whose last, respectively first, letter is
$A$. For a symbol $Z$ that greedy creates at step $j$, let $E_Z=B_Z=g_j$, the number of tokens $Z$ created. Then
\begin{equation}\label{eq:inventory}
\sum_X E_X\;\le\;3U,\qquad \sum_Y B_Y\;\le\;3U,
\end{equation}
the sums over all letters and all symbols created by greedy. The created symbols contribute $\sum_jg_j=G_k\le U$. At
prefix $0$ the shadow consists of single letters, the windows are the pairs of adjacent positions, and such a window is
marked exactly when its gap is a junction: $U$ marked windows (F1). The shadow at prefix $i+1$ is the shadow at prefix
$i$ with the replacements of at most one further letter-pair merge. Each replacement turns two consecutive single letters
$x_j,x_{j+1}$ of the shadow into a dimer; made one at a time, it removes two consecutive entries from the list of single
positions and creates at most one new window, from $x_{j-1}$ to $x_{j+2}$. Whether a window is marked depends only on its
positions and on $\Fo$. So at most $U+M_k$ windows are marked at some prefix, where $M_k\le G_k\le U$ is the number of
dimers of the shadow at prefix $k$ (\cref{lem:lost}). Each window has one first and one last letter, which proves
\eqref{eq:inventory}.

\paragraph{Step 2: freezing the kept windows.} Fix prefixes $0\le i<i+\ell\le k-1$, put $u=g_{i+1}$ and $v=g_{i+\ell+1}$,
assume $v>0$, and let $\Delta=u-v$ and $I=\{i+1,\dots,i+\ell\}$. Freeze the kept windows of prefix $i$: each is the
interval $(\alpha-1,\beta]$ of a kept window on positions $\alpha,\dots,\beta$, with the label $r$ of its mark. By the
kept-window family in the proof of \cref{thm:pointwise} there are at least $U-2G_i$ of them; let $n_r$ be the number of
label $r$. Those of one label spell the same word and lie in the spans of distinct nodes of label $r$, which are disjoint
(F3); their outer cuts $\alpha-1$ and $\beta$ survive in $s_i$. For $i\le j\le i+\ell$ call a frozen interval \emph{live} at
$j$ if both of its outer cuts survive in $s_j$ and it contains at least two tokens of $s_j$. Every frozen interval is
live at $i$, and an interval that dies stays dead, since erased cuts stay erased and tokens are only joined. By
\cref{lem:iso} the live intervals of one label carry the same tokens of $s_j$; their first two tokens give disjoint
occurrences of one pair, so by \cref{lem:cnt} at most $\mu(s_j)=g_{j+1}$ intervals of a label are live at $j$. In
particular $n_r\le u$, and at most $v$ intervals of each label are live at $i+\ell$.

\emph{Charges.} For each label $r$, \emph{charge} its first $(n_r-v)_+$ deaths, ordered by step and, within a step, by
position. Every charged death occurs at a step of $I$; at the start of that step the label has more than $v$ live
intervals; and a label receives at most $(n_r-v)_+\le\Delta$ charges. Let step $j$ merge the pair $(X,Y)$. A death at
step $j$ is a \emph{right} death if the step erases the outer cut $\beta$; otherwise a \emph{left} death if it erases
$\alpha-1$; otherwise a \emph{collapse}: both outer cuts survive and the interval becomes one token. At step $j$ the
charged right deaths, left deaths and collapses number at most, respectively,
\begin{equation}\label{eq:charges}
\min\bigl(g_j,\ E_X\Delta/v\bigr),\qquad \min\bigl(g_j,\ B_Y\Delta/v\bigr),\qquad 2g_j\Delta/v .
\end{equation}
\emph{Right deaths.} Distinct windows have distinct last positions, so distinct frozen intervals have distinct right outer
cuts, and the step erases $g_j$ cuts. Also, a replacement that erases $\beta$ joins the last token of the interval, as
its left operand, to the next token; so just before step $j$ the interval ends with a token $X$. A label charged with a
right death at step $j$ has more than $v$ live intervals, all with the same tokens, so all of them end with a token $X$.
If $X$ is a letter, it is the letter at $\beta$, so each of these windows is counted in $E_X$. If $X$ was created by
greedy, the tokens $X$ that end at the distinct positions $\beta$ are distinct, and there are $E_X$ tokens $X$ in all.
Either way each such label accounts for more than $v$ of the $E_X$ windows or tokens, distinct across labels, so fewer
than $E_X/v$ labels receive right charges at step $j$, each at most $\Delta$ of them. \emph{Left deaths}: the same
argument with first positions and $Y$. \emph{Collapses.} The replacements of one step join disjoint pairs of adjacent
tokens, so an interval whose outer cuts survive becomes one token only if it consisted of exactly two tokens, $X$ and
$Y$. So every live interval of a label charged with a collapse at step $j$ consists of $X$ followed by $Y$, and distinct
intervals give distinct adjacent pairs of tokens $X,Y$ in $s_{j-1}$. There are at most
$\raw_{s_{j-1}}(X,Y)\le2\cnt_{s_{j-1}}(X,Y)=2g_j$ of them (a run of $b$ equal tokens has $b-1\le2\lfloor b/2\rfloor$
adjacencies), so fewer than $2g_j/v$ labels receive collapse charges at step $j$. This proves \eqref{eq:charges}.

\paragraph{Step 3: one interval.} Let $\tau=30\sqrt k\,f$. Call a step $j$ with $g_j>0$ \emph{double} if its pair $(X,Y)$
has $E_X>\tau$ and $B_Y>\tau$, and let $\mathcal D$ be the set of double steps. By \eqref{eq:inventory} fewer than
$3U/\tau$ symbols have $E_X>\tau$, and likewise for $B$. Greedy never merges one ordered pair twice with positive gain:
after the full merge of $(X,Y)$ no token $X$ is followed by a token $Y$ (for $X=Y$, each run of $X$ keeps at most one
token $X$), and later merges create only fresh symbols and never make two old tokens adjacent. Hence
\begin{equation}\label{eq:double}
|\mathcal D|\;\le\;(3U/\tau)^2\;=\;k/100 .
\end{equation}
The total number of charges over $I$ is $\sum_r(n_r-v)_+\ge\sum_rn_r-kv\ge U-2G_i-kv$, since $\sigma$ has $k$ labels,
and $U-2G_i-kv=2\sum_{j\in I}g_j-S_{i+\ell}$. By \eqref{eq:charges} a step $j\in I\setminus\mathcal D$ carries at most
$g_j+\tau\Delta/v+2g_j\Delta/v$ charges, since one of its two outer bounds is at most $\tau\Delta/v$ and the other at most
$g_j$, and a step of $\mathcal D$ at most $2g_j+2g_j\Delta/v$. With $g_j\le u$ on $I$ this gives
\begin{equation}\label{eq:interval}
S_{i+\ell}\;\ge\;\sum_{j\in I\setminus\mathcal D}g_j-\ell\,(\tau+2u)\,\frac{u-v}{v}.
\end{equation}
The inventory and $\mathcal D$ are fixed for the whole run; the frozen intervals and their labels are chosen anew for
each $i$.

\paragraph{Step 4: summing over the middle of the run.} Assume $G_k<\frac9{20}U$ and $k\ge4\cdot10^{10}$, and put
$\ell=\lfloor\sqrt k/200000\rfloor\ge1$. For $j<k$, \cref{cor:rc} and $G_j\le G_k$ give
$g_{j+1}\ge(U-2G_k)/k>f/10$. Take the starts $i=\lceil k/4\rceil,\dots,\lfloor k/2\rfloor-1$, at least $k/5$ of them (as $k\ge25$).
Then $i+\ell\le k-1$, $v>f/10>0$, and $u=g_{i+1}\le G_{i+1}/(i+1)\le4G_k/k\le4f$ by monotonicity. Sum
\eqref{eq:interval} over the starts. The pairs (start, step of its interval) number at least $k\ell/5$; by
\eqref{eq:double} at most $k\ell/100$ of them have the step in $\mathcal D$, and each other pair contributes $g_j>f/10$,
so the positive terms sum to at least $(\frac15-\frac1{100})k\ell f/10=\frac{19}{1000}k\ell f$. For the error terms,
$(\tau+2u)/v\le10(30\sqrt k+8)$, and $\sum_i(g_{i+1}-g_{i+\ell+1})\le4\ell f$, because each difference $g_m-g_{m+1}$
occurs for at most $\ell$ starts and these differences sum to at most the largest gain in the range, which is at most
$4f$. So the error terms sum to at most $40(30\sqrt k+8)\ell^2f\le1520\sqrt k\,\ell^2f\le\frac{1520}{200000}k\ell f$.
As $\frac{19}{1000}-\frac{1520}{200000}>\frac1{100}$, and the indices $i+\ell$ are distinct and every $S_j$ is
nonnegative,
\[
\sum_{j=0}^{k-1}S_j\;\ge\;\frac{k\ell f}{100}.
\]

\paragraph{Step 5: proof of \cref{thm:rate}.} For $k\ge4$, $\theta^k\ge(1/2)^4=\frac1{16}$, since $x\log(1-2/x)$ increases
for $x>2$. In the case of Step~4, \eqref{eq:slack}, with $\theta^{k-1-i}\ge\theta^k$ (as $0\le\theta\le1$) and $S_i\ge0$,
gives
\[
G_k-c_kU\;\ge\;\frac{\theta^k}k\sum_jS_j\;\ge\;\frac{\ell f}{1600}\;=\;\frac{\ell\,U}{1600\,k}\;\ge\;\frac{U}{6.4\cdot10^8\sqrt k},
\]
using $\ell\ge\sqrt k/400000$ (as $\lfloor x\rfloor\ge x/2$ for $x\ge1$). Since $1/(6.4\cdot10^8)>1.5\cdot10^{-9}$, here
$G_k\ge(c_k+1.5\cdot10^{-9}k^{-1/2})U$. If $G_k\ge\frac9{20}U$, then $e^{-2}>\frac18$ gives
$\frac9{20}-\cc>\frac1{80}$, which exceeds both $10^{-9}k^{-1/2}$ and, for $k\ge4\cdot10^{10}$, also
$c_k-\cc+1.5\cdot10^{-9}k^{-1/2}$ (by the bound $c_k-\cc\le e^{-2}/(k-2)$ below). Finally let $2\le k<4\cdot10^{10}$. For
$k>2$, $\log(1-x)\le-x-x^2/2$ gives $\theta^k\le e^{-2-2/k}$, so
\[
c_k-\cc\;=\;\tfrac12\bigl(e^{-2}-\theta^k\bigr)\;\ge\;\tfrac12e^{-2}\bigl(1-e^{-2/k}\bigr)\;\ge\;\frac{e^{-2}}{2k}\;>\;\frac1{16k},
\]
by $1-e^{-x}\ge x/2$ on $[0,1]$; at $k=2$, $c_2-\cc=e^{-2}/2$. As $\sqrt k<2\cdot10^5$, $\frac1{16k}>10^{-9}k^{-1/2}$, and
\cref{thm:main} gives $G_k\ge c_kU$. In all cases $G_k\ge(\cc+10^{-9}k^{-1/2})U$.

\paragraph{Proof of \cref{cor:rate}.} The lower bound and the bound $\rho_k\ge c_k+1.5\cdot10^{-9}k^{-1/2}$ for
$k\ge4\cdot10^{10}$ are Step~5, for every string and greedy run. For the upper bound, if $k<64$ then
$\rho_k\le1<\cc+8k^{-1/2}$. Let $k\ge64$, $d=\lfloor\sqrt k\rfloor\ge8$, $m=d^2$, and let $s$ be the string of
\cref{app:upper} with parameters $d$ and $F=d^3$. A greedy run of $k$ merges on $s$ begins with a greedy run of $m$ merges,
which saves less than $(c_d+d^{-2}+d^{-3})mF$ (proof of \cref{thm:upper}). The largest pair count of $s$ is $F+1$, so by
\cref{lem:mono} each of the remaining $k-m\le2d$ steps saves at most $F+1$; and $\OPT(s,k)\ge\OPT(s,m)\ge mF$. So every
greedy run has ratio at most
\[
c_d+d^{-2}+d^{-3}+\frac{2d(F+1)}{mF}\;\le\;c_d+\frac4d ,
\]
since $d^{-2}+d^{-3}\le1/d$ and $2d(F+1)/(mF)=2/d+2/d^4\le3/d$. Since $-d\log(1-2/d)=2+\sum_{r\ge2}2^r/(r\,d^{r-1})\le2+2/(d-2)$, we have $\theta_d^{\,d}\ge e^{-2}(1-\frac2{d-2})$ for
$\theta_d=1-2/d$, hence $c_d-\cc\le e^{-2}/(d-2)\le2/d$. The ratio is therefore at most $\cc+6/d<\cc+8/(d+1)<\cc+8/\sqrt k$.
The same computation with $k$ in place of $d$ gives the bound $c_k-\cc\le e^{-2}/(k-2)$ used in Step~5.

\subsection{The run at $k=100$}\label{app:m1}

Take $F=\MOneF$ and twenty main letters $X_0,\dots,X_9,Y_0,\dots,Y_9$. The $100$ side pairs are $(X_a,Y_b)$, and the
run's $100$ merges are the middle pairs $(Y_j,X_{(j+t)\bmod10})$ in increasing order of $(t,j)$. The stored data give,
for each middle pair $(Y_j,X_v)$, its gain and two lists of side pairs with integer multiplicities: the left list uses
side pairs $(X_a,Y_j)$ and the right list side pairs $(X_v,Y_b)$, and each list has total multiplicity at most the gain.
Each middle pair yields as many blocks $[X_a]\,Y_j\,X_v\,[Y_b]$ as its gain, the outer letters taken from the two
lists in order and left out once a list is exhausted. Two-letter blocks $X_aY_b$ then bring every side pair to exactly
$F$ occurrences. The blocks are listed in lexicographic order, identical copies consecutive, with a distinct guard
letter between any two consecutive blocks: $\MOneGuards$ guards in all, and the string has length $\MOneLength$. Every pair with a
guard occurs once, and all other counts are sums over the blocks.

The $100$ side merges save $F$ each, and the largest initial pair count is $F$, so $\OPT(s,100)=100F=10^{11}$ by
\cref{lem:cap}. At each of the $100$ steps a program recomputes from the block multiplicities the count of every pair
that can occur (side, middle, a merged middle with an outer letter, and guard pairs) and checks that the middle pair
merged at that step attains the maximum; a second, separately written program repeated these checks from the stored
integers alone (\cref{app:evidence}); at $\MOneTies$ steps another pair attains it as well, so the run uses its freedom to break
ties. The gains sum to $\MOneG$, and the ratio $\MOneRatio$ lies below $\cc+0.01$ by rational Taylor bounds on $e^{-2}$.
The counts depend only on the two lists of each middle pair, not on how they are paired into blocks.

\section{Proofs for word tables}\label{app:table}

\paragraph{The encoding.} For a table $\mathcal T$ with positive integer weights, let $s(\mathcal T)$ be the string that
lists $\phi_j$ copies of $\omega_j$ for every $j$, in any order, with a distinct new letter, a \emph{separator}, between
any two consecutive copies. A rule of $\mathcal T$ never involves a separator, so as long as only such rules have been
applied, each copy of $\omega_j$ is tokenised exactly as row $j$ (a full scan never replaces across a separator); for every
pair of symbols of $\mathcal T$ the string counts $\cnt$ and $\raw$ equal $D_t$ and $R_t$; and every pair that contains a
separator occurs once. Replaying a list of partial rules of $\mathcal T$ copy by copy gives a partial sequence on
$s(\mathcal T)$ with the same utility, so $\OPT(\mathcal T,k)\le\OPT(s(\mathcal T),k)$.

\begin{lemma}[pointwise inequality on tables]\label{lem:tablepw}
Let $p_1,\dots,p_i$ be full rules on a table $\mathcal T$ with positive real weights, with output $t$ and utility $G$,
and let $\sigma$ be a list of at most $k$ partial rules on $\mathcal T$ with utility $U$. Then $U\le2G+k\,\mu(t)$.
\end{lemma}

\begin{proof}
First let the weights be integers. Apply $p_1,\dots,p_i$ as full merges to $s=s(\mathcal T)$, and replay $\sigma$ copy
by copy; the string has utility $G$, the replayed sequence has utility $U$, and its nodes have spans inside copies.
\Cref{thm:pointwise} gives $U-\sum_rm_r\le2G$. The witnessing occurrences of a pair $(X,Y)$ for one label lie in disjoint
spans inside copies, so their number is at most $\sum_j\phi_j\cnt_{t_j}(X,Y)=D_t(X,Y)\le\mu(t)$ (\cref{lem:cnt}); hence
$m_r\le\mu(t)$ for each of the at most $k$ labels. For rational weights multiply all weights by a common denominator:
the rules and choices are unaffected, and $U$, $G$ and $\mu(t)$ scale by the same factor. For real weights fix the rules
$p_1,\dots,p_i$ and the rules and row-by-row choices of $\sigma$. The tokenisation of each row then does not depend on
the weights, so $U$ and $G$ are linear functions of the weight vector and $\mu(t)$ is a maximum of finitely many linear
functions. Both sides are continuous, and the inequality holds at all positive rational weight vectors, which are dense.
\end{proof}

\begin{proof}[Proof of \cref{thm:table}(a)]
Apply \cref{lem:tablepw} to greedy's first $i$ rules and an optimal $\sigma$: $\OPT(\mathcal T,k)\le2G_i+k\,\mu(t_i)=
2G_i+k\,g_{i+1}$, since greedy's next rule makes $\mu(t_i)$ weighted replacements. The recurrence in the proof of
\cref{thm:main} then gives $G_k\ge c_k\OPT(\mathcal T,k)$. For $k=1$, one partial rule of a pair $(X,Y)$ makes at most
$\cnt_{\omega_j}(X,Y)$ replacements in row $j$ (\cref{lem:cnt}), so it saves at most $D(X,Y)\le\mu=g_1$.
\end{proof}

The next lemma transfers any guarantee for runs on strings, including the guarantee for the overlap rule in
\cref{prop:overlap}(a). Its only subtle point is real weights with ties.

\begin{lemma}[transfer]\label{lem:transfer}
Fix a budget $k$ and a ranking rule, replacement counts ($D$) or overlapping counts ($R$). If every legal run of the rule
on every string $s$ saves at least $\alpha\,\OPT(s,k)$, then every legal run of the rule on every word table with
positive real weights saves at least $\alpha\,\OPT(\mathcal T,k)$.
\end{lemma}

\begin{proof}
Let the weights be integers and fix a legal run on $\mathcal T$. If it makes fewer than $k$ nonempty steps, every row ends
as one token and the run saves $\sum_j\phi_j(|\omega_j|-1)\ge\OPT(\mathcal T,k)$. Otherwise, at each step the table has an
adjacency, so its largest score is at least $1$; every pair of symbols of $\mathcal T$ has the same score in
$s(\mathcal T)$ as in $\mathcal T$, and every pair with a separator scores $1$, so the chosen pair is a legal choice in
the string too. The run is therefore a legal run on $s(\mathcal T)$ with the same savings, which are at least
$\alpha\OPT(s(\mathcal T),k)\ge\alpha\OPT(\mathcal T,k)$. Rational weights follow by scaling.

For real weights $\phi$ fix a legal history. Its states depend only on its choices, so its legality says that finitely
many homogeneous linear inequalities with integer coefficients hold at $\phi$ (the chosen pair scores at least as much as
every other pair, at every step). Let $V$ be the subspace cut out by those of them that hold with equality at $\phi$. It
is defined over the rationals, so it has a rational basis, and rational points of $V$ arbitrarily close to $\phi$ exist;
close enough, they keep every strict inequality and the positivity of the weights. At such points the same history is
legal, so its savings are at least $\alpha\OPT$. The savings of the fixed history are linear in the weights. Up to
renaming the fresh symbols there are finitely many lists of $k$ rules, and finitely many choices in each row, so
$\OPT(\mathcal T,k)$ is a maximum of finitely many linear functions of the weights, hence continuous. Letting the
rational points tend to $\phi$ proves the claim.
\end{proof}

\begin{proof}[Proof of \cref{thm:table}(b) and (c)]
Take the parameters $d,F,w_t,h_t,\delta_t$ of \cref{app:upper} with hypothesis (H), the $2d$ letters $X_i,Y_i$, and the rows
\begin{center}\small
\begin{tabular}{@{}lll@{}}
\toprule
Word & Indices & Weight\\
\midrule
$X_i\,Y_j\,X_{j+t}\,Y_i$ & $(t,j,i)\in\{0,\dots,d-1\}^3$ & $w_t$\\
$Y_j\,X_{j+t}$ & $(t,j)\in\{0,\dots,d-1\}^2$ & $\delta_t$\\
$X_i\,Y_j$ & $(i,j)\in\{0,\dots,d-1\}^2$ & $h_d$\\
\bottomrule
\end{tabular}
\end{center}
omitting rows of weight $0$. These are the blocks of \cref{app:upper}, one row per block type with its number of copies
as weight, and without guard letters. The counting and the phase argument of \cref{app:upper} apply word for word, with
the pairs next to a guard removed from the list of competitors: each side pair has weighted count $F$, the side rules
save $kF$ with $k=d^2$, and in phase $t$ every greedy run merges the $d$ middle pairs of level $t$, each saving $h_t+1$,
because a middle token has neighbours only inside a four-letter word, where each pair it forms has count at most
$w_0\le h_{d-1}\le h_t$, and two middle tokens are never adjacent. So every greedy run saves
$G_k=d\sum_{t<d}(h_t+1)$, and for $d\ge8$ and $F\ge8d$ the ratio is below $c_d+(d+1)/F$, as in the proof of
\cref{thm:upper}.

For a budget $K\ge64$ take $d=\lfloor\sqrt K\rfloor\ge8$ and $F=d^3$, so $k=d^2\le K\le d^2+2d$. Greedy's first $k$ steps
do not depend on the budget. By the argument of \cref{lem:cap}, applied row by row, the replacements of one rule erase
distinct original cuts, all with the same pair of letters (the last letter of its left operand and the first of its
right one). No ordered pair of letters is adjacent with total weight above $F+1$ in the table, so every rule saves at
most $F+1$.
Since $\OPT(\mathcal T,K)\ge\OPT(\mathcal T,k)\ge kF$, every greedy run of $K$ merges has ratio at most
$c_d+\frac{d+1}F+\frac{2d(F+1)}{d^2F}$, which is the bound of (b); each of its terms beyond $c_d$ is $O(1/d)$, and so is
$c_d-\cc$, so it is $\cc+O(1/\sqrt K)$. With part (a), which makes every ratio at budget $k\ge2$ at least $c_k>\cc$ and
every ratio at $k=1$ equal to $1$, this proves (b); every greedy run on these tables saves the same amount, so
tie-breaking in greedy's favour does not help.

For (c) take $d=\UpD$, $F=\UpF$. The table has $d^3+2d^2=\WtWordTypes$ rows over $2d=\WtLetters$ letters, every greedy
run saves $\UpG$ as in \cref{app:upper}, and the side rules save $\Uside=\UpOPT$. Conversely, the string of \cref{app:upper}
consists of the same words, each repeated as often as its weight and separated by guard letters, so every list of
partial rules on the table, applied with the same choices in every copy, is a partial sequence on that string with the
same utility. The string's optimum is exactly $\Uside$ (\cref{app:upper}), so $\OPT(\mathcal T,\UpK)=\Uside$.
\end{proof}

\section{String-identified symbols}\label{app:identity}

\paragraph{The model.} The trainers of subword-nmt and Hugging Face identify a symbol
with the string it spells: the merge of $(X,Y)$ produces the symbol $XY$, the same symbol that an earlier merge with a
different split may already have produced, and a later rule acts on every adjacent pair of tokens with the right
spellings. The initial symbols, the \emph{atoms}, are characters, some carrying a marker such as an end-of-word suffix.
Call the input \emph{uniquely decodable} if no two distinct sequences of atoms that are both factors of its rows
(occur consecutively in some row) spell the same string; that is, the string spelled by a factor of the rows
determines its factorisation into atoms. This is our usage, not the standard notion for codes, which compares all
sequences of atoms, not only those that occur in the input. It holds if the atoms are single characters, and also if they are characters together with
word-final variants that carry a marker string never occurring in the corpus text. Every token covers consecutive atoms
of one row, so then the spelling of a token determines its sequence of atoms, the letters of the model above, and cuts,
tokens and the counts $\cnt$, $\raw$, $D$ and $R$ are defined as before. The proof of the following proposition is given
below.

\begin{proposition}[symbol identity]\label{prop:identity}
Fix a uniquely decodable string or word table with positive real weights.
\begin{enumerate}
\item[(a)] The full-merge histories with fresh and with string-identified symbols correspond one to one, with the same
tokens, the same counts $\cnt$ and $\raw$ of corresponding pairs, and the same gains at every step. In particular greedy
under either ranking, with every tie-breaking, has the same runs in both models.
\item[(b)] For every $k$, the tuples of final tokenisations reachable with at most $k$ partial merges are the same in both
models, so $\OPT$ is the same in both; by (a), so is the full-merge optimum $\OPT^m$.
\end{enumerate}
Hence every bound in this paper on the ratio of a greedy run's saving to $\OPT$ or $\OPT^m$, for either ranking rule,
holds verbatim for string-identified symbols.
\end{proposition}

Unique decodability cannot be dropped from the proposition, nor from the shadow inequality (\cref{cor:rc}) for
string-identified symbols: with atoms $\str a$, $\str b$, $\str{ab}$ and the rows $[\str a,\str{ab}]$ and
$[\str b,\str a,\str a,\str b]$ of weight $1$, every pair occurs once, so $\mu=\CollMu$, but two string-identified merges
save $\CollOPT>2\cdot0+2\cdot\CollMu$, while with fresh symbols the optimum is $\CollOPTfresh$ (\cref{app:identity}). A
mid-word token equal to the word-final atom can arise in this way in subword-nmt when the corpus contains its marker
string \texttt{</w>}. Whether \cref{thm:main} itself survives without the hypothesis we do not know; a search of small
non-decodable inputs found no violation of it (\cref{app:evidence}). \Cref{lem:mono} fails along partial
string-identified histories, so it is used only along full-merge histories, where it holds in both models by
\cref{prop:identity}(a); \cref{lem:cap} replaces it for partial merges.

\emph{Examples.} It is not enough for \cref{prop:identity} or for \cref{cor:rc} that no atom string is a concatenation of
other atom strings: with atoms $\str{ab}$,
$\str c$, $\str a$, $\str{bc}$, $\str d$ the rows $[\str{ab},\str c,\str d]$ and $[\str a,\str{bc},\str d]$ have $\mu=\DecMu$,
while three string-identified merges save $\DecOPT$, as both rows become $\str{abcd}$; with fresh symbols the optimum is
$\DecOPTfresh$. In the example of \cref{sec:identity}, the two merges that save $\CollOPT$ are $(\str a,\str b)$ in the
second row and then $(\str a,\str{ab})$ in both rows; the second rule acts on the atom $\str{ab}$ of the first row and on
the merged token $\str{ab}$ of the second as on one symbol. \Cref{lem:mono} fails for string-identified symbols: the state $[\str{abc},\str d,\str a,\str{bc},\str d]$,
reached from $\str{abcdabcd}$ by partial merges, has $\mu=\MonoBefore$, and the full merge of $(\str a,\str{bc})$ raises
it to $\MonoAfter$.

\emph{Setting.} Fix a uniquely decodable input. A token is identified, in the string-identified model, by its spelling, which is a
sequence of atoms; in the fresh model by the step that created it (or by its atom). A rule $(X,Y)$ acts on adjacent
tokens with these identities; a full rule scans each row from left to right, and a partial rule replaces any
non-overlapping occurrences, chosen row by row.

\emph{Isolation.} \Cref{lem:iso} holds in both models, with the same proof: it uses only that each step is a fixed rule
applied by a left-to-right scan whose output identity is determined by the step.

\emph{Unique production along full histories.} In a full-merge history of either model, if replacements at steps $j$ and
$j'$ produce tokens with the same spelling $w$, then $j=j'$. Indeed the end cuts of a token created at step $j$ survive
through step $j$, so by isolation the same list of rules applied to the word $w$ alone has two tokens just before step
$j$ and one token after it; the run on $w$ alone has only one such step. A replacement produces at least two atoms, so it
never recreates an atom.

\emph{Proof of (a).} Couple the two models step by step, starting with the atoms, and suppose that after a coupled prefix
the tokens agree as intervals and the map from fresh symbols to spellings is a bijection between the symbols in use and
their spellings. Then every ordered pair has the same occurrences in both models, including overlaps and the left-to-right
choices of a full scan, so the two scans make the same replacements and have the same counts $\cnt$ and $\raw$. The output
spelling of the step was never produced before, by unique production applied to the extended history, so the new fresh
symbol extends the bijection. By induction the full-merge trees of the two models coincide, with their gains; in
particular so do the runs of either greedy rule and their sets of ties.

\emph{Proof of (b).} A fresh partial history becomes a string-identified one by mapping each symbol to its spelling: every
replaced occurrence still spells the two operands, the occurrences stay disjoint, and extra matches may be skipped.
Conversely fix a string-identified partial history whose rules produce at most $k$ distinct output spellings (for
instance, one with at most $k$ nonempty rules), and its final tokenisations. Let $\mathcal D$ be the set of these
spellings, $|\mathcal D|\le k$, and choose for every $w\in\mathcal D$ one split $w=u_wv_w$ that some rule of the history
used; each part is an atom or lies in $\mathcal D$ and is shorter than $w$. Unfolding these splits gives a binary tree
over every final token, whose leaves are its atoms in order. Process the spellings of $\mathcal D$ by increasing length,
and at $w$ apply one fresh rule that joins the current tokens for $u_w$ and $v_w$ at exactly the nodes of the trees that
spell $w$. These nodes are disjoint, since a proper ancestor is longer, their children exist and are adjacent, so this
is a legal partial merge. After $|\mathcal D|\le k$ rules the final tokenisations are reproduced. Both conversions keep
the tuple of final tokenisations, so the reachable sets coincide, and $\OPT$ is the same in both models for every weight
vector.

\section{Proofs for overlapping counts}\label{app:overlap}

Throughout, greedy means overlap greedy with arbitrary tie-breaking, its gains are $o_j$, and $O_i=o_1+\dots+o_i$. All
statements are proved for strings. \Cref{lem:transfer} carries part~(a) to word tables; a string is a one-row table, so
part~(b) holds for tables too. Part~(c) also holds for word tables, with $U$ the weighted utility of $\sigma$ and each
final token weighted by its row. For integer weights write the table as the string $s(\mathcal T)$ of \cref{app:table}.
As in the proof of \cref{lem:transfer}, a legal run on $\mathcal T$ either ends with every row one token, and then saves
$\sum_j\phi_j(|\omega_j|-1)\ge U$, or is a legal run on $s(\mathcal T)$ with the same savings. Replaying $\sigma$ copy by
copy gives a partial $k$-sequence on $s(\mathcal T)$ with the same utility $U$ and, copy for copy, the same final
tokens, so part~(c) on $s(\mathcal T)$ gives it on $\mathcal T$. Rational weights follow by scaling, and real weights by
the continuity argument of \cref{lem:transfer}: once the legal history and the rules and row-by-row choices of $\sigma$
are fixed, both sides of (c) are linear in the weights. Finally \cref{prop:identity} carries (a), (b) and (c) to
string-identified symbols on uniquely decodable inputs; for (c) note that it depends on $\sigma$ only through its final
tokenisation. Let $r_j$ be the raw
count of the pair chosen at step $j$, the largest raw count of $s_{j-1}$, and let $\bar r$ be the largest raw count of
$s_k$. A maximal run of $l$ tokens $X$ contributes $l-1$ to $\raw(X,X)$ and $\lfloor l/2\rfloor$ to
$\cnt(X,X)$, and for $X\ne Y$ the two counts agree. So $r_{i+1}$, which is at least the largest $\cnt$, satisfies
\begin{equation}\label{eq:raw}
\mu(s_i)\;\le\;r_{i+1}\;\le\;2o_{i+1},
\end{equation}
and \cref{cor:rc}, applied to the full merges of the run, gives
\begin{equation}\label{eq:ovrc}
\OPT(s,k)\;\le\;2O_i+k\,\mu(s_i)\;\le\;2O_i+k\,r_{i+1}\qquad\text{for every }0\le i<k .
\end{equation}
At a step whose pair has equal symbols (a \emph{self step}) put $\eta_j=2o_j-r_j$; by the count above it is the number
of maximal runs of even length, so $0\le\eta_j\le o_j$. At any other step (an \emph{unequal step}) put $\eta_j=o_j=r_j$.
The largest raw count does not increase along full merges: a pair of old symbols keeps a subset of its adjacencies, and a
pair with the fresh symbol of a merge of $o$ replacements has at most $o$ adjacencies. So $r_1\ge\dots\ge r_k\ge\bar r$.
If $\bar r=0$, then $s_k$ is one token, every cut is erased, and all three parts hold trivially; we assume $\bar r\ge1$
below.

\paragraph{Junctions.} For a partial $k$-sequence $\sigma$ call the junctions of its nodes the \emph{junctions} of
$\sigma$; there are $U=\mathrm{util}(\sigma)$ of them (F1). A cut is \emph{intact} while the two positions next to it are
single-letter tokens; a cut that stops being intact never becomes intact again. The junctions of label $r$ all have the
same pair of letters $\pi_r$ on their two sides (F2), so the intact junctions of a label, or of all labels with one pair,
are distinct adjacencies of that pair of letters: before step $j$ there are at most $r_j$ of them, and at the end at most
$\bar r$. A replacement that joins two tokens across a cut $c$ changes the tokens next to exactly three cuts: $c$,
its \emph{centre}, and the cuts just outside its two tokens, its left and right \emph{sides}. Only there can a cut stop
being intact.

\subsection{The suffix inequality}

\begin{lemma}[suffix inequality]\label{lem:suffix}
Let $\bar r\ge1$ and $0\le i\le k$, and call the tokens of $s_i$ \emph{atoms}. Let $\mathcal W$ be a set of distinct cuts,
each between two consecutive atoms and each with one of $K$ labels, such that the cuts of each label $r$ are pairwise
disjoint occurrences in $s_i$ of one ordered pair $\pi_r$ of atoms. Then
\[
|\mathcal W|\;\le\;\tfrac76K\bar r+\sum_{j=i+1}^{k}\bigl(2o_j+\eta_j-\bar r\bigr).
\]
In particular, if $\sigma$ is a partial $k$-sequence with utility $U$ and $R=r_1+\dots+r_k$, then
$U\le4O_k-R+\frac16k\bar r$.
\end{lemma}

\begin{proof}
Only the steps $j=i+1,\dots,k$ occur in this proof. Call the cuts of $\mathcal W$ \emph{gaps}, and a symbol
\emph{compound} if it is not an atom. A cut between two atoms is \emph{intact} while both atoms are still tokens. For
$i=0$ the atoms are the letters and this is the notion of the paragraph on junctions, and its three facts hold for atoms
with the same proofs: a cut that stops being intact never becomes intact again; only the centre or a side of a
replacement can stop being intact; and the intact gaps of a label, or of all labels with one pair, are distinct
adjacencies of that pair of atoms, at most $r_j$ of them before step $j$ and at most $\bar r$ at the end.

\emph{Marks.} At a self step on an atom $x$ with fresh symbol $Z$, a maximal run of $2m+1\ge3$ tokens $x$ becomes
$Z\cdots Z\,x$; \emph{mark} the cut between its last $Z$ and its final $x$. A mark is \emph{live} until one of its two
tokens is replaced. Live marks occupy distinct cuts, and no live mark is an intact cut, since its left token is
compound. After the full scan of $(x,x)$ no two tokens $x$ are adjacent, and a merge never makes two old tokens adjacent, so
$x$ takes part in no later self step. Hence a live mark breaks either at a side or the centre of a replacement of an
unequal step, or at a self step on $Z$, where its $Z$, the last token of its run, is replaced only if the run is even,
and then the mark is the right side of the run's last replacement.

\emph{Credits.} Consider an unequal step $j$ with pair $(X,Y)$, and let $\kappa_j\le r_j$ count the intact gaps or live
marks at the centres of its replacements. If $X,Y$ are atoms, the centres carry intact gaps of the labels with
$\pi_r=(X,Y)$; the step makes all of their intact gaps non-intact, and no adjacency $(X,Y)$ reappears, so these
labels have no intact gap at the end. Call $\min(\kappa_j,\bar r)$ of the gaps lost at the centres
\emph{credited}; across these labels the credits total at most $\bar r$. If instead $(X,Y)=(Z,x)$ for the fresh symbol $Z$ of an earlier
self step on an atom $x$, credit in the same way $\min(\kappa_j,\bar r)$ of the marks of that step broken at the centres; that step
then has no live mark at the end, and $(Z,x)$ is never chosen again. The two cases exclude each other.

\emph{Discarded gaps.} Call a gap \emph{discarded} if it stops being intact, without credit, at an unequal step
or at the right side of an even run of a self step. Let $d_j$ count the marks born at the self step $j$ that break
without credit. Then
\begin{equation}\label{eq:discard}
\#\{\text{discarded gaps}\}+\sum_{\text{self }j}d_j\;\le\;\sum_{\text{unequal }j}(3r_j-\bar r)+\sum_{\text{self }j}\eta_j .
\end{equation}
Indeed, at an unequal step each of the $r_j$ replacements has two sides and a centre, each carrying at most one discarded
gap or broken mark, and at most $(\kappa_j-\bar r)_+$ of the items at the centres are uncredited;
$2r_j+(\kappa_j-\bar r)_+\le3r_j-\bar r$ because $\bar r\le r_j$ and $\kappa_j\le r_j$. At a self step gaps are discarded,
and marks broken, only at the right sides of even runs, at most one per run.

\emph{Grouping.} A gap that is not discarded is credited, or intact at the end, or stops being intact at a self step
on an atom $x$ at an internal cut of a run, where $\pi_r=(x,x)$, or at the left side of a run, where $\pi_r=(y,x)$ with
$y\ne x$ (a self step on a compound symbol touches no intact cut). Assign each label with $\pi_r=(y,x)$, $y=x$ allowed,
to the self step on $x$ if there is one (there is at most one); it is \emph{matching} if $y=x$. Let $\bar N_r$ count the
gaps of label $r$ that are not discarded. A label that is not assigned has $\bar N_r\le\bar r$: credits go only to a
label with no intact gap at the end, at most $\bar r$ of them.

Fix a self step $j$ on an atom $x$, with $v\le o_j$ runs of length at least $2$, and with $p$
non-matching and $h$ matching labels assigned. For a non-matching label let $L$ count its gaps lost at this step.
If its pair $(y,x)$ was chosen at an earlier step, it has no intact gap now, so $L=0$ and $\bar N_r\le\bar r$, its
credits; otherwise its gaps that are neither discarded nor lost at this step are intact after it (credited later,
or intact at the end), at most $\bar r$ of them, and together with the $L$ they were intact before it, at most $r_j$ in
all. So $\bar N_r\le\bar r+\min(L,r_j-\bar r)$. The $L$ sum to at most $v$, one per left side. A
matching label has no intact gap after the step and is never credited, so its gaps that are not discarded are
lost at internal cuts of runs: distinct cuts, at most $r_j$ in all, and disjoint occurrences of $(x,x)$, at most $o_j$ per
label. Finally $v\le\eta_j+\bar r+d_j$: the even runs, plus the odd runs of length at least $3$, each of which leaves a
mark that is live at the end (at most $\bar r$ of them, adjacencies of $(Z,x)$), or credited (then none is live at the
end, and at most $\bar r$ are credited), or broken without credit. Since $\min(a+d,b)\le\min(a,b)+d$, the labels
assigned to this step have
\[
\sum \bar N_r\;\le\;\tfrac76(p+h)\bar r+d_j+\Xi_j,\qquad
\Xi_j=\min\bigl(o_j,\eta_j+\bar r,p(r_j-\bar r)\bigr)-\tfrac16p\bar r+\min(r_j,ho_j)-\tfrac76h\bar r .
\]
We claim $\Xi_j\le2o_j-\bar r=r_j+\eta_j-\bar r$. If $h\ge2$, then $\Xi_j\le\eta_j+\bar r+r_j-\frac73\bar r$. If
$h=1$, then $\Xi_j\le o_j+o_j-\frac76\bar r$. If $h=0$, put $u=r_j/\bar r-1\ge0$; since
$o_j=\frac12(1+u)\bar r+\frac12\eta_j$,
\[
\Xi_j\;\le\;\bar r\min\bigl(\tfrac12(1+u),\,pu\bigr)+\tfrac12\eta_j-\tfrac16p\bar r\;\le\;\bar ru+\tfrac12\eta_j\;\le\;r_j+\eta_j-\bar r ,
\]
using $\min(\frac12(1+u),pu)\le u+\frac p6$: for $p\le1$ this is clear, and for $p\ge2$ the left side minus $u$ is at most
its value $(p-1)/(2p-1)$ at $u=1/(2p-1)$, and $(p-1)/(2p-1)\le p/6$ is $(2p-3)(p-2)\ge0$.

Adding up over the $K$ labels (labels without gaps included), with $\frac76\bar r$ for each label, $d_j+\Xi_j$ for each
self step on an atom, and \eqref{eq:discard} for the discarded gaps,
\[
|\mathcal W|\;\le\;\tfrac76K\bar r+\sum_{\text{unequal }j}(3r_j-\bar r)+\sum_{\text{self }j}(\eta_j+2o_j-\bar r),
\]
where self steps without assigned labels, including those on compound symbols, contribute $\eta_j\le\eta_j+2o_j-\bar r$.
At an unequal step $3r_j=2o_j+\eta_j$, which proves the inequality. For the last claim take $i=0$, $K=k$ and for
$\mathcal W$ the junctions of $\sigma$, labelled by their nodes: they are $U$ distinct cuts (F1), those of label $r$ have
one pair of letters (F2), and they are disjoint occurrences of it because their nodes have disjoint spans (F3). Finally
$\sum_j\eta_j=2O_k-R$.
\end{proof}

The marks come from the odd-run argument of \prevOV, reproved here.

\subsection{Proof of part (a)}\label{app:parta}

\paragraph{Roadmap.} Let $U=\OPT(s,k)>0$, fix an optimal partial $k$-sequence $\sigma$, and recall $\bar r\ge1$. \Cref{cor:rc}
bounds $U$ at one cut through greedy's next score, and at a self step that score can be twice the gain. The proof below
writes several upper bounds on $U$ at every \emph{level} $\gamma\ge\bar r$, the cut where greedy's scores fall to $\gamma$,
and averages them over the levels. Each bound is an allowance plus one cost per step of the run. The cost of a step
depends only on its score, its gain, the kind of its operands, and how many of its raw adjacencies are junctions of
$\sigma$. Three facts about a self step on a letter $\str x$ keep the costs small.
\begin{enumerate}
\item \emph{Refunds} (\cref{lem:refund}). If few raw gaps of the step are junctions, the step consumes letters that sit
between two non-junctions, and the shadow argument at every later cut gains one unit per such letter.
\item \emph{Detection} (\cref{lem:detect}). If an earlier shadow sees one junction of the step as a window of one gap, a
whole label of $\sigma$ has the pair $(\str x,\str x)$, and that label cannot pack its occurrences fully before the step.
\item \emph{Alternating paths} (\cref{lem:paths}). The junctions that the earlier shadow fails to see number at most the
mixed replacements made so far, which the bounds already pay for.
\end{enumerate}
\Cref{lem:levelbounds,lem:segment} turn these facts into five bounds at each level, with the kind codes \kd S, \kd V,
\kd C, \kd H and \kd J of the certificate files; the segment bounds \kd H and \kd J are a signed form of
\cref{lem:suffix}. \Cref{lem:types} shows that only six types of steps matter, and \cref{lem:average} reduces the theorem
to a check in one real variable, the score. For $\lambda_{\rm h}$ the check is a finite calculation that can be
checked by hand from \cref{tab:ovhand,tab:ovsign}; both certificates are checked in exact arithmetic by
\texttt{scripts/overlap\_cert}. \Cref{tab:ovlocal} lists the symbols of this subsection.

\begin{table}[!htb]
\centering\small
\begin{tabular}{@{}>{\raggedright}p{4.1cm}p{11.4cm}@{}}
\toprule
Symbol & Meaning\\
\midrule
letter, long symbol & a symbol that is a letter of $s$; any other symbol\\
letter--letter, mixed, long--long & a step whose two operands are letters; one letter and one long symbol; two long symbols\\
$\Nmix_i$, $\Nlong_i$ & numbers of mixed and of long--long replacements among the first $i$ steps\\
letter self step, raw gaps, $t_j$ & a self step on a letter; its $r_j$ adjacencies, which are original cuts; the number of
them that are junctions of $\sigma$\\
$\Pi_j$; $\Psi_j$, $\Psi'_j$ & prefix cost of step $j$, \eqref{eq:prefixcost}; its middle and later costs (\cref{lem:segment})\\
level $\gamma$, cut $i(\gamma)$ & a real $\gamma\ge\bar r$; the number of steps with $r_j>\gamma$\\
$\mathcal W_i$ & the shadow family at cut $i$ (\cref{lem:refund})\\
$\gb$, $\varpi$, $\iota$ & stop level, price and switch of the segment bounds (\cref{lem:segment})\\
$\Lx(r,o)$, $\Mx(o,t)$, $\Lx_0$, $\Lx_{\rm c}$ & left and matching excess \eqref{eq:excess}; $\Lx(r,\frac r2)$ and its
upper bound \eqref{eq:ellc}\\
$\ty{lu}$, $\ty{mx}$, $\ty{sn}$, $\ty{sh}$, $\ty{sa}$, $\ty{ls}$ & the six scalar step types (\cref{tab:costsSVC})\\
\bottomrule
\end{tabular}
\caption{Symbols local to \cref{app:parta}.}
\label{tab:ovlocal}
\end{table}

\paragraph{Conventions.} Throughout, $p_1,\dots,p_k$ are the full merges of the fixed overlap-greedy run, so for each
$0\le i\le k$, $s_i$ is the output of $p_1,\dots,p_i$, and the shadow, windows, marks and kept windows of \cref{sec:proof} are those of
$p_1,\dots,p_i$; \cref{sec:proof} allows any full merges, so its results apply. A symbol is a \emph{letter} if it is a
letter of $s$, and \emph{long} otherwise. All replacements of one step have operands of the same kinds, so every step is
letter--letter (both operands letters), \emph{mixed} (one letter and one long symbol) or long--long. We write $\Nmix_i$
and $\Nlong_i$ for the numbers of mixed and of long--long replacements among the first $i$ steps. A \emph{letter self
step} is a self step on a letter $\str x$; its $r_j$ adjacencies $\str x\str x$ are its \emph{raw gaps}, which are
original cuts, and $t_j$ is the number of them that are junctions of $\sigma$. Every
step has $o_j\ge r_j/2\ge\frac12$, so all $k$ steps are nonempty. A pair is chosen at most once: after its full merge no
occurrence of it remains, and adjacencies of old symbols never reappear. In particular different letter self steps have
different letters. We write $[\,\cdot\,]$ for the indicator of a condition. At a self step a maximal run of $l\ge2$ tokens has $l-1$ raw adjacencies and $\lfloor l/2\rfloor$
replacements. So a self step has at most $o_j$ runs, of which $\eta_j$ are even, and $r_j-o_j=\sum(\lceil l/2\rceil-1)$
over its runs. Scores do not increase, so every raw count of $s_{i(\gamma)}$ is at most $\gamma$: it is at most
$r_{i+1}\le\gamma$ if $i=i(\gamma)<k$, and at most $\bar r$ if $i=k$. The \emph{prefix cost} of step $j$ is
\begin{equation}\label{eq:prefixcost}
\Pi_j=\begin{cases}2o_j&\text{at an unequal letter--letter step or a mixed step},\\
0&\text{at a long--long step},\\
2o_j-(r_j-o_j-t_j)_+&\text{at a letter self step}.\end{cases}
\end{equation}

\begin{lemma}[alternating paths]\label{lem:paths}
Fix $i$, and let $M$ and $L$ be the numbers of dimers and of absorbed letters of the shadow of $p_1,\dots,p_i$
(\cref{def:shadow}). Call a position a \emph{stray} if it is a single letter of $s_i$ but lies in a dimer of the shadow.
Then there are at most $\Nmix_i$ strays, and $M+L\le O_i-\Nlong_i$.
\end{lemma}

\begin{proof}
A position stops being a single letter only through a letter--letter or a mixed replacement. Call a position
\emph{deleted} if a mixed replacement among the first $i$ steps absorbs it as its letter operand; each mixed replacement
absorbs one letter, and distinct mixed replacements absorb distinct positions, since an absorbed letter is no longer a
single-letter token; so there are $\Nmix_i$ deleted positions. By \cref{lem:matching} the shadow's dimers are the cuts
accepted by the matching process, in the order $(\tau(c),c)$. Run the same process again, but never accept a cut with a
deleted position. We claim that its accepted cuts are exactly the cuts erased by letter--letter replacements of the
actual run. Take the eligible cuts in order, and let $c$ be one whose positions are not deleted. If neither position of
$c$ belongs to an accepted cut so far, both are still single letters at step $\tau(c)$: they are not deleted and, by
induction, took part in no letter--letter replacement. That step scans the single letters of its pair from left to
right, and it replaces the pair at $c$ unless it has just consumed the left position of $c$, which would make $c-1$ an
accepted cut earlier in the order. So $c$ is replaced exactly when it is accepted. A cut with a deleted position is
never replaced, since that position takes part in no letter--letter replacement. As in \cref{lem:matching}, a later merge
of the same pair finds no occurrence.

So the shadow and the actual letter--letter replacements are two matchings of the path of positions, built by the same
greedy rule, the second with the deleted positions removed. Every component of their symmetric difference is a path
whose edges alternate between the two matchings, and a deleted position, having no actual edge, can only be an
endpoint. Every nonempty component has a deleted endpoint. Otherwise let $e$ be its earliest edge in the order. Both
positions of $e$ are present in both runs of the process, so in the run that rejects $e$ it is blocked by an earlier
accepted edge $e'$ that shares a position with it. If $e'$ were in both matchings it would block $e$ in the other run
too, so $e'$ lies in the same component and is earlier than $e$, a contradiction.

A stray is an endpoint of its component, with a shadow edge and no actual edge. The other endpoint of a component with a
stray is deleted, since every component has a deleted endpoint; so different strays have different deleted partners, and
there are at most $\Nmix_i$ strays. A component starts and ends with shadow edges at a deleted endpoint and at a stray,
so it has one more shadow edge than actual edges; every other component has a deleted endpoint, which carries a shadow
edge, so it has at least as many shadow edges as actual ones. Hence $M-N^{\rm ll}\ge N^{\rm st}$, where $N^{\rm ll}$ counts the
letter--letter replacements and $N^{\rm st}$ the strays. Counting positions that are not single letters of $s_i$: letter--letter
replacements absorb $2N^{\rm ll}$ of them, mixed ones $\Nmix_i$, and they consist of the $L$ absorbed letters and the
$2M-N^{\rm st}$ positions of dimers that are not strays. So $L=2N^{\rm ll}+\Nmix_i-2M+N^{\rm st}$, and with $O_i=N^{\rm ll}+\Nmix_i+\Nlong_i$,
\[
M+L\;=\;2N^{\rm ll}+\Nmix_i-M+N^{\rm st}\;\le\;N^{\rm ll}+\Nmix_i\;=\;O_i-\Nlong_i .\qedhere
\]
\end{proof}

\begin{lemma}[shadow family with refunds]\label{lem:refund}
For every $0\le i\le k$ there is a set $\mathcal W_i$ of cuts, each carrying a label of $\sigma$, such that:
\begin{enumerate}
\item[(i)] every cut of $\mathcal W_i$ lies between two consecutive tokens of $s_i$, the right one a letter, and is a
junction of $\sigma$;
\item[(ii)] the cuts of one label are pairwise disjoint occurrences in $s_i$ of one ordered pair of symbols of $s_i$;
\item[(iii)] $|\mathcal W_i|\ge U-\sum_{j\le i}\Pi_j$.
\end{enumerate}
\end{lemma}

\begin{proof}
Apply the proof of \cref{thm:pointwise} to $p_1,\dots,p_i$ and $\sigma$, and for each kept window, on positions
$\alpha,\dots,\beta$ with its mark on a node of label $r$, take the cut $\beta-1$ with label $r$. This is the mirror
image of the kept-window family defined at the end of that proof, and the properties (i)--(iv) stated there hold with the same proofs: the cut $\beta-1$
is a gap of its window, all kept windows of a label carry the same tokens of $s_i$, so their last two tokens form one
pair, and the occurrences lie in the spans of distinct nodes of one label. The right token of the cut is the single
letter at $\beta$, and the cut is a junction because the window is marked. This proves (i) and (ii).

For (iii) we count kept windows more carefully. Call a \emph{refund pair} two consecutive cuts $c-1,c$, neither of them a
junction, whose common position $c$ is not a single letter of $s_i$. Let $\mathcal B$ be a set of pairwise disjoint
refund pairs; distinct pairs then have distinct common positions. If the common position $c$ lies in a dimer of the
shadow, both cuts $c-1$ and $c$ lie between the same two consecutive single letters of the shadow, so they are gaps of
one window, or both lie before the first or after the last single letter. In the proof of \cref{lem:windows}(a) a window
with $\nu$ dimers contributes at most $2\nu$ to $U-P$; if it holds $\nu'\ge1$ of these pairs, it has at least
$2\nu'$ gaps that are not junctions, so it is not marked and contributes at most $2\nu+1-2\nu'\le2\nu-\nu'$. The cuts
before the first single letter contribute at most $2\nu-2\nu'$, and likewise after the last. So $U-P\le2M-|\mathcal B_{\rm d}|$, where
$\mathcal B_{\rm d}$ holds the pairs whose common position lies in a dimer. If instead $c$ is a single letter of the
shadow, it is absorbed, and the two windows that end at $c$ contain the gaps $c-1$ and $c$, so neither is marked. The
proof of \cref{thm:pointwise} loses at most two marked windows per absorbed letter, and this letter loses none, so at
least $P-2L+2|\mathcal B_{\rm s}|$ marked windows are kept, with $\mathcal B_{\rm s}=\mathcal B\setminus\mathcal B_{\rm d}$.
Together,
\[
|\mathcal W_i|\;\ge\;U-2M-2L+|\mathcal B| .
\]

Now take refund pairs from the letter self steps $j\le i$. At such a step on $\str x$, pair the raw gaps of each run
from its left end: the first two, the next two, and so on. A run of $l$ letters gives $\lceil l/2\rceil-1$ pairs, so the
step gives $r_j-o_j$ pairs. The common position of each pair is the second letter of a replacement of step $j$, so it is
not a single letter of $s_i$. At most $t_j$ of the pairs contain a junction, so at least $(r_j-o_j-t_j)_+$ are refund
pairs. Pairs of one step are disjoint, and pairs of different steps consist of cuts between different letters. By
\cref{lem:paths},
\[
|\mathcal W_i|\;\ge\;U-2(O_i-\Nlong_i)+\sum_{j\le i}(r_j-o_j-t_j)_+\;=\;U-\sum_{j\le i}\Pi_j,
\]
the sum over letter self steps; the equality is \eqref{eq:prefixcost}, since a long--long step saves only long--long
replacements.
\end{proof}

\begin{lemma}[detection]\label{lem:detect}
Let $0\le i<j\le k$, where step $j$ is a letter self step on $\str x$ with $t_j>0$. Call step $j$ \emph{detected at
$i$} if one of its raw gaps that is a junction has both of its letters single letters of the shadow of
$p_1,\dots,p_i$.
\begin{enumerate}
\item[(a)] If step $j$ is detected at $i$, some label of $\mathcal W_i$ has the pair $(\str x,\str x)$. For every $i'$
with $i\le i'<j$, a set of pairwise disjoint occurrences of $(\str x,\str x)$ in $s_{i'}$ has at most
$\raw_{s_{i'}}(\str x,\str x)-(r_j-o_j)$ elements.
\item[(b)] The sum of $t_j$ over the letter self steps $j>i$ that are not detected at $i$ is at most $\Nmix_i$.
\item[(c)] If every raw count of $s_i$ is at most $\gamma$, then $|\mathcal W_i|\le k\gamma-\sum_{j>i}\min(t_j,r_j-o_j)
+\Nmix_i$, the sum over letter self steps.
\end{enumerate}
\end{lemma}

\begin{proof}
The raw gaps of step $j$ are adjacencies of two single letters $\str x$ just before step $j$. Single letters only
disappear, so they are adjacencies of single letters in $s_{i'}$ for every $i'<j$.

(a) Let $c$ be the raw gap in the definition. Its two letters are consecutive single letters of the shadow, so $c$ is
the only gap of a window. The window is marked, since $c$ is a junction, and kept, since its ends are single letters of
$s_i$. So $c$ belongs to $\mathcal W_i$, and every cut of its label is an occurrence of $(\str x,\str x)$. For the packing
bound, a set of disjoint occurrences of $(\str x,\str x)$ uses at most $\lfloor l/2\rfloor$ raw gaps of each run of step
$j$, at most $o_j$ in all, and at most all of the other $\raw_{s_{i'}}(\str x,\str x)-r_j$ adjacencies.

(b) If step $j$ is not detected, each of its $t_j$ raw gaps that are junctions has a letter that lies in a dimer of the
shadow and is a single letter of $s_i$, a stray. The dimer is not $\str x\str x$, because the pair $(\str x,\str x)$ is
chosen at step $j>i$ and a pair is chosen once. So of the two cuts next to the stray, the one inside the dimer is not an
adjacency $\str x\str x$, and the stray meets at most one raw gap of step $j$. Strays of different steps carry different
letters. So the sum is at most the number of strays, and \cref{lem:paths} bounds it by $\Nmix_i$.

(c) The cuts of a label are disjoint occurrences of one pair in $s_i$, at most $\gamma$ of them. A label found in (a) for
a detected step $j$ has at most $\gamma-(r_j-o_j)$, and labels found for different steps have different pairs. So
$|\mathcal W_i|\le k\gamma-\sum(r_j-o_j)$ over the detected steps, and $\min(t_j,r_j-o_j)\le t_j$ for the others, whose
sum is at most $\Nmix_i$ by (b).
\end{proof}

\begin{lemma}[three bounds at a level]\label{lem:levelbounds}
Let $\gamma\ge\bar r$ and $i=i(\gamma)$, and call the steps $j\le i$ \emph{before} the level and the others \emph{after}
it. Then
\begin{align*}
&(\kd S)& U&\le k\gamma+\textstyle\sum_{j\le i}\Pi_j,\\
&(\kd V)& U&\le k\gamma+\textstyle\sum_{j\le i}\bigl(\Pi_j+[\text{$j$ mixed}]\,o_j\bigr)-\sum_{j>i}\min(t_j,r_j-o_j),\\
&(\kd C)& U&\le k\gamma+\textstyle\sum_jC_j ,
\end{align*}
where in (\kd V) the last sum is over letter self steps, and in (\kd C) a step before the level costs $C_j=3r_j-\gamma$ if
it is an unequal letter--letter step, $r_j$ if mixed, $0$ if long--long, and
$o_j+\eta_j+t_j-([t_j>0]+[t_j>o_j])\gamma$ if it is a letter self step; a letter self step after the level with
$t_j>0$ costs $C_j=-(r_j-t_j)-[t_j>o_j]\gamma$, and every other step after it costs $0$.
\end{lemma}

\begin{proof}
Every raw count of $s_i$ is at most $\gamma$. (\kd S) The cuts of each label of $\mathcal W_i$ are disjoint occurrences
of one pair, so $|\mathcal W_i|\le k\gamma$; apply \cref{lem:refund}(iii). (\kd V) Combine \cref{lem:refund}(iii) with
\cref{lem:detect}(c), and note $\Nmix_i=\sum_{j\le i}[\text{$j$ mixed}]\,o_j$.

(\kd C) Use the intact junctions of the paragraph on junctions. Every junction is either intact at cut $i$ or stopped
being intact at a step $j\le i$. Give each of the $k$ labels a budget of $\gamma$ intact junctions at cut $i$; the intact
junctions of a label are distinct adjacencies of its pair of letters, at most $\gamma$ of them. A step $j\le i$ loses
intact junctions only at its centres and sides. An unequal letter--letter step loses at most $2r_j$ at its sides and
$r_j$ at its centres. If its pair of letters is the pair of some label, that label has no intact junction afterwards and
returns its budget $\gamma$; if not, no centre is a junction, and $2r_j<3r_j-\gamma$ since $r_j>\gamma$. A mixed step
loses only the cuts on the far side of its letter operand, at most $r_j$, because every other cut it touches has a long
symbol next to it; a long--long step loses none. A letter self step on $\str x$ loses at most $o_j$ cuts at the left ends
of its runs, $\eta_j$ at the right ends of even runs and $t_j$ inside the runs. Afterwards no label with the pair
$(\str x,\str x)$ has an intact junction. There are at least $[t_j>0]+[t_j>o_j]$ such labels, because the $t_j$
junctions inside the runs belong to labels with that pair and one label has at most $o_j$ of them, which are disjoint;
all of them return their budgets. Finally take a letter self step $j>i$ on $\str x$ with $t_j>0$, and the
$n\ge1+[t_j>o_j]$ labels with the pair $(\str x,\str x)$. Their intact junctions at cut $i$ are distinct adjacencies
$\str x\str x$ of $s_i$, at most $\gamma$ in all, and the $r_j-t_j$ raw gaps of step $j$ that are not junctions are
adjacencies $\str x\str x$ of $s_i$ too. So these labels hold at most $\gamma-(r_j-t_j)$ intact junctions against a budget
of $n\gamma$, a saving of at least $(r_j-t_j)+[t_j>o_j]\gamma$. All these returns and savings concern labels with
different pairs, since a pair is chosen once.
\end{proof}

The segment bounds follow one cut to a later one, as \cref{lem:suffix} does, and charge a price $\varpi$ per label. For a
stop level $\gb$ and a price $\varpi>1$, write
\begin{equation}\label{eq:excess}
\Lx(r,o)=\max_{p\in\{0,1,2,\dots\}}\Bigl(\min\bigl(o,\,p(r-\gb)\bigr)-p(\varpi-1)\gb\Bigr),\qquad
\Mx(o,t)=\max\bigl(\min(t,o)-\varpi\gb,\ t-2\varpi\gb\bigr).
\end{equation}

\begin{lemma}[segment bounds]\label{lem:segment}
Let $\gamma\ge\gb\ge\bar r$, $\varpi>1$ and $\iota\in\{0,1\}$ (kind \kd H is $\iota=0$, kind \kd J is $\iota=1$). Let
$i=i(\gamma)$, and let $i'=i(\gb)$ if $\gb>\bar r$ and $i'=k$ if $\gb=\bar r$. Call the steps $j\le i$ \emph{before},
$i<j\le i'$ \emph{middle} and $j>i'$ \emph{later}. Then
\[
U\;\le\;k\varpi\gb+\sum_{j\le i}\bigl(\Pi_j+\iota[\text{$j$ mixed}]\,o_j\bigr)+\sum_{i<j\le i'}\Psi_j+\sum_{j>i'}\Psi'_j ,
\]
where a middle step costs $\Psi_j=3r_j-\gb$ if it is an unequal letter--letter step, $2r_j$ if mixed, $r_j$ if long--long and
unequal, $\eta_j$ if long--long and a self step, and
\[
\Psi_j=\eta_j+\Lx(r_j,o_j)+\max\bigl(\Mx(o_j,t_j),-\iota t_j\bigr)
\]
if it is a letter self step; a later letter self step costs $\Psi'_j=-\min\bigl((\varpi-1)\gb+r_j-o_j,\ \iota t_j\bigr)$, and
every other later step costs $0$.
\end{lemma}

Middle steps have $r_j\ge\gb$, every raw count of $s_{i'}$ is at most $\gb$, and later steps have $r_j\le\gb$. At
$r_j=\gb$ the two cases are told apart by $i'$: if $\gb>\bar r$, a step with $r_j=\gb$ is later, since $i'=i(\gb)$ counts
only the steps with $r_j>\gb$; if $\gb=\bar r$, then $i'=k$ and every step after the level is middle, including those with
$r_j=\bar r$. So when $\gb=\bar r$ there are no later steps; this is the \emph{terminal horizon} used by the hand
certificate.

\begin{proof}
Start from $\mathcal W_i$ of \cref{lem:refund} and follow it through the middle steps. Call the tokens of $s_i$ atoms and
the cuts of $\mathcal W_i$ gaps, and use the notions of the proof of \cref{lem:suffix} for these atoms: a gap is intact
while both its atoms are tokens, it stops being intact only at the centre or a side of a replacement, and it never
becomes intact again; the intact gaps of a label, or of all labels with one pair, are distinct adjacencies of that pair.
By \cref{lem:refund}(i) the right atom of every gap is a letter. So a gap can stop being intact at the left side of a
replacement only if the left operand is a letter, and at a centre only if both operands are atoms and the right one is a
letter.

\emph{Survivors and credits.} At cut $i'$ every raw count is at most $\gb$, so each label has at most $\gb$ intact gaps
there. At a middle unequal step whose pair $(X,Y)$ consists of a long atom or a letter $X$ and a letter $Y$, the gaps lost
at its centres are all the intact gaps of the labels with the pair $(X,Y)$. If $X$ is a letter, call $\min(\kappa,\gb)$
of them \emph{credited}, where $\kappa$ is their number. These labels have no intact gap afterwards, and a pair is chosen
once, so every label has at most $\gb$ gaps that are credited or intact at cut $i'$.

\emph{Discarded gaps.} A gap is \emph{discarded} if it stops being intact at a middle step without being credited, either
at an unequal step or at the right end of an even run of a self step. An unequal letter--letter step discards at most
$2r_j$ gaps at its sides and $(\kappa-\gb)_+\le r_j-\gb$ at its centres, at most $3r_j-\gb$ in all, as $r_j\ge\gb$. A
mixed step discards at most $2r_j$: with a letter on the left it loses nothing at its centres, whose right symbol is long,
and with a long symbol on the left nothing at its left sides. An unequal long--long step loses gaps only at its right
sides, at most $r_j$, and a self step discards at most $\eta_j$, one per even run. At a self step on a long symbol every
lost gap is at the right end of an even run, since all other cuts it touches have a long symbol on their right.

\emph{Charging.} Give each of the $k$ labels a budget of $\varpi\gb$. The gaps that are not discarded are credited,
intact at cut $i'$, or lost at a middle letter self step on some $\str x$, inside a run (pair $(\str x,\str x)$) or at
its left end (pair $(y,\str x)$ with $y\ne\str x$, a letter or a long atom). A label of the last kind is \emph{assigned}
to that step; it is \emph{matching} if its pair is $(\str x,\str x)$. A label that is not assigned keeps at most
$\gb<\varpi\gb$ gaps, so it has no excess over its budget, except as noted for later steps below. Fix a middle letter
self step $j$ on $\str x$.
\begin{itemize}
\item \emph{Left groups.} Let $p$ labels with pairs $(y,\str x)$, $y\ne\str x$, lose gaps at the left ends of the runs.
Their pairs were not chosen before step $j$, since they still had intact gaps, so all their gaps that are not discarded
were intact just before step $j$, at most $r_j$ per label. They lose at most one gap per run, at most $o_j$ in all, and
keep at most $\gb$ each otherwise. Their excess is at most $\min(o_j+p\gb,\,pr_j)-p\varpi\gb\le\Lx(r_j,o_j)$.
\item \emph{Matching labels.} Let $n$ labels have the pair $(\str x,\str x)$. Such a pair is never chosen at an unequal
step and has no adjacency after step $j$, so their gaps that are not discarded are lost inside the runs of step $j$.
These are raw gaps of step $j$ that are junctions (\cref{lem:refund}(i)), at most $t_j$ in all and at most $o_j$ per
label, being disjoint. So their excess is at most $\min(t_j,no_j)-n\varpi\gb$, which is at most $\Mx(o_j,t_j)$ if
$n\ge1$ and $0$ if $n=0$. If step $j$ is detected at $i$, then $n\ge1$ by \cref{lem:detect}(a), and the excess is at
most $\Mx(o_j,t_j)\le\max(\Mx(o_j,t_j),-\iota t_j)$. If it is not detected and $n\ge1$, the excess is at most
$\Mx(o_j,t_j)\le\max(\Mx(o_j,t_j),-\iota t_j)+\iota t_j$. If it is not detected and $n=0$, the excess $0$ equals
$-\iota t_j+\iota t_j$. In either case the excess is at most $\max(\Mx(o_j,t_j),-\iota t_j)$, plus $\iota t_j$ if step $j$
is not detected at $i$.
\end{itemize}
For a later letter self step $j$ on $\str x$ detected at $i$, the label of \cref{lem:detect}(a) loses no gap at a middle
self step and is never credited, so its gaps that are not discarded are intact at cut $i'$: by \cref{lem:detect}(a) at
$i'$, at most $\gb-(r_j-o_j)$. Its excess is at most $-((\varpi-1)\gb+r_j-o_j)\le \Psi'_j$. If step $j$ is not detected, the true
excess is at most $0\le \Psi'_j+\iota t_j$. All labels named here have different pairs.

Adding up, $|\mathcal W_i|$ is at most the discarded gaps, plus $k\varpi\gb$, plus the excesses above, plus $\iota t_j$
for every letter self step $j>i$ that is not detected at $i$. By \cref{lem:detect}(b) the last term is at most
$\iota\Nmix_i=\iota\sum_{j\le i}[\text{$j$ mixed}]\,o_j$. With \cref{lem:refund}(iii) this is the claim.
\end{proof}

\paragraph{Six step types.} Measure scores, gains and levels in units of $\bar r$, so $\bar r=1$. Every bound above has
the form ``$U\le$ allowance $+\sum_j$ cost of step $j$'', with allowance $k\gamma$ (kinds \kd S, \kd V, \kd C) or
$k\varpi\gb$ (kinds \kd H, \kd J); give each step one $k$-th of it. Whether a step is before, after, middle or later
depends only on its score. \Cref{tab:costsSVC,tab:costsHJ} list the costs of six scalar types of step at score $r$,
without the allowance. In them $\Lx_0$ is $\Lx(r,\frac r2)$ or any upper bound on it that is at most $\frac r2$, such as
$\Lx_{\rm c}$ of \eqref{eq:ellc}.

\begin{table}[!htb]
\centering\small
\renewcommand{\arraystretch}{1.12}
\begin{tabular}{@{}llccccc@{}}
\toprule
Type & Gain & \kd S before & \kd V before & \kd V after & \kd C before & \kd C after\\
\midrule
$\ty{lu}$: unequal letter--letter & $r$ & $2r$ & $2r$ & $0$ & $3r-\gamma$ & $0$\\
$\ty{mx}$: mixed & $r$ & $2r$ & $3r$ & $0$ & $r$ & $0$\\
$\ty{sn}$: letter self, $t=0$ & $r/2$ & $r/2$ & $r/2$ & $0$ & $r/2$ & $0$\\
$\ty{sh}$: letter self, $t=r/2$ & $r/2$ & $r$ & $r$ & $-r/2$ & $r-\gamma$ & $-r/2$\\
$\ty{sa}$: letter self, $t=r$ & $r/2$ & $r$ & $r$ & $-r/2$ & $3r/2-2\gamma$ & $-\gamma$\\
$\ty{ls}$: long self & $r/2$ & $0$ & $0$ & $0$ & $0$ & $0$\\
\bottomrule
\end{tabular}
\caption{Costs of the six step types at score $r$ in the bounds of \cref{lem:levelbounds}, without the allowance
$\gamma$ per step. ``Before'' means $r>\gamma$; the costs of kind \kd S after the level are $0$. The self types have gain
$o=r/2$; $t$ is the number of raw gaps that are junctions.}
\label{tab:costsSVC}
\end{table}

\begin{table}[!htb]
\centering\small
\renewcommand{\arraystretch}{1.12}
\begin{tabular}{@{}llll@{}}
\toprule
Type & Before & Middle & Later\\
\midrule
$\ty{lu}$ & $2r$ & $3r-\gb$ & $0$\\
$\ty{mx}$ & $(2+\iota)r$ & $2r$ & $0$\\
$\ty{sn}$ & $r/2$ & $\Lx_0$ & $0$\\
$\ty{sh}$ & $r$ & $\Lx_0+\max(r/2-\varpi\gb,\,-\iota r/2)$ & $-\min(r/2+(\varpi-1)\gb,\,\iota r/2)$\\
$\ty{sa}$ & $r$ & $\Lx_0+\max(r/2-\varpi\gb,\,r-2\varpi\gb,\,-\iota r)$ & $-\min(r/2+(\varpi-1)\gb,\,\iota r)$\\
$\ty{ls}$ & $0$ & $0$ & $0$\\
\bottomrule
\end{tabular}
\caption{Costs of the six step types at score $r$ in the segment bounds of \cref{lem:segment}, without the allowance
$\varpi\gb$ per step; gains as in \cref{tab:costsSVC}.}
\label{tab:costsHJ}
\end{table}

\begin{lemma}[six types]\label{lem:types}
For every step $j$ there are weights $\xi_{j\upsilon}\ge0$ on the six types $\upsilon$, summing to $1$, such that
$o_j=\sum_\upsilon\xi_{j\upsilon}o_\upsilon(r_j)$ and, for every kind, every choice of $\gamma,\gb,\varpi,\iota$ and each position
(before, after, middle or later), the cost of step $j$ is at most $\sum_\upsilon\xi_{j\upsilon}$ times the cost of type $\upsilon$ at
score $r_j$.
\end{lemma}

\begin{proof}
Write $r=r_j$, $o=o_j$, $t=t_j$. An unequal letter--letter step and a mixed step have exactly the costs of $\ty{lu}$ and
$\ty{mx}$ (\cref{lem:levelbounds,lem:segment}, with $o=\eta=r$). An unequal long--long step has gain $r$ and costs $0$
except $r\le2r$ in the middle, so $\ty{mx}$ dominates it. A long self step with gain $o$ has cost $\eta=2o-r$ in the
middle and $0$ elsewhere; weight $\eta/r$ on $\ty{mx}$ and $1-\eta/r$ on $\ty{ls}$ gives gain $o$ and middle cost
$2\eta\ge\eta$.

Letter self steps have $(o,t)$ in the trapezoid $\frac r2\le o\le r$, $0\le t\le r$. Cut it by the lines $t=r-o$ and
$t=o$, which meet at $(\frac r2,\frac r2)$, into three triangles:
\[
\triangle_1=\mathrm{conv}\{(\tfrac r2,0),(\tfrac r2,\tfrac r2),(r,0)\},\ \
\triangle_2=\mathrm{conv}\{(\tfrac r2,\tfrac r2),(r,0),(r,r)\},\ \
\triangle_3=\mathrm{conv}\{(\tfrac r2,\tfrac r2),(\tfrac r2,r),(r,r)\},
\]
below $t=r-o$, between the two lines, and above $t=o$. The points $(\frac r2,0)$, $(\frac r2,\frac r2)$, $(\frac r2,r)$
are the types $\ty{sn},\ty{sh},\ty{sa}$. Give step $j$ the barycentric weights of $(o,t)$ in a triangle that contains it, with the weight of
$(r,0)$ and $(r,r)$ moved to $\ty{lu}$. Points on a common side get the same weights from both triangles, and the gain $o$
is affine, so the gains agree. It remains to bound each cost on each triangle by a convex function whose values at the
vertices are at most the costs of the corresponding types.
\begin{itemize}
\item The costs before (kinds \kd S, \kd V, \kd H, \kd J), $\Pi_j=2o-(r-o-t)_+$ plus $0$, and the cost $-\min(t,r-o)$ of
kind \kd V after, are affine on $\triangle_1$ and on $\triangle_2\cup\triangle_3$.
\item Kind \kd C before (so $r>\gamma$). The cost has three affine pieces. On the points of $\triangle_1\cup\triangle_2$
with $t>0$ (there $t\le o$) it is $3o-r+t-\gamma$, with vertex values $r/2-\gamma$ at $\ty{sn}$, $r-\gamma$ at $\ty{sh}$,
$2r-\gamma$ at $(r,0)$ and $3r-\gamma$ at $(r,r)$. On the points of $\triangle_3$ with $t>o$ it is $3o-r+t-2\gamma$, with
values $r-2\gamma$ at $\ty{sh}$, $3r/2-2\gamma$ at $\ty{sa}$ and $3r-2\gamma$ at $(r,r)$. On the side $t=0$ of
$\triangle_1$ it is $3o-r$, with values $r/2$ at $\ty{sn}$ and $2r$ at $(r,0)$. A point is given the weights of a triangle
on whose closure its piece is defined (points with $t=o$ lie in $\triangle_2$, points with $t=0$ use only the two
vertices of that side), so it suffices that each vertex value is at most the cost of its type: at most $r/2$ at
$\ty{sn}$, $r-\gamma$ at $\ty{sh}$, $3r/2-2\gamma$ at $\ty{sa}$, and $3r-\gamma$ (type $\ty{lu}$) at $(r,0)$ and $(r,r)$;
the last uses $2r\le3r-\gamma$, that is, $r>\gamma$. After the level ($r\le\gamma$)
the same argument applies to $-(r-t)$, $-(r-t)-\gamma$ and $0$.
\item Kind \kd H or \kd J in the middle. Each minimum in \eqref{eq:excess} grows by at most the increase of $o$, so
$\Lx(r,o)\le\Lx_0+(o-\frac r2)$ and the cost is at most $\Lx_0+\frac32(2o-r)+\max(\Mx(o,t),-\iota t)$. Here $\Mx(o,t)$ is
$t-\varpi\gb$ on $\triangle_1\cup\triangle_2$ and a maximum of two affine functions on $\triangle_3$, so the bound is convex on each triangle.
At $o=\frac r2$ it equals the costs of $\ty{sn},\ty{sh},\ty{sa}$. At $o=r$ it is at most
$\frac r2+\frac32r+(r-\varpi\gb)_+\le3r-\gb$, because $\Lx_0\le\frac r2$ and $r\ge\gb$ in the middle.
\item Kind \kd H or \kd J later: $-\min((\varpi-1)\gb+r-o,\iota t)$ is a maximum of two affine functions, so convex; its
values are those of the types at $o=\frac r2$ and at most $0$ at $o=r$.
\item Kind \kd S after the level costs $0$.\qedhere
\end{itemize}
\end{proof}

\begin{lemma}[averaging]\label{lem:average}
Let finitely many \emph{rows} be given, each with a kind, an interval $[a,b]$ with $1\le a<b$, a stop level
$1\le\gb\le a$ and a price $\varpi>1$ for kinds \kd H and \kd J, and a weight $x\ge0$. For a row and a type $\upsilon$ let
$\mathcal C_\upsilon(r)$ be the expectation, over $\gamma$ uniform on $[a,b]$, of the allowance per step plus the cost of
$\upsilon$ at score $r$. If
\begin{equation}\label{eq:certcheck}
o_\upsilon(r)\;\ge\;\sum_{\text{rows}}x\,\mathcal C_\upsilon(r)\qquad\text{for all six types $\upsilon$ and all real }r\ge1,
\end{equation}
then every overlap-greedy run on every string has $O_k\ge(\sum x)\,\OPT(s,k)$, at every budget $k$.
\end{lemma}

\begin{proof}
For each row and each $\gamma\in[a,b]$, \cref{lem:levelbounds} or \cref{lem:segment} gives $U\le\sum_j(\text{allowance per
step}+\text{cost of }j)$, and by \cref{lem:types} the cost of $j$ is at most $\sum_\upsilon\xi_{j\upsilon}$ times the cost of type
$\upsilon$ at $r_j$, with weights that do not depend on the row or on $\gamma$. All these inequalities hold on the same finite
run. Average over $\gamma$, multiply by $x$, sum over the rows, and apply \eqref{eq:certcheck}:
\[
\Bigl(\sum x\Bigr)U\;\le\;\sum_j\sum_\upsilon\xi_{j\upsilon}\sum_{\rm rows}x\,\mathcal C_\upsilon(r_j)\;\le\;
\sum_j\sum_\upsilon\xi_{j\upsilon}\,o_\upsilon(r_j)\;=\;O_k .\qedhere
\]
\end{proof}

The expectations are explicit. Let $\hat r=\min(b,\max(a,r))$; then $\Pr[\gamma<r]=(\hat r-a)/(b-a)$ and
$\mathbb E[\gamma\,;\gamma<r]=\Pr[\gamma<r]\,(a+\hat r)/2$, and these two numbers and $\mathbb E\gamma=(a+b)/2$ evaluate every
$\mathcal C_\upsilon$. For example, kind \kd C gives type $\ty{sa}$ the averaged cost
$\frac{a+b}2+\frac{3r}2\Pr[\gamma<r]-2\,\mathbb E[\gamma;\gamma<r]-\bigl(\frac{a+b}2-\mathbb E[\gamma;\gamma<r]\bigr)$.

\paragraph{The hand certificate.} All segment rows of the hand certificate use the terminal horizon $\gb=1$, so they have
no later steps, and they use in place of $\Lx_0$ the larger charge
\begin{equation}\label{eq:ellc}
\Lx_{\rm c}(r)=\begin{cases}0,&r\le\varpi,\\[2pt] \dfrac{r(r-\varpi)}{2(r-1)},&r>\varpi,\end{cases}
\end{equation}
the maximum in \eqref{eq:excess} over all real $p\ge0$ at $o=\frac r2$; its two branches meet at $p=r/(2(r-1))$, and
$\Lx_0\le\Lx_{\rm c}\le\frac r2$. Split $[1,\infty)$ at $1$ and $3$, at the ends $a,b$ of the rows, at each price
$\varpi$ and $2\varpi$, and, for \kd J rows, at $2\varpi/3$ and $2(\varpi-1)$; these are the points where a maximum or
minimum in \cref{tab:costsHJ} or in \eqref{eq:ellc} changes branch. On each cell between consecutive points, every slack
$o_\upsilon(r)-\sum x\,\mathcal C_\upsilon(r)$ is a polynomial of degree at most $2$ plus terms with the denominator $r-1$ from
$\Lx_{\rm c}$. Above the smallest price, multiply it by $r-1>0$; the result $\mathcal P$ has degree at most $3$. If $u,v,v',u'$
are the values of $\mathcal P$ at the four equally spaced points of a closed cell, from left to right, its cubic Bernstein
coefficients are
\begin{equation}\label{eq:bernstein}
u,\qquad\tfrac16(-5u+18v-9v'+2u'),\qquad\tfrac16(2u-9v+18v'-5u'),\qquad u' ,
\end{equation}
and $\mathcal P$ is at least their minimum on the cell, because the Bernstein basis polynomials are nonnegative and sum to $1$.
Beyond the last point the slacks are affine.

\emph{A two-row warm-up.} Take only kind \kd C on $[1,\frac52]$ with weight $\frac9{70}$, and kind \kd H on
$[\frac65,\frac{12}5]$ with $\varpi=\frac{11}{10}$ and weight $\frac14$. The cells end at $\frac{11}{10},\frac65,\frac{11}5,
\frac{12}5,\frac52,3$; on them the smallest coefficients \eqref{eq:bernstein} over the six types are \OvTwoRowMinima,
and beyond $3$ the slacks are affine with slopes at least $\frac2{35}$ and nonnegative values at $3$. So every run has
$O_k\ge(\frac9{70}+\frac14)U=\frac{53}{140}U\approx\OvTwoRowDec\,U$.

\emph{The certificate.} \Cref{tab:ovhand} lists the $\OvHandRows$ rows, and \cref{tab:ovsign} the $\OvHandCells$ cells
and, for each, the floor of $10^6$ times the smallest coefficient \eqref{eq:bernstein} over the six types; all are
positive. At $r=1$ the six slacks, in the order of \cref{tab:costsSVC}, are \OvHandOneSlacks{} divided by
$2\cdot10^8$, and beyond $3$ they are affine with slopes whose floors, in millionths, are \OvHandSlopes. So
\eqref{eq:certcheck} holds, and \cref{lem:average} gives $O_k\ge\lambda_{\rm h}U$ with
$\lambda_{\rm h}=\OvLBhand=\OvLBhandDec$; and $\lambda_{\rm h}-\frac7{18}=\frac{586}{9\cdot10^6}>0$.

\begin{table}[!htb]
\centering\small
\begin{tabular}{@{}rrrrl@{}}
\toprule
$a$ & $b$ & \kd V & \kd C & segment row: kind, $\varpi$, weight\\
\midrule
\TabOvHand
\bottomrule
\end{tabular}
\caption{The hand certificate: rows with $\gamma$ uniform on $[a,b]$ (units of $\bar r$) and weights in millionths;
a blank is $0$. Segment rows have $\gb=1$. The weights sum to $\OvLBhandDec$.}
\label{tab:ovhand}
\end{table}

\begin{table}[!htb]
\centering\scriptsize
\setlength{\tabcolsep}{2.2pt}
\begin{tabular}{@{}l*{16}{r}@{}}
\toprule
\TabOvHandSign
\bottomrule
\end{tabular}
\caption{Sign table of the hand certificate. The cells start at $1$; for each the right end and the floor of $10^6$ times
the smallest Bernstein coefficient \eqref{eq:bernstein} over the six types.}
\label{tab:ovsign}
\end{table}

\paragraph{The larger certificate.} The $\OvFullRows$ rows of \texttt{weights.json}, printed in \cref{tab:ovfull}, also use
kind \kd S, stop levels
$\gb>1$ and the exact charge $\Lx$ with integer $p$. For $r\le\varpi\gb$ that charge is $0$. Otherwise the function of $p$
in \eqref{eq:excess} is concave, so it is enough to compare $p=0$ with the two integers next to $r/(2(r-\gb))$, and every
averaged slack is piecewise quadratic. Its pieces are separated by the points listed for the hand certificate (with
$\gb\varpi$ for $\varpi$), by $\gb$, and, above $\varpi\gb$, by $2p\gb/(2p-1)$ and $2\gb(p-\varpi+1)/(2p-1)$ for
$1\le p\le\lfloor\varpi/(2(\varpi-1))\rfloor+2$, where the best integer changes; this gives $\OvFullCells$ finite cells.
On each cell the quadratic through the slacks at the two ends and the midpoint has an exact minimum on the cell. At a
stop level $\gb>1$ both one-sided values are checked, since a step with $r_j=\gb$ is later. Beyond the last point the
slacks are affine with nonnegative values and slopes. The floors of $10^6$ times the minima over the cells are, by type,
\OvFullMinima, and the slopes, in millionths, are \OvFullSlopes. So \eqref{eq:certcheck} holds and
$O_k\ge\lambda_*U$ with $\lambda_*=\OvLBfull=\OvLBfullDec$. The rows are printed so that the claim can be re-checked
from the paper alone; the sign check over the $\OvFullCells$ cells is done by the released program.

\begin{table}[!htb]
\centering\scriptsize
\setlength{\tabcolsep}{3.2pt}
\begin{tabular}{@{}lrrrrr@{\hspace{14pt}}lrrrrr@{}}
\toprule
kind & $a$ & $b$ & $\gb$ & $\varpi$ & weight & kind & $a$ & $b$ & $\gb$ & $\varpi$ & weight\\
\midrule
\TabOvFull
\bottomrule
\end{tabular}
\caption{The $\OvFullRows$ rows of the larger certificate (\texttt{weights.json}): kind, interval $[a,b]$ of the level
$\gamma$, stop level $\gb$ and price $\varpi$ (used by kinds \kd H and \kd J only), all in units of $\bar r$, and the weight
in units of $10^{-8}$. The weights sum to $\OvLBfullDec$.}
\label{tab:ovfull}
\end{table}

\begin{sloppypar}
\paragraph{Running the checks.} The folder \texttt{scripts/overlap\_cert} holds the two certificates
(\texttt{hand.json} and \texttt{weights.json}, lists of rows: kind, $a$, $b$, $\gb$, $\varpi$, and the weight in units
of $10^{-8}$), the checker \texttt{certificate.py}, its recorded output \texttt{certificate.json}, the relaxation check
below, and \texttt{SHA256SUMS}. The command \texttt{python3 certificate.py}, run in that folder (Python 3.10 or later,
standard library only, about one second), verifies both certificates in exact rational arithmetic as described above:
the coefficients \eqref{eq:bernstein} on every cell of the hand certificate, the exact quadratic minima on every cell of
the larger one, the value at $r=1$, the values and slopes of the affine tails, the weight sums $\lambda_{\rm h}>\frac7{18}$
and $\lambda_*$, and a negative control (all weights raised by $5\%$ must make a slack at $r=1$ negative). It stops with
an error if any check fails, and otherwise prints \texttt{certificate.json}. \texttt{build.sh} reruns it and compares the
output with the recorded file, and \texttt{make\_numbers.py} compares the recorded numbers with a separately written
verifier (\cref{app:evidence}). The SHA-256 hashes are listed below.
\end{sloppypar}
\begin{center}\scriptsize
\begin{tabular}{@{}ll@{}}
\TabOvShas
\end{tabular}
\end{center}

\paragraph{A limit of these rows.} The program \texttt{relaxation.py} puts mass on the seven points (type, score) of
\cref{tab:ovrelax}: it
evaluates the averaged costs of all $\OvRelaxRows$ rows of the two certificates there, checks in exact arithmetic that
each row's total is at least $U$, and that the total gain is $\OvRelaxGain\,U<\frac25U$. If weights on these rows
satisfied \eqref{eq:certcheck}, summing \eqref{eq:certcheck} against the seven masses would bound their sum by
$\OvRelaxGain$. So no reweighting of these rows proves $\frac25$. Whether the seven points are realised by a string is
open, and other rows may exclude them.

\begin{table}[!htb]
\centering\small
\begin{tabular}{@{}lrr@{}}
\toprule
type & score $r$ & mass ($10^{-4}$)\\
\midrule
\TabOvRelax
\bottomrule
\end{tabular}
\caption{The seven points of the relaxation (\texttt{relaxation.py}). A mass is the number of steps of that type and score
times $\bar r/U$; scores are in units of $\bar r$. The masses sum to $\OvRelaxMass$, and the total gain is
$\OvRelaxGain\,U$.}
\label{tab:ovrelax}
\end{table}

\paragraph{Word tables and string-identified symbols.} The bound $O_k\ge\lambda_*\OPT(s,k)$ holds for every legal run on
every string at each fixed budget $k$, which is the hypothesis of \cref{lem:transfer}. Applied with $\alpha=\lambda_*$
(or $\lambda_{\rm h}$) at each budget, that lemma gives part~(a) for word tables with positive real weights, and
\cref{prop:identity} gives it for string-identified symbols on uniquely decodable inputs. This completes the proof of
part~(a).

\subsection{Proof of part (b): a four-phase construction}\label{app:fourphase}

In this subsection the \emph{score} of an ordered pair is its raw count $\raw$, which overlap greedy maximises; a merge
replaces from left to right, and its saving is $\cnt$. A run of three equal letters scores $2$ and saves $1$. We build
one string on which every legal overlap-greedy run, under every tie order, has the same saving after $k$ steps.

\paragraph{The construction.} Fix integers $4\le d\le H$ with $3\mid H$, and $h\ge1$. Subscripts $u,v$ are residues
modulo $H$ (inside a floor, take the representative in $\{0,\dots,H-1\}$), and $0\le l<d$. Put
\[
R=(1-1/d)^h,\qquad \Gamma=\tfrac56R,\qquad m=\min\{i\ge0:R(1-2/d)^i\le\Gamma\},\qquad S=R(1-2/d)^m ,
\]
so $m\ge1$. Let $k=2Hd$, let $n_q=Hd/3$ be the number of $q$ letters below, $z=k-H(h+m)-n_q$ and $F=12d^{h+m}$, and
require $h+m\le H$, $1/d\le\Gamma$ and $z\ge0$. Let $f_j=(1-1/d)^j$ for $0\le j<h$ and $f_j=R(1-2/d)^{j-h}$ for
$h\le j\le h+m$, so $f_0=1$, $f_h=R$, $f_{h+m}=S$, and put $e_j=Ff_j/d$ for $j<h+m$. Letters in different named families
are distinct, and $q(v,l)=q_{\lfloor v/3\rfloor,l}$, so each $q$ letter serves exactly three pairs $(v,l)$. The $k$
\emph{side} pairs, half of each class, and the $H(h+m)$ \emph{middle} pairs are
\[
A_{u,l}=b_{u+l}\,a_u,\qquad B_{v,l}=b_v\,q(v,l),\qquad M_{j,u}=a_u\,b_{u+j}\quad(0\le j<h+m);
\]
$d\le H$ makes the $A_{u,l}$ distinct and $h+m\le H$ the $M_{j,u}$. The string consists of the following blocks.
\begin{itemize}
\item For each $j,u,l$, with $v=u+j$: $e_j$ copies of $b_{u+l}\,a_u\,b_v\,q(v,l)$ if $j<h$ (first series), and of
$b_{u+l}\,a_u\,b_v\,a_{v-l}$ if $h\le j<h+m$ (second series).
\item $FS$ copies of each $A_{u,l}$, $F\Gamma$ copies of each $B_{v,l}$, and, for each $B_{v,l}=b_vq$, $FR/6$ copies of
$b_v\,qqq$.
\item For each of $z$ private letters $x$, $F\Gamma/2$ copies of $xxx$.
\item The \emph{bonus} blocks: one block $a_ub_{u+j}$ per middle, two blocks $qq$ per $q$ letter, one block $xx$ per $x$.
\end{itemize}
All blocks are joined by distinct separator letters, in any order. The multiplicities are integers:
$e_j=12d^{h+m-j-1}(d-1)^j$ for $j<h$ and $e_j=12(d-1)^h(d-2)^{j-h}d^{m-1-(j-h)}$ for $h\le j<h+m$, $FS=12(d-1)^h(d-2)^m$,
$FR/6=2(d-1)^hd^m$ and $F\Gamma/2=5(d-1)^hd^m$. Write $[a_ub_v]$, $[qq]$, $[xx]$ for the fresh tokens of the middle and
self-pair merges.

\begin{lemma}[order inside a group does not matter]\label{lem:grouporder}
Apply any set $P$ of distinct rules from the middles, the pairs $qq$ and the pairs $xx$, in any order. Each block is then
changed by at most its own rule of $P$: a level block becomes $b_{u+l}\,[a_ub_v]\,t$, with $t$ its last letter, a block
$b_vqqq$ becomes $b_v\,[qq]\,q$, a block $xxx$ becomes $[xx]\,x$, and a bonus block becomes one token; all other blocks
are unchanged. So the string after $P$ depends only on the set $P$, up to the names of fresh tokens.
\end{lemma}

\begin{proof}
A middle $a_ub_{u+j}$ occurs only in the level blocks of level $j$ with that $u$ (another level $j'$ would need
$j'\equiv j\pmod H$, and $j,j'<h+m\le H$) and in its bonus block; $qq$ only in the blocks $b_vqqq$ of that $q$ and its
bonus blocks; $xx$ only in the blocks of $x$. A merged block contains no further occurrence of any rule of the list, and
separators keep every merge inside its block.
\end{proof}

\paragraph{The scores.} Let $I_{j,u}$, $I_q$, $I_x$ be the indicators that the rule $M_{j,u}$, $qq$, $xx$ is not in
$P$. By \cref{lem:grouporder} the positive scores after $P$ are exactly the following.
\begin{enumerate}
\item The rules of the list score $I_{j,u}(Ff_j+1)$, $I_q(FR+2)$ and $I_x(F\Gamma+1)$: a middle has $d$ groups of $e_j$
level blocks and its bonus; each $q$ has three groups of $FR/6$ triples, two adjacencies each, and two bonuses; each
$x$ has two adjacencies per triple and one bonus.
\item The sides score
\begin{align*}
\raw(A_{u,l})&=FS+\sum_{j=0}^{h+m-1}e_jI_{j,u}+\sum_{j=h}^{h+m-1}e_jI_{j,u+l-j},\\
\raw(B_{v,l})&=F\Gamma+\tfrac{FR}6\,I_{q(v,l)}+\sum_{j=0}^{h-1}e_jI_{j,v-j}.
\end{align*}
The first sum counts $A_{u,l}$ at the left end of level blocks. At the right end of a second-series block $A_{u,l}$
needs last letter $a_u$ after $b_{u+l}$, which forces the same $l$ and the middle $M_{j,u+l-j}$; when this is the same
middle as in the first sum ($j\equiv l$), the block is $b_{u+l}a_ub_{u+l}a_u$ and its two occurrences are disjoint. For
$B_{v,l}$ the right end of a first-series block fixes $v$ and $l$.
\item Pairs of a fresh token with a neighbour: $b_{u+l}[a_ub_{u+j}]$ and $[a_ub_{u+j}]\,t$ score $e_j$ (the fresh token
fixes $j,u$ and the neighbour fixes $l$, also when $t=a_{u+j-l}$); $[qq]\,q$ scores $FR/2$ and each of the three
$b_v[qq]$ scores $FR/6$; $[xx]\,x$ scores $F\Gamma/2$. All are at most $F\Gamma$, since $e_j\le F/d\le F\Gamma$ and
$R/2<\Gamma$.
\item Every pair containing a separator scores $1$.
\end{enumerate}
Every block has at most four letters, and the letter pairs inside a block are side, middle or self pairs, so this
exhausts all adjacencies. Initially both sums telescope: $e_j=F(f_j-f_{j+1})$ for $j<h$ and $2e_j=F(f_j-f_{j+1})$ for
$j\ge h$, so every side scores $FS+F(1-S)=F$ and $F\Gamma+FR/6+F(1-R)=F$. Every side score is nonincreasing in $P$.

\emph{The comparator.} The designated side occurrences are pairwise disjoint (positions $1,2$ and $3,4$ of a level
block, $1,2$ of $b_vqqq$), so merging the $k$ sides saves $kF$, and $\OPT(s,K)\ge kF$ for every $K\ge k$. The largest
initial score is $F+1$ (item~1 at $j=0$). A replacement by any rule, even a partial rule on compound tokens, erases one
original gap whose two letters are the last letter of its left operand and the first of its right operand, and distinct
replacements erase distinct gaps; so every rule saves at most $F+1$.

\paragraph{Every maximising choice.} The groups, in order, are
\[
(M_{0,u})_u,\ \dots,\ (M_{h-1,u})_u,\quad (qq)_q,\quad (M_{h,u})_u,\ \dots,\ (M_{h+m-1,u})_u,\quad (xx)_x ,
\]
where an empty last group is skipped. We show by induction on the step that the maximisers are exactly the unused rules
of the first unfinished group. Suppose the claim holds so far, so the prefix is a set $P$ of earlier groups plus part of
the current one. Subtracting the finished groups from the initial side scores $F$ gives the table below. Its rows are the
first-series levels $j<h$, the $qq$ group, the second-series levels $h\le j<h+m$ and the $xx$ group; each row holds at
any point inside its group, since side scores are bounded by their value at its start and only decrease.
\begin{center}\small
\begin{tabular}{@{}lccccccc@{}}
\toprule
group & current & $A_{u,l}$ & $B_{v,l}$ & later middles & $qq$ & $xx$ & other pairs\\
\midrule
level $j<h$ & $Ff_j+1$ & $\le Ff_j$ & $\le Ff_j$ & $\le Ff_{j+1}+1$ & $FR+2$ & $F\Gamma+1$ & $\le F\Gamma$\\
$qq$ & $FR+2$ & $\le FR$ & $\le FR$ & $\le FR+1$ & -- & $F\Gamma+1$ & $\le F\Gamma$\\
level $j\ge h$ & $Ff_j+1$ & $\le Ff_j$ & $F\Gamma$ & $\le Ff_{j+1}+1$ & $0$ & $F\Gamma+1$ & $\le F\Gamma$\\
$xx$ & $F\Gamma+1$ & $FS$ & $F\Gamma$ & $0$ & $0$ & -- & $\le F\Gamma$\\
\bottomrule
\end{tabular}
\end{center}
The $B$ entries follow because each $B$ loses $e_j$ at the first-series level $j$ and $FR/6$ in the $qq$ group, ending at
$F\Gamma$; the $A$ entries because each $A$ loses $e_j$ at a first-series level and $2e_j$ at a second-series level,
reaching $Ff_j$ at the start of level $j$ and $FS$ at the end. In every row the first column is strictly largest:
\begin{itemize}
\item first series: $f_j>f_{j+1}$, and $Ff_j\ge Ff_{h-1}$ with
\[
F(f_{h-1}-R)=Ff_{h-1}/d=12\,d^m(d-1)^{h-1}>1,
\]
so $Ff_j+1>FR+2$; also $FR+2>F\Gamma+1$;
\item $qq$: $FR+2>FR+1>F\Gamma+1$, since $F(R-\Gamma)=FR/6\ge2$;
\item second series: $f_j>\Gamma$ for $h\le j<h+m$ by the minimality of $m$, so $Ff_j+1>F\Gamma+1$;
\item $xx$: $F\Gamma+1>F\Gamma\ge FS$.
\end{itemize}
The last column is at most $F\Gamma$ by items~3 and~4, as $F\Gamma\ge F/d>1$. So the maximisers are exactly the unused
rules of the current group, whatever order was used inside it, and by \cref{lem:grouporder} the next row of the table is
reached however the group was ordered. This covers every tie history, and the groups use $H(h+m)+n_q+z=k$ rules.

\paragraph{Savings.} A middle of level $j$ saves $Ff_j+1$, a $qq$ saves $FR/2+2$ (one replacement per triple, one per
bonus block) and an $xx$ saves $F\Gamma/2+1$. By telescoping the two middle series save $HFd(1-R)$ and $HFd(R-S)/2$, and
with $n_q/k=1/6$ and $z/k=\frac56-\frac{h+m}{2d}$ every run saves
\begin{equation}\label{eq:Dd}
O_k=kF\,D_d+k+n_q,\qquad
D_d\;=\;\underbrace{\frac{1-R}{2}}_{\text{phase 1}}\;+\;\underbrace{\frac{R}{12}}_{\text{phase 2}}
\;+\;\underbrace{\frac{R-S}{4}}_{\text{phase 3}}\;+\;\underbrace{\frac{\Gamma}{2}\Bigl(\frac{5}{6}-\frac{h+m}{2d}\Bigr)}_{\text{phase 4}} ,
\end{equation}
where $k+n_q$ counts the bonus blocks. Without the bonus blocks the largest initial score is $F$, the optimum is still
$kF$, the same order is legal by the weak forms of the comparisons and saves $kFD_d$, but other tie orders can do better;
the bonuses remove those ties.

\paragraph{The limit.} Take $H=3d$, $\beta=\frac{13}{15}-\frac12\log\frac65$ and $h=\lfloor\beta d\rfloor$. As
$d\to\infty$, $R\to e^{-\beta}$, $m=\lceil\log(5/6)/\log(1-2/d)\rceil$ satisfies $m/d\to\frac12\log\frac65$, and
$\Gamma(1-2/d)<S\le\Gamma$. So $(h+m)/d\to\frac{13}{15}$, $\Gamma,S\to\Gamma_*=\sqrt{5/6}\,e^{-13/15}$ and
$R\to\frac65\Gamma_*$; in particular $h\ge1$, $h+m\le H$, $\Gamma\ge1/d$ and $z/k\to\frac56-\frac{13}{30}>0$ for all large
$d$. Collecting terms in \eqref{eq:Dd},
\[
D_d=\frac12-\frac R6-\frac S4+\frac{5\Gamma}{12}-\frac{h+m}{d}\,\frac\Gamma4
\;\longrightarrow\;\frac12+\Gamma_*\Bigl(-\frac15-\frac14+\frac5{12}-\frac{13}{60}\Bigr)=\frac12-\frac{\Gamma_*}4=\Lambda_4 .
\]
(For general $\beta$ the limit is $\frac12+\frac\Gamma4\log\Gamma+\frac\Gamma2(\frac14\log\frac65-\frac1{15})$ with
$\Gamma=\frac56e^{-\beta}$, a convex function of $\Gamma$ minimised at $\Gamma_*$; this is how $\beta$ was chosen.) For a
large budget $K$ let $d=\lfloor\sqrt{K/6}\rfloor$, so $k=6d^2\le K<6(d+1)^2$. Every legal run has the saving above after
$k$ steps and saves at most $F+1$ per later step, and $\OPT(s,K)\ge kF$, so
\[
\bar\rho^{\mathrm{ov}}_K\;\le\;\max_{\text{runs on }s}\frac{O_K}{\OPT(s,K)}\;\le\;D_d+\frac{k+n_q}{kF}+\frac{K-k}{k}\Bigl(1+\frac1F\Bigr),
\]
and the last two terms tend to $0$, as $0\le K-k<12d+6$. This proves \cref{prop:overlap}(b), with ties broken in
greedy's favour. For the enclosure let $P_n=\sum_{i=0}^n(-13/15)^i/i!$; the alternating series gives
$P_9<e^{-13/15}<P_{10}$, squaring gives $0.912870<\sqrt{5/6}<0.912871$, and
$\OvFourClo<\frac12-0.912871\,P_{10}/4<\Lambda_4<\frac12-0.912870\,P_9/4<\OvFourC$.

\subsection{Proof of part (c): a potential on original junctions}\label{app:twofifths}

Part~(c) follows from a statement about any set of original cuts. Fix a set of $W$ cuts of $s$, the \emph{designated}
cuts, and call the pair of letters on the two sides of a cut its \emph{type}; suppose at most $k$ types occur. For a
greedy run of $k$ merges let $\ell_j$ count the designated cuts that stop being intact at step $j$.

\begin{lemma}[two fifths]\label{lem:twofifths}
If $\ell_j\le2o_j$ for every $j$, then $O_k\ge\frac25W$. In general $O_k\ge\frac25\bigl(W-\sum_j(\ell_j-2o_j)_+\bigr)$.
\end{lemma}

\begin{proof}[Proof of \cref{prop:overlap}(c) from \cref{lem:twofifths}]
Designate the $U$ junctions of $\sigma$; their types are the pairs $\pi_r$, at most $k$ of them. A replacement changes
the tokens next to three cuts, so $\ell_j\le2o_j+\tau_j$, where $\tau_j$ counts the replacements of step $j$ whose centre and
two sides are all intact junctions. Such a replacement joins two single letters at positions $p,p+1$, and the cuts
$p-1,p,p+1$ are junctions. By (F1) a junction of a tree lies strictly inside the span of its root, and the spans of
different roots are disjoint, so the positions $p-1,\dots,p+2$ lie in one final token $T$ of $\sigma$, with $p,p+1$
interior to it. After the replacement $p$ and $p+1$ are no longer single letters, so different such replacements, at any
steps, use disjoint pairs of interior positions, at most $\lfloor(|T|-2)/2\rfloor$ in $T$. Hence
$\sum_j(\ell_j-2o_j)_+\le\sum_j\tau_j\le\sum_T\lfloor(|T|-2)/2\rfloor$, and \cref{lem:twofifths} gives (c).
\end{proof}

The general form of the lemma follows from the first: at each step $j$ remove from the designated set
$(\ell_j-2o_j)_+$ of the cuts that stop being intact at step $j$. The other cuts keep their times, so the new counts are
$\min(\ell_j,2o_j)$, at most $k$ types remain, and the first statement applies to $W-\sum_j(\ell_j-2o_j)_+$ cuts.

\begin{proof}[Proof of the first statement of \cref{lem:twofifths}]
Assume $\bar r\ge1$ (if $\bar r=0$, every cut is erased and $O_k=|s|-1\ge W$), and measure numbers of cuts, raw counts and gains in units of $\bar r$, so that $r_j\ge1$ and
$o_j\ge\frac12$ at every step. Let
\[
\Phi(y)=\begin{cases}0,&y\le1,\\ \tfrac12(y-1)-\tfrac14\ln y,&y>1,\end{cases}\qquad\text{so}\qquad
\Phi'(y)=\begin{cases}0,&y<1,\\ \tfrac12-\tfrac1{4y},&y>1.\end{cases}
\]
$\Phi$ is convex, $\Phi(y)\le\frac12(y-1)$ for $y>1$, and for $r\ge1$, $\frac14\le\Phi'(r)<\frac12$ and $\Phi'\le\Phi'(r)$
on $(0,r)$; also $2r\Phi'(r)=r-\frac12$. Number the types $1,\dots,k$ (adding empty ones), let $x_i$ be the initial
number of designated cuts of type $i$ and $m=W/k$ their mean. The \emph{mass} of a type is its number of intact
designated cuts; it is at most $r_j$ before step $j$ and at most $1$ at the end, since these cuts are distinct
adjacencies of its two letters. The potential is $\sum_i\Phi(\text{mass}_i)$: it starts at $\sum_i\Phi(x_i)$ and ends at
$0$, so the drops $\Delta_j$ of the steps sum to $\sum_i\Phi(x_i)$. When a type's mass falls from $y$ to $y-\lambda$, we
say that its lost cuts occupy the \emph{heights} $(y-\lambda,y]$; the drop is $\int\Phi'$ over the lost heights, and a
lost cut at a height at most $r_j$ costs at most $\Phi'(r_j)$.

\emph{Unequal steps}, $r=o$. Let $\ell^{\rm L},\ell^{\rm C},\ell^{\rm R}$ count the lost cuts at left sides, centres and
right sides (a cut that is a side of two replacements counted once); $\ell^{\rm L},\ell^{\rm R}\le r$, and if the pair
$(X,Y)$ is a pair of letters, $\ell^{\rm C}$ is the whole mass of type $(X,Y)$, else $\ell^{\rm C}=0$. As
$\ell=\ell^{\rm L}+\ell^{\rm C}+\ell^{\rm R}\le2r$, the slack
\begin{equation}\label{eq:Sunequal}
S_j=o_j-\tfrac12-\Delta_j=(2r-\ell)\Phi'(r)+\int_{\text{lost}}\bigl(\Phi'(r)-\Phi'(u)\bigr)du
\end{equation}
is nonnegative. \emph{Self steps on a compound symbol} touch no intact cut; put $S_j=o_j-\frac12\ge0$.

\emph{Self steps on a letter $x$.} The touched cuts are the internal cuts of the runs, of type $(x,x)$, whose whole mass
$i_j$ is lost; the left sides, one per run, of types $(y,x)$ with $y\ne x$; and the right sides of even runs, at most
$\eta=\varphi r$ of them, $0\le\varphi\le1$. If $i_j>1$, then $\Phi(i_j)\le\frac12(i_j-1)$ and every other lost cut costs
less than $\frac12$, so $\Delta_j\le\frac12(\ell_j-1)\le o_j-\frac12$. If $i_j\le1$, let $n$ be the number of types
with a loss at a left side. A loss $\lambda$ from a mass $y\le r$ costs at most $\Phi(r)-\Phi(r-\lambda)$, which is
concave and nondecreasing in $\lambda$, the left losses total at most $o=\frac12(1+\varphi)r$, and right losses cost at
most $\Phi'(r)$ each. In detail: let $\lambda_1,\dots,\lambda_n$ be the left losses of the $n$ types. Jensen's
inequality for the concave function $\lambda\mapsto\Phi(r)-\Phi(r-\lambda)$ bounds their total cost by $n$ times its
value at the mean $\frac1n\sum_i\lambda_i$, and since the function is nondecreasing and the mean is at most $o/n=r\vartheta$,
by $n\bigl[\Phi(r)-\Phi(r(1-\vartheta))\bigr]$; the at most $\eta$ right losses add at most $\eta\Phi'(r)$, and the
internal cuts, of total mass $i_j\le1$, cost nothing since $\Phi$ vanishes on $[0,1]$. So $\Delta_j-(o_j-\frac12)$ is at most
\[
H(r)=n\bigl[\Phi(r)-\Phi\bigl(r(1-\vartheta)\bigr)\bigr]+\eta\Phi'(r)-o+\tfrac12,\qquad \vartheta=\frac{1+\varphi}{2n}.
\]
For $n=0$, $H=-\frac12(r-1)-\eta/(4r)\le0$. At fixed $\varphi$, $dH/dr=n\Phi'(r)-\frac12$ below the breakpoint
$r_0=1/(1-\vartheta)$ and $dH/dr=-\frac14(1-\varphi)\le0$ above it. For $n=1$ both are at most $0$ and $H(1)=-\varphi/4$,
so $H\le0$. For $n\ge2$, $\vartheta\le\frac12$ and $n\Phi'\ge\frac12$, so $H\le H(r_0)=nV(\nu,\vartheta)$ with $\nu=1/(2n)$,
$\vartheta\in[\nu,2\nu]$ and
\[
V(\nu,\vartheta)=\tfrac32\nu-\tfrac12\vartheta-\frac{2\nu-\vartheta}{2(1-\vartheta)}+\tfrac14\ln(1-\vartheta).
\]
$\partial V/\partial\vartheta$ has the sign of $-1+5\vartheta-2\vartheta^2-4\nu$, increasing on $[\nu,2\nu]$, so on any
subinterval $V$ is largest at an endpoint. $V(\nu,2\nu)=\frac\nu2+\frac14\ln(1-2\nu)\le0$, and
$V(\nu,\nu)=\chi_n=\nu-\frac{\nu}{2(1-\nu)}+\frac14\ln(1-\nu)$. As a function of $\nu$, $\chi$ has derivative of the sign
of $1-7\nu+4\nu^2$, whose root lies in $(\frac18,\frac16)$, and $\chi_3-\chi_4=\frac{11}{840}-\frac14\ln\frac{21}{20}>0$.
So in all cases $H\le\max(0,n\chi_n)\le n\chi$, with $\chi=\chi_3=\frac1{15}+\frac14\ln\frac56>\frac1{50}$.

\emph{Groups and slacks.} The \emph{group} of a letter $x$ is the set of nonempty types $(y,x)$, $y=x$ allowed; let
$d_x$ be its size, so $\sum_xd_x\le k$; the $n$ types above are types of the group of $x$ other than $(x,x)$, so
$n\le d_x$. For the
self step on $x$, if any, put $S_x=o-\frac12+d_x\chi-\Delta\ge0$; if there is none, $S_x=d_x\chi$. Put
$S_{\rm pad}=(k-\sum_xd_x)\chi$ and $J=\sum_i\Phi(x_i)-k\Phi(m)\ge0$ (Jensen). Summing the drops,
\begin{equation}\label{eq:Gslack}
O_k=k\bigl(\tfrac12-\chi+\Phi(m)\bigr)+S,
\end{equation}
where $S\ge0$ is the sum of $J$, $S_{\rm pad}$ and all the $S_j$ and $S_x$. Suppose $O_k<\frac25W=\frac25km$. As
$O_k\ge\frac k2$, $m>\frac54$, and \eqref{eq:Gslack} gives $0\le S/k<\psi(m)=\chi-\frac m{10}+\frac14\ln m$. The function
$\psi$ increases up to $\frac52$ and decreases after it, and $\psi(\frac{12}5)\approx\OvPsiTwelveFifths<0$,
$\psi(\frac{13}5)\approx\OvPsiThirteenFifths<0$ and $\psi(\frac52)\approx\OvPsiFiveHalves<\frac1{6000}\approx\OvInvSixThousand$. So $\frac{12}5<m<\frac{13}5$ and $S<k/6000$. Since $\sum_i(x_i-m)=0$, $J$ is the sum of the
tangent gaps $\Phi(x_i)-\Phi(m)-\Phi'(m)(x_i-m)\ge0$, and for $x_i>4$ the gap is at least
$\int_4^{x_i}(\Phi'(u)-\Phi'(m))du\ge(x_i-4)(\Phi'(4)-\Phi'(\frac{13}5))>(x_i-4)/30$. So at most $30J$ lost cuts lie at
heights above $4$.

\emph{Exceptional groups.} Call the group of $x$ \emph{exceptional} if $d=d_x\ge10$, or if the self step on $x$ has
$o>d/3$. We show $S_x\ge d/150$ for these. With no self step, or with $i_j>1$, $S_x\ge d\chi>d/50$. Otherwise
$S_x\ge d\chi-H$. If $d\ge10$: $\ln(1-\nu)\le-\nu$ gives $n\chi_n\le\frac18$, so $S_x\ge d\chi-\frac18\ge
d(\frac1{50}-\frac1{80})>d/150$. If $d\le9$ and $o>d/3$, it suffices that $H\le d/75$, since $d\chi-d/150>d/75$. For
$n\le1$, $H\le0$; for $n=2$, $H\le2\chi_2<\frac2{75}$, as $\chi_2=\frac1{12}+\frac14\ln\frac34<\frac1{75}$. For $3\le
n\le9$ and $\varphi\ge\frac13$, $\vartheta\ge\frac43\nu$, so $V\le\max\{V(\nu,\frac43\nu),0\}$, and with
$\ln(1-\vartheta)\le-\vartheta$, $V(\nu,\frac43\nu)\le\frac32\nu-\frac34\vartheta-\frac{2\nu-\vartheta}{2(1-\vartheta)}
\le\frac1{75}$ at $\vartheta=\frac43\nu$ for $n=3,\dots,9$ (by direct evaluation); so $H\le n/75\le d/75$. For $n\ge3$
and $\varphi<\frac13$, $r=2o/(1+\varphi)>\frac32o>\frac d2$, while $r_0\le1/(1-\frac2{3n})\le\frac97$, and above $r_0$
the slope of $H$ is below $-\frac16$; so $H\le n\chi-\frac16(\frac d2-\frac97)$ and $S_x\ge\frac16(\frac
d2-\frac97)\ge\frac d{84}>\frac d{150}$, as $n\le d$ and $d\ge3$. In a group that is not exceptional, the type $(x,x)$, if
present, is not among the $n$ types, so $S_x\ge\chi$.

\emph{Unequal steps.} On $(0,r)$, $\Phi'(r)-\Phi'(u)\ge(r-u)/(4r^2)$. The centre type loses all its mass, at heights
$(0,\ell^{\rm C}]$, and below height $1$ the integrand is $\Phi'(r)\ge\frac14$, so it contributes at least
$\frac14\min(\ell^{\rm C},1)$ to \eqref{eq:Sunequal}. The left losses are of types $(w,X)$ in the group of $X$, of size
$d$; left losses of total length $\lambda$ of one type, at any heights in $(0,r]$, contribute at least
$\int_{r-\lambda}^r(r-u)/(4r^2)du=\lambda^2/(8r^2)$, so together at least $(\ell^{\rm L})^2/(8dr^2)$: by Cauchy--Schwarz,
losses $\lambda_1,\dots,\lambda_d$ of the at most $d$ types of the group satisfy $\sum_i\lambda_i^2\ge(\sum_i\lambda_i)^2/d$. Suppose $r\le5$. Then $\ell^{\rm C}\le r\le5$ gives $\ell^{\rm C}\le20S_j$; and as
$2r-\ell\ge r-\ell^{\rm L}-\ell^{\rm C}$, the first term of \eqref{eq:Sunequal} and the centre give at least
$\frac14\bigl((r-\ell^{\rm L}-\ell^{\rm C})_++\min(\ell^{\rm C},1)\bigr)\ge\frac1{20}(r-\ell^{\rm L})$. Adding the left part, which concerns other types, with $y=\ell^{\rm L}/r$,
$S_j\ge\ell^{\rm L}\bigl[\frac{1-y}{20}+\frac y{8dr}\bigr]\ge\ell^{\rm L}/360$ if $d\le9$. If $r>5$, every lost cut at a
height at most $4$ contributes at least $\Phi'(5)-\Phi'(4)=\frac1{80}$, so there are at most $80S_j$ of them. In every
case, the centre losses and the left losses in groups that are not exceptional, at heights at most $4$, number at most
$380S_j$.

\emph{Counting.} At most $k$ designated cuts are intact at the end, so $W\le k+\sum_j\ell_j$. Every lost cut is
counted at least once below.
\begin{itemize}
\item Cuts of types in exceptional groups, and of types $(x,x)$ in other groups, at heights at most $4$: at most $4$ per
type, so at most $600\sum_xS_x$ (as $4d\le600\cdot d/150$ and $4<600\chi$).
\item Right sides at unequal steps, at most $r_j=o_j$, and right sides of even runs at self steps, at most
$\eta_j\le o_j$: at most $O_k$.
\item Left sides at the self step of a group that is not exceptional: at most $o\le d_x/3$, and each group has at most
one self step: at most $k/3$.
\item Centres and left sides at unequal steps not counted above, at heights at most $4$: at most $380\sum_jS_j$.
\item Cuts at heights above $4$: at most $30J$.
\end{itemize}
Internal cuts of runs have type $(x,x)$, and steps on compound symbols lose nothing, so this covers every lost cut. Hence
$W-k-O_k\le\frac k3+600S<\frac k3+\frac k{10}=\frac{13}{30}k$. But $W-k-O_k>km-k-\frac25km=(\frac35m-1)k>\frac{11}{25}k$,
and $\frac{11}{25}>\frac{13}{30}$, a contradiction.
\end{proof}

The class of comparators covered by the hypothesis of the first statement includes, besides those with tokens of at most
three letters, every set of disjoint original pairs in at most $k$ types. The only replacements that the general form
charges are those that consume two letters strictly inside one token of the comparator: on five separated copies of
\str{abcd} with $k=3$, a legal run takes \str{bc} first and makes all $15$ junctions of the optimum
$\str{ab},\str{cd},(\str{ab})(\str{cd})$ non-intact with a gain of $5$, although greedy still attains the optimum there.

\section{Two further worked examples}\label{app:perstep}

\subsection{One replacement can remove three witnesses}\label{app:threewit}
Take $s=\str{abaaa}$, $k=5$, and the optimal partial sequence that builds the prefixes $\str{ab}$, $\str{aba}$,
$\str{abaa}$ and $\str{abaaa}$, one node each, $U=4=n-1$. In $s$ every pair has count $1$, so greedy may merge $\str{aa}$, then
$\tok{aa}\str a$, then $\str{ab}$, then the two remaining tokens; each step saves one, and every greedy run here saves
$4=\OPT(s,5)$. Along this run:
\begin{center}\small
\begin{tabular}{@{}cllcc@{}}
\toprule
$i$ & tokens of $s_i$ & nodes with a witness & $\sum_rm_r$ & $2G_i-(U-\sum_rm_r)$\\
\midrule
$0$ & $\str a\,\str b\,\str a\,\str a\,\str a$ & all four, pair $(\str a,\str b)$ & $4$ & $0$\\
$1$ & $\str a\,\str b\,\tok{aa}\,\str a$ & all four, pair $(\str a,\str b)$ & $4$ & $2$\\
$2$ & $\str a\,\str b\,\tok{aaa}$ & all four, pair $(\str a,\str b)$ & $4$ & $4$\\
$3$ & $\tok{ab}\,\tok{aaa}$ & only the root & $1$ & $3$\\
$4$ & $\tok{abaaa}$ & none & $0$ & $4$\\
\bottomrule
\end{tabular}
\end{center}
At step $3$ one replacement, the merge of $\str{ab}$, removes three witnesses and adds only $2$ to $2G$, so the slack of
\eqref{eq:pointwise} drops from $4$ to $3$. The inequality survives on slack accumulated at steps $1$ and $2$. So the
natural charging, two units per replacement for the witnesses it removes, fails; this is why our proof does not follow
the run.

\paragraph{The curvature of Zouhar et al.} The total backward curvature $\sigma_{\mathrm Z}$ of \cite{zouhar2023formal} can exceed $2$:
with their Definition~3.9 as we read it, on $\str{abab}$ with budget $2$ the optimal sequence
$\langle[\str a,\str b],[[\str a,\str b],[\str a,\str b]]\rangle$ and the prefix $\langle[\str b,\str a]\rangle$ give
$\sigma_{\mathrm Z}\ge3$, although greedy is optimal there. So \cref{thm:main} is not a bound on $\sigma_{\mathrm Z}$.

\section{Computational evidence}\label{app:evidence}

No proof in this paper rests on a computer search. The computer enters through exact arithmetic only: rational
enclosures of the constants, the explicit numbers of the $k=\UpK$ certificates and of the run of \cref{rem:m1}, and the finite sign check behind the
constant $\lambda_*$ of \cref{prop:overlap}(a) (\cref{app:parta}; its smaller constant $\lambda_{\rm h}$ rests on a finite calculation that can be checked by hand,
and the same program checks it too). The checks below are evidence, not part of the proofs;
they were run by programs written separately from the proofs, in exact arithmetic (\texttt{make\_numbers.py} also
re-evaluates transcendental comparisons at $60$ digits, as a further check), and all numbers in this
appendix are generated from the recorded output files.

\paragraph{Released scripts and how to run them.} The supplement contains the folder \texttt{scripts/} of this paper.
Each script is pure Python~3 (\texttt{make\_numbers.py} also uses \texttt{mpmath}), runs in seconds on a laptop, and
exits with status~$1$ if any of its pre-registered checks fails; \texttt{bash build.sh} runs them all, stops at the first
failure, and then compiles the paper.
\begin{itemize}
\item \texttt{python3 scripts/examples.py} recomputes, in exact integer arithmetic, every number of the examples of
\cref{sec:labz}, \cref{app:threewit} and \cref{fig:shadow}, the isolation examples after \cref{lem:iso} and the
curvature example of \cref{app:threewit}, including the optimum of the first two by exhaustive search.
\item \texttt{python3 scripts/editorial/check\_figures.py} recomputes the two-merge example of \cref{sec:k2} as drawn in
\cref{fig:k2} (the optimum by exhaustive search, both greedy runs, junctions, windows, marks and $m_r$), the facts shown
in \cref{fig:overview}(b) and (c), the equality instance $\str{aabaab}$ of \cref{sec:overview} (its optimum by
exhaustive search), the constants quoted in \cref{sec:intro,sec:overview}, and the plotted curve of
\cref{fig:overview}(a): every plotted point is $(\log_{10}k,c_k)$ for an integer $k$, against its own exact $c_k$, and the
plotted values strictly decrease.
\item \texttt{python3 scripts/collisions.py} recomputes, by exhaustive search in both models, the examples of
\cref{sec:identity} and \cref{app:identity} on string-identified symbols.
\item \texttt{cd scripts/overlap\_cert; python3 certificate.py} verifies, in exact rational arithmetic, the two
certificates of \cref{prop:overlap}(a), and \texttt{python3 relaxation.py} checks the seven-point relaxation; what they
check, and the SHA-256 hashes of the folder, are given in \cref{app:parta}.
\item \texttt{python3 scripts/make\_numbers.py} reads the recorded outputs of the checks described below, compares the
outputs of separately written programs where both exist, checks the scalar comparisons quoted in the proofs (rational
ones exactly, transcendental ones at $60$ digits with a margin far above the working precision), and writes
\texttt{numbers.tex}, from which the paper takes every computed number.
\item \texttt{python3 scripts/make\_supplement.py} builds the supplement and verifies it from a clean unzip: every file
against \texttt{SHA256SUMS}, \texttt{make\_numbers.py} regenerating \texttt{numbers.tex} byte for byte, and the scripts
above reproducing their recorded outputs.
\end{itemize}
The supplement also contains the larger, separately written exhaustive checkers behind the tables below, each next to
its recorded outputs and under the folder name it has in our repository (its \texttt{README.md} lists them by family);
for each family below we say what the program checks.
\paragraph{The pointwise inequality and the lemmas.} Each instance is a string, a partial merge sequence $\sigma$ and a
sequence of full merges. The program checks \cref{thm:pointwise}, \cref{lem:lost}, both parts of \cref{lem:windows}, and the
three steps of the proof of \cref{thm:pointwise}: at least $P-2L$ kept windows, identical actual tokens in the kept windows
of each label, and at most $m_r$ kept windows per label. Exhaustive rows cover every string up to renaming of the letters,
every partial merge sequence with the stated number of rules, and every full-merge prefix with the stated number of rules
(over present pairs and over all letter pairs). Planted instances repeat a few random words with one random tree per word;
the hard instances are four strings with repeated blocks, built to stress the inequality.
\begin{center}\small
\begin{tabular}{@{}p{8.2cm}rrr@{}}
\toprule
Family & Instances & Violations & Equality cases\\
\midrule
\TabPointwise
\bottomrule
\end{tabular}
\end{center}
Equality in \cref{thm:pointwise} occurs in every family. Two control runs show that the test has teeth. With partial
merges allowed in the prefix (\cref{sec:labz}), the inequality failed in \CtrlPartialHits{} of \CtrlPartialN{} random
instances. With $2G-1$ in place of $2G$ it failed in \CtrlTightHits{} of \CtrlTightN{} planted instances; these have
equality with $G>0$, so the factor $2$ cannot be lowered.

\paragraph{The ratio and the shadow inequality.} For every string up to renaming over the stated alphabet and length,
every budget in the stated range and every greedy tie-break, the program computes $\OPT(s,k)$ exactly and checks
$\BPE(s,k)\ge c_k\OPT(s,k)$ and $\OPT(s,k)\le2G_i+kg_{i+1}$ for every $0\le i\le k$. Strings are over at most the stated number of
letters. The last columns give the smallest ratio $\BPE/\OPT$ found at each budget.
\begin{center}\small
\setlength{\tabcolsep}{4pt}
\begin{tabular}{@{}lrrcrr*{\RatioCols}{c}@{}}
\toprule
Letters & Length & Strings & Budgets & Ratio viol. & RC viol. & \RatioKhead\\
\midrule
\TabRatio
\bottomrule
\end{tabular}
\end{center}
The smallest ratio at $k=2$ is $\RatioMinTwo$, the exact worst case at that budget, well above $c_2=1/2$. For $k\le6$
every ratio is at least $c_6=665/1458>0.456$ by \cref{thm:main}; the constructions that approach the limit use budgets of
a hundred or more (\cref{sec:upper}).

\paragraph{The matching construction.} The construction's program builds the literal string, follows every greedy
history (merging states reached by different histories), and replays the side merges; for $k=\UpK$ it works with exact
block multiplicities and checks the maximum count at every step against the guard cap. A separately written program
rebuilt every string from the description in \cref{app:upper} alone, matched all SHA-256 hashes, and repeated each
count; at $k=\UpK$ it also checked the ingredients of the exact optimum. The two programs agree on every entry below.
The column $G/\Uside$ is the ratio against the side merges, an upper bound on the ratio for the small cases (whose exact
optimum was not computed) and the exact ratio at $k=\UpK$.
\begin{center}\small
\setlength{\tabcolsep}{4pt}
\begin{tabular}{@{}rrrrrrrrr@{}}
\toprule
$d$ & $F$ & $k$ & Length & Letters & Histories & $G$ & $\Uside$ & $G/\Uside$\\
\midrule
\TabUpper
\bottomrule
\end{tabular}
\end{center}

\paragraph{The finite-budget theorem.} The proof of \cref{thm:rate} is by hand; the programs below only test it. A
position-level checker, written from the definitions of \cref{sec:proof} and separately from the proof, replays Steps
1--3 of \cref{app:rate} on random strings with random greedy ties and arbitrary partial merge sequences (not only
optimal ones), together with hand-built instances that force composite operands and the error term of
\eqref{eq:interval}. Over $\RateInst$ instances (in $\RateFlat$ of them every greedy merge joins two letters), $\RateIntervals$ intervals
with $v>0$ and $\RateCharges$ charged deaths, it found no violation of \eqref{eq:inventory}, \eqref{eq:charges},
\eqref{eq:interval}, of the bounds $n_r\le u$ and at most $v$ live intervals at $i+\ell$, or of \eqref{eq:slack}. Each
of $\RateMutants$ weakened variants of the proof (no collapse term, no inventory for created symbols, collapses bounded
by $\cnt$ instead of $\raw$, \eqref{eq:interval} without its error term, and $n_r\le v$) fails on every seed. A second
script checks the constants of Steps~3--5 in exact rational arithmetic at five budgets from $4\cdot10^{10}$ to above
$10^{24}$, the small-budget branch for every $k\le3000$, and the upper bound $8k^{-1/2}$ of \cref{cor:rate} exactly at
$k=2,\dots,63$ and at three budgets in each interval $[d^2,(d+1)^2)$ for $d=8,\dots,60$, $100$ and $200$. The run of \cref{rem:m1} is replayed
from the stored integers by \texttt{audit/finite-rate/m1\_replay.py}, which also rejects five altered versions of the
data; the gains, the optimum and the ratio quoted in \cref{rem:m1} are generated from its output and from the stored
data.

\paragraph{Overlap greedy.} Two separately written programs enumerated every tie path of the string of
\cref{app:fourphase} at $(d,H,h)=(4,6,3)$ and $(5,6,5)$: the minimum and maximum savings are $64{,}280$ and
$4{,}816{,}110$, equal to $kFD_d+k+n_q$, and the side merges save $kF$. One of them also enumerated eight further cases
with $d\le6$, with the same outcome, and followed random tie paths for $4\le d\le16$, including cases with two
second-series levels, checking at every step that the maximisers are exactly the current group. The controls fail as they should:
without the bonus blocks the savings of tie paths range up to $kF$, and with one $qq$ bonus block instead of two the
per-step check fails. A separately written program recomputed the limit of the family. For \cref{prop:overlap}(a) a
separately written verifier, with its own derivation of the costs in \cref{tab:costsSVC,tab:costsHJ}, a finer set of
$\OvAudBreaks$ breakpoints and exact sign decisions by real-root isolation, confirmed both certificates and the
seven-point relaxation, and found that raising all weights by $0.5\%$ breaks both certificates.

\paragraph{String-identified symbols.} The checker of \cref{prop:identity} implements both models separately and compares
the reachable final partitions and the exact optima; over $\SidInputs$ inputs (binary strings of length at most $9$ and strings over three letters of length at most $7$ with
budgets up to $4$, and pairs of binary rows with budgets up to $3$) it found them equal, and $\SidShadow$ checks of the
shadow inequality and $\SidRatio$ checks of the ratio of \cref{thm:main} over every tie in the string-identified model
found no violation. A second, separately written program, with both models implemented again, agreed. It also searched
$\NonUDInputs$ small inputs that are \emph{not} uniquely decodable (one or two rows over sets of atoms of one to three
characters, weights $1$ to $3$, budgets up to $3$, every tie): the shadow inequality failed in $\NonUDShadowViol$ of
$\NonUDShadowChecks$ checks, but the ratio of \cref{thm:main} held in all $\NonUDRatioChecks$ checks, with smallest ratio
$\NonUDMinRatio$. The examples of \cref{sec:identity} are recomputed by exhaustive search by a short script.

\paragraph{Word tables.} Two programs, written separately, work directly on weighted rows. They compute the exact
partial-merge optimum by branching over every pair and every set of disjoint occurrences in every row, and they follow
every greedy tie history under both ranking rules. The first covered $\WtGridBN$ (table, budget) instances made of two
binary words of lengths $1$ to $4$, the second $\WtGridN$ instances, including non-integer weights and random ternary
tables, and also checked the separator encoding on $\WtEncN$ integer instances. No instance violated
\cref{thm:table}(a) or the bound of \cref{prop:overlap}(a). The smallest ratios found at $k=1,2,3$ were \WtMinD{} for
replacement counts and \WtMinR{} for overlapping counts. The second program replayed the table of \cref{thm:table}(c)
along a run, checking the set of maximisers at every step.

\paragraph{The examples.} The scripts \texttt{examples.py} and \texttt{editorial/check\_figures.py}, separate from the
programs above and from each other, recompute every number in \cref{sec:warmup}, \cref{app:threewit}, \cref{fig:shadow,fig:k2,fig:overview}
and the examples after \cref{lem:iso}, including the optima by exhaustive search; the build of this paper stops if any
of them differs from the text.

\section{Fidelity to deployed trainers}\label{app:fidelity}

\begin{sloppypar}
\paragraph{Counting and replacing.} subword-nmt \cite[Algorithm~1]{sennrich2016bpe} and Hugging Face rank a pair by its
occurrences counted with overlaps and then replace from left to right (\texttt{get\_pair\_statistics} and
\texttt{replace\_pair} in \cite{subwordnmt}; \texttt{count\_pairs} and \texttt{Word::merge} in \cite{hftokenizers}). Our
model recounts the pairs exactly after every merge, while subword-nmt updates its counts incrementally
(\texttt{update\_pair\_statistics}, with special cases for overlapping patterns such as \str{ABCBC}); by our reading of
the code, on uniquely decodable inputs these updates reproduce the exact overlapping counts of every pair, and the pair
it selects after pruning small counts is a maximiser of them; we have not tested this by a program.

\paragraph{Vocabulary-size budgets.} Hugging Face's budget is a vocabulary size, so a merge whose output string is
already in the vocabulary costs nothing. On uniquely decodable inputs whose text spells no special token's string
(special tokens are outside the model) this changes nothing: no step of a full-merge history produces a spelling that
already exists (\cref{app:identity}), so a greedy run that adds $k$ entries makes $k$ merges; and the proof of
\cref{prop:identity}(b) uses only that the rules produce at most $k$ distinct spellings, so the best partial merges that
add at most $k$ entries save exactly $\OPT$.

\paragraph{Options outside the model.} Hugging Face's option \texttt{continuing\_subword\_prefix}, off by default,
strips a prefix from the right operand, so the merged symbol does not spell the concatenation of its operands; it is
outside the model, as are \texttt{max\_token\_length}, \texttt{limit\_alphabet} (which changes the input) and boundary
policies other than ``no merge across a row'' (SentencePiece's BPE, for example).
\end{sloppypar}

\end{document}